\documentclass[journal]{IEEEtran}
\usepackage{amsmath,amsfonts}
\usepackage{algorithmic}
\usepackage{array}
\usepackage{textcomp}
\usepackage{url}

\usepackage{verbatim}
\usepackage{graphicx}
\def\BibTeX{{\rm B\kern-.05em{\sc i\kern-.025em b}\kern-.08em
    T\kern-.1667em\lower.7ex\hbox{E}\kern-.125emX}}
\usepackage{balance}

\usepackage{amssymb,amsmath,amsthm,enumerate,threeparttable,bm,graphicx,hyperref,subfigure}
\usepackage{algorithm}
\usepackage{caption}
\usepackage{algorithmic}
\usepackage{booktabs}
\usepackage{multirow}
\usepackage{threeparttable}
\newtheorem{lemma}{Lemma}
\newtheorem{theorem}{Theorem}

\newtheorem{definition}{Definition}

\newtheorem{proposition}{Proposition}
\newtheorem{remark}{Remark}

\usepackage{color}

\makeatletter
\newcommand{\triangledComment}[1]{%
  \hfill$\triangleright$\ \textit{#1}%
}
\makeatother

\long\def\comment#1{}

\def\ie{$i.e.$}
\def\eg{$e.g.$}
\def\etal{\textit{et al.} }

\usepackage{tikz}

\begin{document}

\title{Cert-SSBD: Certified Backdoor Defense with Sample-Specific Smoothing Noises}

\author{Ting~Qiao,
        Yingjia~Wang,
        Xing Liu,
        Sixing~Wu,
        Jianbin~Li,
        and Yiming~Li
 \thanks{This work was supported by National Key Research and Development Plan of China (No. 2022YFB3103304), Beijing Natural Science Foundation (No. L251061), and the Industrial Foundation Reengineering and High-Quality Manufacturing Development Project (No. ZC25T320057/100).}       
\thanks{Ting~Qiao, Yingjia~Wang, Sixing~Wu and Jianbin~Li are with School of Control and Computer Engineering, North China Electric Power University, Beijing 102206, China (e-mail: qiaoting@ncepu.edu.cn, wyj@ncepu.edu.cn, wusx@ncepu.edu.cn, lijb87@ncepu.edu.cn).}%
\thanks{Xing Liu is with Research Institute, China Unicom, Beijing 100048, China, and also with National Engineering Research Center of Next Generation Internet Broadband Service Application, Beijing 100037, China (e-mail: liux737@chinaunicom.cn).}
\thanks{Yiming~Li is with College of Computing and Data Science, Nanyang Technological University, Singapore 639798 (e-mail: liyiming.tech@gmail.com).}
\thanks{Corresponding Author(s): Jianbin~Li and Yiming~Li.}
}

\markboth{IEEE Transactions on Information Forensics and Security}%
{IEEE Transactions on Information Forensics and Security}


\maketitle
\begin{abstract}

Deep neural networks (DNNs) are vulnerable to backdoor attacks, where an attacker manipulates a small portion of the training data to implant hidden backdoors into the model. The compromised model behaves normally on clean samples but misclassifies backdoored samples into the attacker-specified target class, posing a significant threat to real-world DNN applications. Currently, several empirical defense methods have been proposed to mitigate backdoor attacks, but they are often bypassed by more advanced backdoor techniques. In contrast, certified defenses based on randomized smoothing have shown promise by adding random noise to training and testing samples to counteract backdoor attacks. In this paper, we reveal that existing randomized smoothing defenses implicitly assume that all samples are equidistant from the decision boundary. However, it may not hold in practice, leading to suboptimal certification performance. To address this issue, we propose a certified backdoor defense method with sample-specific smoothing noises, termed Cert-SSBD. Cert-SSBD first employs stochastic gradient ascent to optimize the noise magnitude for each sample, ensuring a sample-specific noise level that is then applied to multiple poisoned training sets to retrain several smoothed models. After that, Cert-SSBD aggregates the predictions of multiple smoothed models to generate the final robust prediction. In particular, in this case, existing certification methods become inapplicable since the optimized noise varies across different samples. To conquer this challenge, we introduce a storage-update-based certification method, which dynamically adjusts each sample’s certification region to improve certification performance. We conduct extensive experiments on multiple benchmark datasets, demonstrating the effectiveness of our proposed method. Our code is available at \url{https://github.com/NcepuQiaoTing/Cert-SSBD}.

\end{abstract}


\section{Introduction}

\begin{figure}[ht]
    \centering
    \includegraphics[width=0.45\textwidth]{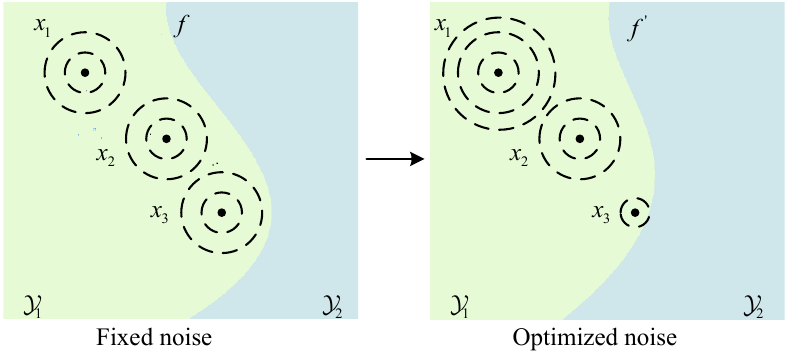}
    \vspace{-0.5em}
    \caption{An overview of existing randomized smoothing-based certified backdoor defenses and our Cert-SSBD, providing an abstract illustration of the input space and decision boundaries. Existing methods apply fixed noise to smooth classifiers for all inputs, ignoring sample diversity, which often leads to suboptimal certification performance. In contrast, Cert-SSBD optimizes the noise, enabling the smoothing strategy to adapt to different inputs (as shown in the right figure), thereby achieving more robust certified backdoor defenses.}
    \label{fig:introduction}
    \vspace{-1em}
\end{figure}

\IEEEPARstart{R}{ecently}, deep neural networks (DNNs) have been widely and successfully adopted in various domains, including mission-critical applications, such as face recognition \cite{yang2021larnet,deng2019mutual,luo2022memory}. However, training high-performance models typically requires large amounts of data and computational resources, which can be costly. Consequently, researchers often rely on third-party resources, such as publicly available datasets, cloud computing platforms, and pre-trained models, to reduce the training burden. Arguably, this reliance introduces security risks, with backdoor attacks \cite{fan2021text,li2022backdoor,cai2024toward, Zhang2024BadCM} being among the most severe threats. In a backdoor attack, adversaries inject predefined trigger patterns into a subset of the training data, causing the model to misclassify any input containing the trigger according to the attacker's intent. These attacks are both stealthy and highly detrimental, making them a key concern in both academia and industry. An industry report \cite{kumar2020adversarial} highlights that backdoor attacks rank as the fourth most significant security threat faced by enterprises. Government agencies also recognize the severity of this issue. For instance, the U.S. intelligence community \cite{RN16} has launched a dedicated funding program to counter backdoor attacks and related threats. To prevent models from becoming compromised due to backdoor attacks, developing effective defense mechanisms has become an urgent priority.

\looseness=-1
To mitigate backdoor threats, researchers have proposed a wide range of backdoor defense strategies, including backdoor detection \cite{xiang2023umd,xiang2020revealing,xiang2023cbd} and mitigation-based approaches \cite{qiu2023towards,yi2025probe,chen2025refine}.  However, advanced backdoor attacks \cite{Li2021Invisible,nguyen2021wanet,duan2024conditional} can still easily bypass existing defenses, leading to an ongoing arms race between defenders and attackers. To address this issue, some studies have proposed certified backdoor defense methods, primarily categorized into deterministic certification \cite{levine2020deep,wang2022improved,rezaei2023run,zhang2023pecan} and probabilistic certification \cite{wang2020certifying, Weber2023RAB}. These methods aim to provide theoretical guarantees, ensuring that the classification results of testing samples remain consistent regardless of whether the model is trained on clean or backdoor data, as long as the perturbation induced by the trigger remains within an $\ell_p$ norm ball of radius $r$. However, deterministic methods face scalability challenges when applied to large-scale neural networks. Consequently, probabilistic certification approaches based on randomized smoothing have emerged as a more practical alternative and have demonstrated robustness on large-scale datasets such as ImageNet \cite{Deng2009ImageNet}. Randomized smoothing was initially developed to certify robustness against adversarial examples. Its principle is to introduce random noise into the input data, ensuring that the classification results remain consistent within a specified region (\eg, an $\ell_p$ norm neighborhood), thereby achieving robustness. Notably, pioneering studies \cite{wang2020certifying, Weber2023RAB} showed that certified backdoor defenses based on random smoothing, which are robust against bounded backdoor patterns (\ie, constrained pixel-level perturbation), can be achieved by introducing isotropic Gaussian noise into a tuple consisting of a testing instance and the training set to mitigate the impact of attacker-injected triggers, effectively neutralizing backdoor attacks during the training phase.

In this paper, we revisit existing randomized smoothing-based certified backdoor defenses. We find that these methods typically apply a fixed (\ie, identical) magnitude of Gaussian noise to each sample to smooth the base classifier (\ie, the decision boundary), thereby producing the final robust predictions. In other words, this approach (implicitly) assumes that all samples are equidistant from the decision boundary. However, inspired by \cite{gao2023not}, we recognize that this assumption may not hold in practice and could even degrade defense performance, as it may not be optimal for every sample. For example, as shown in the left part of Figure \ref{fig:introduction}, adding an overly large noise magnitude to samples near the decision boundary can lead to misclassification, whereas increasing the noise magnitude for samples farther from the decision boundary can potentially enhance their certification performance. Based on this observation, we further analyze the intrinsic characteristics of samples, particularly their distances to the decision boundary. We find that these distances vary significantly among samples, and regardless of whether they belong to the training or testing set, their certification radius under a fixed noise magnitude is influenced by their individual properties. Therefore, an ideal strategy should be: applying smaller noise to samples near the decision boundary while assigning larger noise to those farther away, thereby better balancing classification performance and robustness, as illustrated in the right part of Figure \ref{fig:introduction}. This finding raises a key question: \emph{How can we exploit the intrinsic properties of samples to adjust the noise magnitude for each sample to design more effective certified backdoor defenses?} 

Fortunately, the answer to the above question is affirmative. Arguably, the most direct approach is to optimize the noise at each sample by maximizing the confidence margin between the top-1 and top-2 predicted classes of the classifier (\ie, the certification radius). However, the certification radius does not admit a closed-form analytical expression, which renders direct analytical optimization or deterministic gradient-based methods inapplicable. To overcome this limitation, we optimize a Monte Carlo–estimable surrogate objective that is tightly coupled with the certification radius. Inspired by the approach of \cite{zhai2020macer}, we adopt stochastic gradient ascent to optimize this surrogate, enabling the learning of an optimal noise level on a per-sample basis. Nevertheless, dynamically adjusting the noise during optimization inevitably alters the underlying data distribution, which increases the variance of gradient estimates and undermines optimization stability. To address this issue, we propose an advanced certified backdoor defense method with sample-specific smoothing noises, termed Cert-SSBD. In general, Cert-SSBD consists of two main stages: training and inference. In the first stage, we train multiple smoothed models using the optimized noise, which is obtained through stochastic gradient ascent to maximize the certification radius. Generally, the certification radius is computed based on the predictions of classifiers trained with fixed noise. Besides, we adopt a reparameterization technique to reduce gradient variance and enhance optimization stability. In the inference stage,  we aggregate multiple smoothed classifiers trained in the first stage to generate the final smoothed prediction. However, since the optimized noise results in different noise magnitudes for each sample, existing certification methods, which typically assume a fixed noise level, are no longer directly applicable. To resolve this issue, we propose a storage-update-based certification method, which dynamically adjusts the certification region (\ie, the space covered by the certification radius) for each sample. This ensures that certification regions do not overlap between different samples and that predictions remain consistent within each certified region.

\vspace{0.3em}

Our main contributions can be summarized as follows:

\begin{itemize}
    \item We revisit existing randomized smoothing-based certified backdoor defenses and reveal that their use of fixed noise results in suboptimal certification performance for samples, affecting both training and testing samples.
    
    \item We propose a sample-specific certified backdoor defense method (\ie, Cert-SSBD) to dynamically adjust the smoothing noise magnitude for different samples to optimize certification performance.
    
    \item We introduce a storage-update-based certification method to dynamically update each sample's certification region, ensuring non-overlapping certified regions across different samples and improving certification robustness.
    
    \item We conduct extensive experiments on benchmark datasets to validate Cert-SSBD's effectiveness, demonstrating its superior certification performance over existing methods.
\end{itemize}


\section{Related Works}
\label{sec:related_works}

\subsection{Backdoor  Attacks}
Backdoor attacks \cite{li2022Backdoors,ding2023backdoor,gao2023imperceptible,gao2024backdoor} represent an emerging threat during the DNN training phase. Based on the adversarial objective, they can be categorized into two types: \emph{all-to-one attacks}, which misclassify all triggered samples into a single fixed target label and are relatively straightforward; and \emph{all-to-all attacks}, which map samples to specific target classes based on their original categories, making them more complex.
Besides, backdoor attacks can also be categorized based on the threat scenario into three major types: \textbf{(1)} poison-only attacks \cite{Gu2017BadNets,Li2021Invisible}, \textbf{(2)} training-controlled attacks \cite{li2020invisible,li2022few}, and \textbf{(3)} model-modified attacks \cite{tang2020embarrassingly, bai2022hardly}. Specifically, poison-only attacks restrict the adversary to modifying the training dataset; training-controlled attacks allow the adversary to fully control the training process, including both the training data and algorithms. In contrast to these approaches, model-modified attacks mainly focus on the deployment phase rather than the training phase, embedding hidden backdoors by directly modifying model weights or introducing malicious DNN modules. In this paper, we mainly focus on poison-only backdoor attacks, which represent the most classical setting and pose the broadest threat scenarios. Recently, there are also a few works exploring how to exploit backdoor attacks for positive purposes \cite{li2023black,guo2023domain,wei2024pointncbw,li2025reliable,li2025move,shao2025explanation}, which is out of the scope of this paper.

\subsection{Backdoor Defense}

 In general, existing backdoor defense methods can be categorized into empirical defenses \cite{yi2025probe,chen2025refine}, which rely on heuristic approaches to counter specific types of attacks, and certified defenses \cite{zhang2023pecan,Weber2023RAB}, which provide theoretical guarantees for classifier robustness against adversarial perturbations.

\subsubsection{Empirical defenses}  
Existing empirical defense methods can be classified into five main categories: \textbf{(1)} the detection of poisoned training samples \cite{wang2019neural,hayase2021spectre}, \textbf{(2)} poison suppression \cite{li2021anti,tang2023setting}, \textbf{(3)} backdoor removal \cite{liu2018fine-pruning,li2024nearest}, \textbf{(4)} the detection of poisoned testing samples \cite{chou2020sentinet,hou2024ibd}, and \textbf{(5)} the detection of attacked models \cite{xu2021detecting,xu2024towards}. Specifically, the detection of poisoned training samples aims to identify and filter out malicious samples from the training set. Poison suppression prevents the model from learning poisoned samples by modifying the training process, thereby inhibiting the formation of hidden backdoors. Backdoor removal focuses on eliminating hidden backdoors from pre-trained (third-party) models. Detection of poisoned testing samples is designed to identify and block poisoned inputs during the testing phase. Lastly, the detection of attacked models determines whether a given model has been compromised by analyzing certain model properties. However, \cite{Koh2022Stronger}  and \cite{Wang2020Attack} revealed that new attack strategies could circumvent these empirical defenses, highlighting the ongoing arms race between attack and defense techniques. 

\subsubsection{Certified defenses} 
\label{Cert:defense}

 \looseness=-1

Existing certified defense methods can be categorized into \emph{deterministic defenses} \cite{levine2020deep,wang2022improved,rezaei2023run,zhang2023pecan}, which provide guaranteed outcomes but face scalability issues, and \emph{probabilistic defenses} \cite{wang2020certifying,Weber2023RAB}, ensure a `certified' result with a certain probability (\eg, 99.9\%), where the randomness is independent of the input sample. In this work, we focus on probabilistic certified defenses.

\looseness=-1
Probabilistic certification offers better scalability. Previous methods primarily relied on intrinsic mechanisms \cite{jia2021intrinsic, jia2022certified} or randomized smoothing techniques \cite{wang2020certifying, Weber2023RAB} to achieve robust predictions. For example, Jia \etal \cite{jia2021intrinsic, jia2022certified} leveraged ensemble techniques in bagging or majority voting in $k$-nearest, but these remain unsuitable for backdoor defense as they do not consider trigger size. Subsequently, Wang \etal \cite{wang2020certifying} first applied randomized smoothing to backdoor defense, yet the approach lacked comprehensive evaluation and high robustness bounds. Recently, the RAB framework \cite{Weber2023RAB} established a theoretical foundation for provable defenses in this field (see Section \ref{sec:random smoothing} for more details). However, current methods implicitly assume that all samples are equidistant from the decision boundary, which may not hold in practice. This leads to suboptimal certification and highlights the urgent need for adaptive approaches that account for sample-specific characteristics.

\subsection{ Randomized Smoothing and RAB}
\label{sec:random smoothing}
Randomized Smoothing (RS) \cite{Cohen2019Certified} is a probabilistic defense method that enhances classifier robustness by smoothing predictions. Specifically, given an input $\bm{x}$, the smoothed classifier $g(\bm{x},\sigma)$ selects the most probable class predicted by the base classifier $f$ under isotropic Gaussian noise. Formally:
\begin{equation}
\label{eq:RS_base}
			g(\bm{x},\sigma) \triangleq \arg\max_{y \in \mathcal{Y}} \mathcal{P}_{\bm{\epsilon }}(f(\bm{x}+\bm{\epsilon })=y), 
	\end{equation} 
where $\bm{\epsilon } \sim \mathcal{N}(0,\sigma^2 I)$. The noise level $\sigma$ is a hyperparameter that controls the trade-off between robustness and accuracy; it does not change with the input $\bm{x}$. Using the Neyman-Pearson lemma \cite{neyman1933on}, Cohen \textit{et al.} \cite{Cohen2019Certified} proved that $g(\bm{x},\sigma)$ is certifiably robust to adversarial perturbations under the $\ell_2$ norm constraint. Define $y_A=\arg\max_y \mathcal{P}_{\bm{\epsilon }}(f(\bm{x}+\bm{\epsilon })=y)$, and assume that when classifying a perturbed input $ \bm{x}+\bm{\epsilon } $, the base classifier $f$ assigns the most probable class $y_A$ with probability $P_A=\mathcal{P}_{\bm{\epsilon }}(f(\bm{x} + \bm{\epsilon }) = y_A)$, and the second most probable class $y_B$ with probability $P_B=\max_{y_B \neq y_A} \mathcal{P}_{\bm{\epsilon }}(f(\bm{x} + \bm{\epsilon }) = y)$. Then, it is always true that $g(\bm{x}+\bm{\Delta},\sigma)=y_A$ as long as $\|\bm{\Delta}\|_2<r$, where the certified robust radius $r$ is given by:
\begin{equation}
\label{eq:RS_radius}
			r(\bm{x},\sigma) \triangleq \frac{\sigma }{2} \left ( \Phi ^{-1}(P_A (\bm{x},\sigma) )-\Phi ^{-1}(P_B(\bm{x},\sigma)  ) \right ),
	\end{equation}
where $\Phi ^{-1}$ represents the inverse Gaussian cumulative distribution function (CDF).

In general, RS techniques are primarily designed to certify adversarial robustness by adding noise to testing instances. Most recently, a few pioneering research \cite{wang2020certifying,Weber2023RAB} showed that we can achieve certified backdoor defenses that are robust against bounded backdoor patterns by introducing isotropic Gaussian noise to a tuple consisting of a testing instance and the training set to neutralize backdoor effects. Among these approaches, the most notable is RAB \cite{Weber2023RAB}. In the following, we briefly describe the implementation details of RAB.

\vspace{0.1em}
\noindent \textbf{Overview of RAB \cite{Weber2023RAB}.} Given a dataset $\mathcal{D}$ and a testing instance $\bm{x}$, the base classifier $f$ induces a predictive distribution over class labels under random perturbations: $p_f(y | \bm{x}, \mathcal{D})
\triangleq \mathcal{P}_{\bm{\epsilon}}\big(f(\bm{x}, \mathcal{D}) = y\big)$. The predicted label is then given by $f(\bm{x}, \mathcal{D}) \triangleq \arg\max_y p_f(y | \bm{x}, \mathcal{D})$.
A smoothed classifier $g(\bm{x},\mathcal{D},\sigma)$ returns whichever class the base classifier $f(\bm{x},\mathcal{D})$ is most likely to predict when $\bm{x}$ is perturbed by smoothing distributions $X=(Z,D)$:
\begin{equation}\label{eq:f-g}
			g(\bm{x},\mathcal{D}, \sigma)= \arg\max_y \mathcal{P}_{\bm{\epsilon }(Z,D)}(f(\bm{x}+Z , \mathcal{D}+D)=y ),
	\end{equation}
where $Z\sim \mathcal{N}(0,\sigma^2 I)$ is assumed to be independent, and $D\sim \mathcal{N}(0,\sigma^2 I) $ consists of $n$ independent and identically distributed random variables $D^{(i)}$, each added to a training instance in $\mathcal{D}$. 
Let $\bm{\delta}=(\bm{\Delta}_1,\ldots,\bm{\Delta}_n)$ denote the collection of
training-set perturbations, where $\bm{\Delta}_i=\bm{0}$ for benign training samples and $\bm{\Delta}_i \neq \bm{0}$ only for poisoned training samples, and let $ \mathcal{B}_{\bm{x}} $ denote the backdoor trigger added to the testing instance $ \bm{x} $. Define $y_A=\arg\max_y \mathcal{P}_{\bm{\epsilon}(Z,D)}(f(\bm{x} + \mathcal{B}_{\bm{x}}+Z,\mathcal{D} + \bm{\delta}+D)=y )$,  assume that when classifying a point $\mathcal{N} (\bm{x},\sigma ^{2} \mathrm {I})$, the base classifier $f(\bm{x},\mathcal{D})$ assigns the most probable class $y_A$  with probability $P_A(\bm{x},\mathcal{D},\sigma)=\mathcal{P}_{\bm{\epsilon}(Z,D)}(f(\bm{x} + \mathcal{B}_{\bm{x}}+Z,\mathcal{D} + \bm{\delta}+D) = y_A)$, and the ``runner--up" class $y_B$ with probability $P_B(\bm{x},\mathcal{D},\sigma)
= \max_{y_B \neq y_A}\;
\mathcal{P}_{\bm{\epsilon}(Z,D)}\!\left(
f(\bm{x} + \mathcal{B}_{\bm{x}}+Z,\mathcal{D} + \bm{\delta}+D) = y
\right)$. Then, it is always true that  $g(\bm{x} + \mathcal{B}_{\bm{x}},\mathcal{D},\sigma )=g(\bm{x} + \mathcal{B}_{\bm{x}},\mathcal{D} + \bm{\delta}, \sigma)=y_A$ as long as the training-set perturbations satisfy $\sqrt{\sum_{i=1}^{n} \left \| \bm{\Delta} _i  \right \|_2^2 }\le r$, where 
\begin{equation}\label{eq:robust_radius}
				r  = \frac{\sigma}{2} \left ( \Phi ^{-1}(P_A(\bm{x},\mathcal{D},\sigma)  )-\Phi ^{-1}(P_B (\bm{x},\mathcal{D},\sigma) ) \right ).
		\end{equation}

By analyzing Eq. \eqref{eq:robust_radius}, we find that increasing the hyperparameter $\sigma$ enlarges the certified radius $r$, thereby enhancing the model's robustness. However, excessively increasing the noise magnitude may degrade classification accuracy (\ie, incorrect predictions), which reflects the trade-off between robustness and accuracy. Therefore, a key challenge remains: how to determine the optimal noise level $\sigma$ for each input.

\section{Revisiting Certified Backdoor Defenses}
\label{sec:revisiting}

Existing randomized smoothing-based certified backdoor defense methods \emph{implicitly} assume that all samples are equidistant from the decision boundary, \ie, they apply a fixed noise magnitude to each sample to smooth the classifier and obtain the final robust prediction. In this section, we analyze the variations in sample-to-decision-boundary distances from an intrinsic sample property perspective and further explore the limitations of using fixed Gaussian noise in existing methods. 

\subsection{Preliminaries}
\label{sec:pre}

\vspace{0.3em}
\noindent \textbf{The Main Pipeline of (Poisoning-based) Backdoor Attacks.} 
Let $\mathcal{D}=\{(\bm{x}_i,y_i)\}_{i=1}^n$ represents the benign dataset consisting of $n$ samples, where $\bm{x}_i \in \mathcal{X}$ is the $i$-th image, $y_i \in \mathcal{Y}=\{1, 2, \cdots, K\}$ is its corresponding label, and $K$ denotes the total number of classes. In general, adversaries create a poisoned dataset $\mathcal{D}_p$ to train the target model using either a standard loss function or a customized one specified by the attacker. Specifically, $\mathcal{D}_p$ consists of two main parts: \textbf{1)} the modified version of a selected subset (\ie, $\mathcal{D}_s$) of $\mathcal{D}$, and \textbf{2)} the remaining benign subset $\mathcal{D}_b$. Formally, $\mathcal{D}_p=\mathcal{D}_m(\bm \delta,\hat{y}) \cup \mathcal{D}_b$, where $\mathcal{D}_m(\bm \delta,\hat{y})=\left \{\bm{x}_i +\bm{\Delta} _i,\hat{y} \right \} _{i=1}^{\tilde{r}}$, $\mathcal{D}_b=\mathcal{D} \backslash \mathcal{D}_s=\left \{\bm{x}_i ,y_i \right \}_{i=\tilde{r} +1}^{n}$, $\bm \delta$ denotes the collection of unique trigger patterns $\bm{\Delta}_i$ injected into the selected training instances, and $\hat{y}=G_Y(y)$. Here, $\lambda \triangleq \frac{|\mathcal{D}_m|}{|\mathcal{D}|}$ is the \emph{poisoning rate}, and $G_Y$ is adversary-specified poisoned label generator. For example, in Badnets \cite{Gu2017BadNets}, $G_Y(y)=y_t$ for all-to-one attacks, where $y_t \in  \mathcal{Y}$ is the target label, and  $G_Y(y)=y+1$ mod $K$ for all-to-all attacks.  The attack succeeds if the classifier predicts the target label $\hat{y}$ for a testing example $\bm{x}$ modified with the backdoor pattern $\mathcal{B}_{\bm{x}}$: $f(\bm{x}+\mathcal{B}_{\bm{x}},\mathcal{D}_p)=\hat{y}$.

\begin{definition}[Boundary Samples and Closest Boundary Samples] 
\label{def:boundary_sample}
Consider the logit margin of model $f:\mathcal{X} \to [0,1]^K$ with respect to the label $y$, defined as: $\phi_y(\bm{x};\bm{w}) =f_y(\bm{x};w)-\max_{y'\ne y} f_{y'}(\bm{x};\bm{w})$. A sample $\bm{x}$ is classified as $y$ by the model $f(\cdot;\bm{w})$ if and only if $\phi_y(\bm{x};\bm{w})\ge 0$. The set of \textbf{boundary samples} belonging to class $y$ can be expressed as $\mathcal{T}(y;\bm{w}) =\{\bm{x}^*:\phi_y(\bm{x}^*;\bm{w})= 0\}$. 
Following the prior work \cite{guo2024zero}, the \textbf{closest boundary sample} for $\bm{x}$ is defined as: 
\begin{equation}\label{eq:closest boundary sample2}
		\bar{\bm{x}}^* \triangleq \arg \min_{\bar{\bm{x}} } \left \| \bm{x}^*-\bm{x} \right \|_p ,\quad s.t. \quad \phi_y(\bm{x}^*,\bm{w}) =0,
	\end{equation}
where $\left \| \cdot \right \|_{1\le p\le \infty } $ is the $\ell_p$ norm.
 \end{definition}

 \begin{figure*}[ht!]
 \vspace{-2em}
		\centering
		\subfigure[\,MNIST]{
			\includegraphics[scale =0.245]{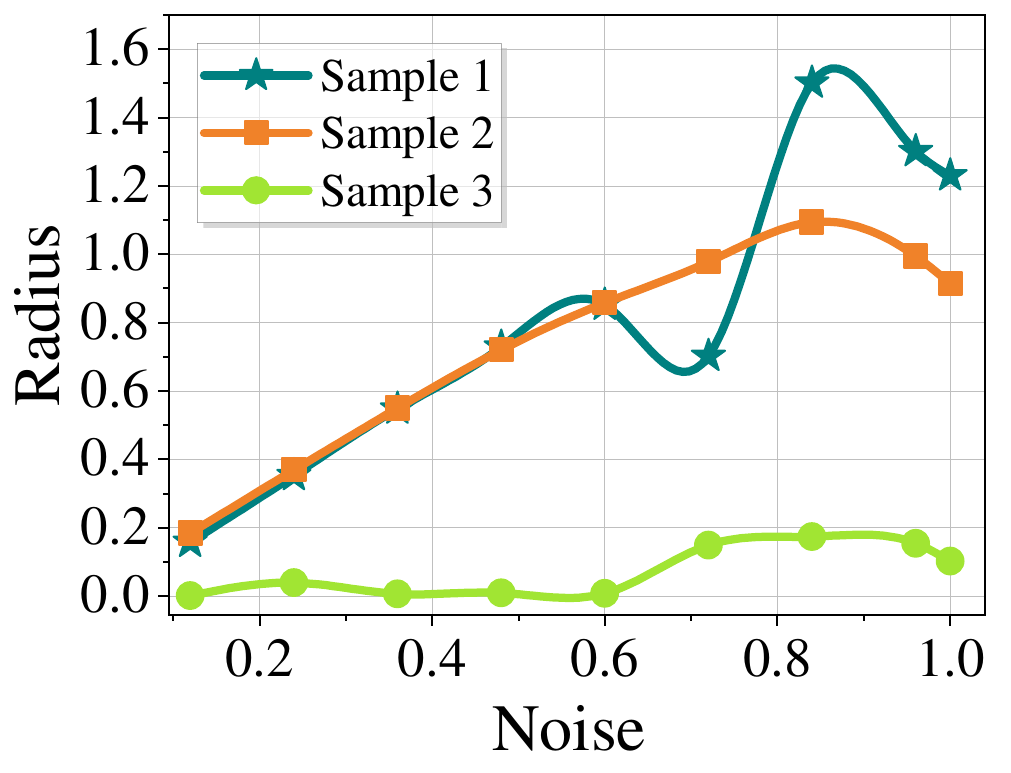}\label{fig:test_sample_mnist} }
		\subfigure[\,CIFAR-10]{
			\includegraphics[scale =0.245]{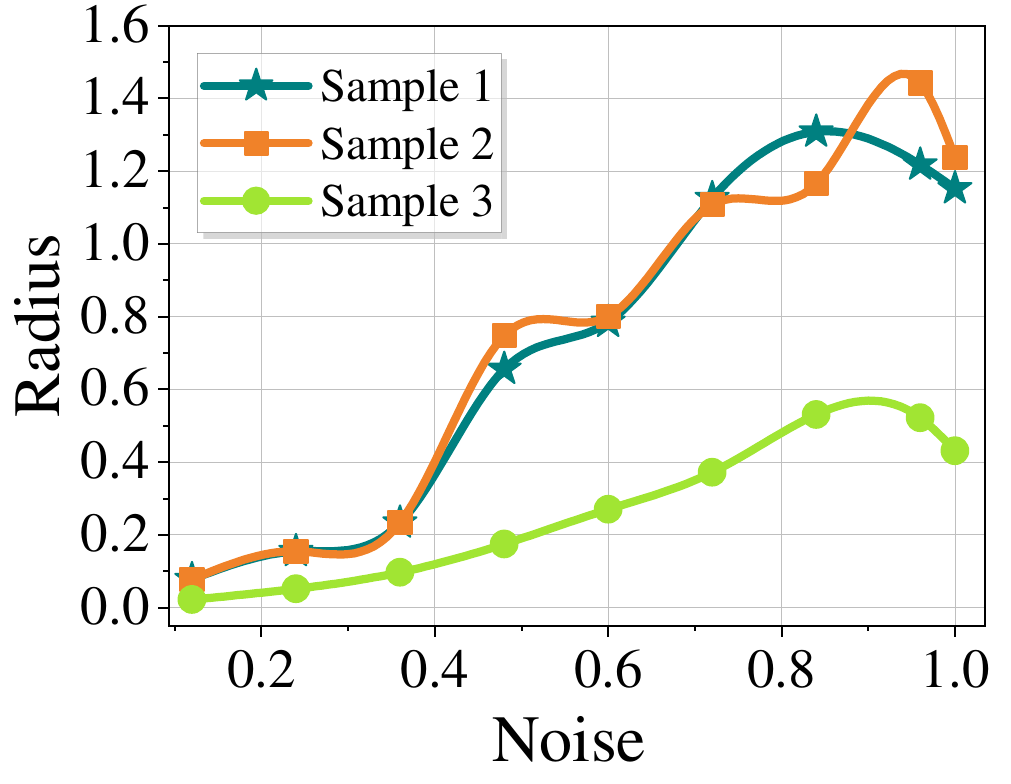}\label{fig:test_sample_cifar} }
		\subfigure[\,MNIST]{
			\includegraphics[scale =0.245]{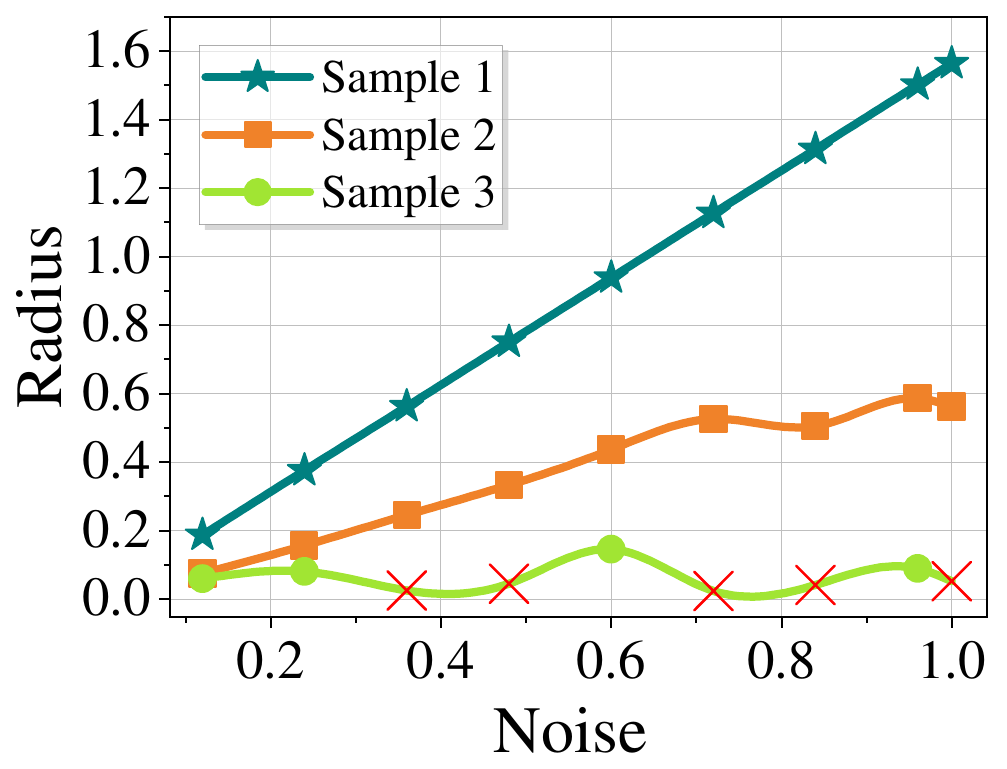}\label{fig:train_sample_mnist} }
            \subfigure[\,CIFAR-10]{
			\includegraphics[scale =0.245]{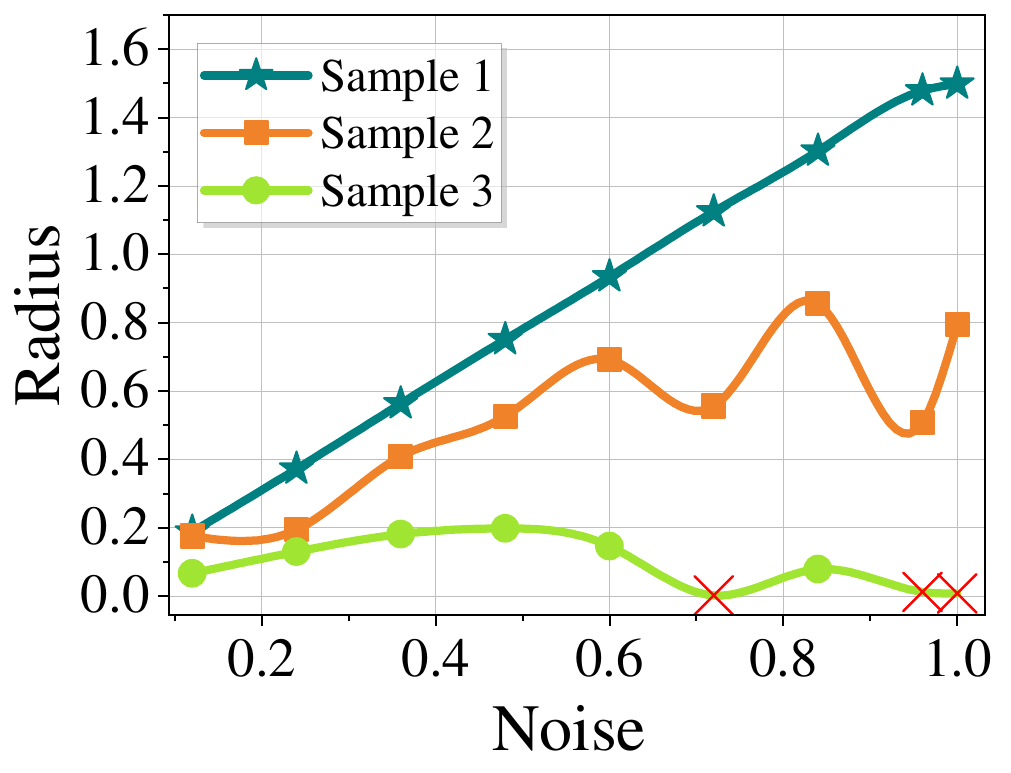}\label{fig:train_sample_cifar} }
		 \vspace{-0.5em}
		\caption{Effect of different noise levels on the certified radius for MNIST and CIFAR-10 datasets. The first two subfigures show results for testing samples, while the last two show results for training samples.}
        \vspace{-1em}
		\label{fig:test_train_sample}
	\end{figure*}

\vspace{0.3em}

\noindent \textbf{Generating the Closest Boundary Samples.}  To compute the closest boundary sample, we leverage the fast adaptive boundary attack (FAB) \cite{croce2020minimally}. Specifically, we modify FAB to implement an iterative algorithm using gradient ascent with $\nabla_{\bm{x}} \phi_y(\bm{x},\bm{w})$, updating the boundary sample at the ($t+1$)-th iteration as follows:
\begin{equation}\label{eq:closest boundary sample3}
		\bm{x}^*_{t+1}=\beta _t\cdot \bm{x}_0+(1-\beta _t)\left \{ \bm{x}^*_t+\alpha_t \frac{\bigtriangledown_{\bm{x}} \phi_y(\bm{x}^*_t;\bm{w})}{\left \| \bigtriangledown_{\bm{x}} \phi_y(\bm{x}^*_t;\bm{w}) \right \| }  \right \} ,
	\end{equation}
where $\alpha _t$ is a positive step size, $\bm{x}_0$ is an initial point \emph{s.t.} $\phi_y(\bm{x}_0;\bm{w})\le 0$ and $\beta _t \in [0,1]$ is a line search parameter \emph{s.t.} $\phi_y(\bm{x}^*_{t+1}; \bm{w}) = 0$. In practice, $\bm{x}_0$ is randomly selected from the validation set, ensuring its label differs from $y$.

\subsection{Analysis of Sample's Distance to Decision Boundary}

\label{sec:sample_analy}

We hereby analyze how the distance from a poisoned sample to the decision boundary of the target class varies across different inputs. Here, the closest decision boundary refers to the minimal perturbation required to change the prediction from the target label $y_t$ to any non-target class $y \neq y_t$. Specifically, this distance is estimated using the `closest boundary sample' defined in Definition~\ref{def:boundary_sample} to avoid the inaccurate estimation using a random boundary one since there are multiple of them.

\vspace{0.3em}

\noindent \textbf{Setting.} We hereby use BadNets \cite{Gu2017BadNets} attack with a ResNet model \cite{zagoruyko2016wide} on the CIFAR-10 \cite{krizhevsky2009learning} datasets for discussion. Specifically, we set the target label $y_t$ as `0' and the poisoning rate as  5\%. Following the previous work \cite{Weber2023RAB}, we use a one-pixel patch located at the lower right corner of the image as the trigger pattern. We randomly select 2,000 poisoned testing samples (\ie, samples containing the trigger and predicted as the target label $y_t$ by the backdoored model) and use Eq. (\ref{eq:closest boundary sample3}) to generate their closest boundary samples for the target label $y_t$. We then compute the $\ell_2$ norm between each poisoned sample and its closest boundary sample. Samples with a small distance are referred to as \textit{easy poisoned samples}, while those with a larger distance are referred to as \textit{hard poisoned samples.}

\begin{figure}[!t]
    \vspace{-0.6em}
    \centering
    \includegraphics[width=0.25\textwidth]{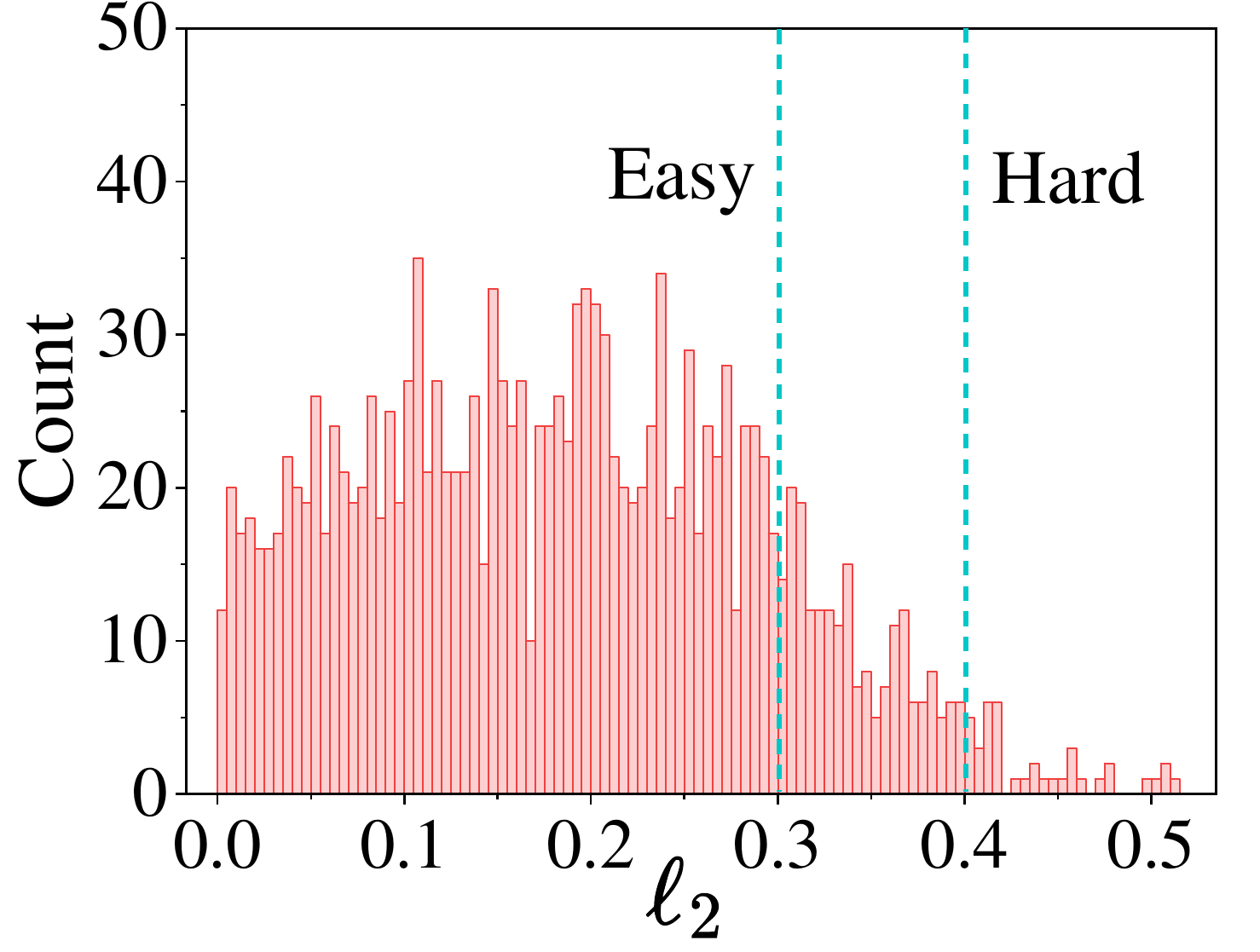}
    \vspace{-0.5em}
    \caption{Distribution of $\ell_2$ norm distances between poisoned samples and their closest boundary samples. Samples closer to the boundary are `easy' poisoned samples, while those farther away are `hard' poisoned samples.}
    \vspace{-1em}
    \label{fig:mov_distance}
\end{figure}

\vspace{0.3em}

\noindent \textbf{Result.} As shown in Figure~\ref{fig:mov_distance}, the $\ell_2$ distances to the closest boundary samples vary significantly among different samples in the poisoned dataset. Specifically, although most samples have relatively small distances (\eg, $\ell_2 \le 0.3$), a considerable number of hard samples exhibit larger distances to their closest boundary samples. Therefore, these hard poisoned samples would require a larger magnitude of noise to effectively suppress the backdoor effect. In contrast, for easy poisoned samples with smaller distances, only a smaller magnitude of noise is needed to achieve the desired defense effect. For samples that fall between easy and hard poisoned samples, \emph{a trade-off and adjustment in noise selection are necessary}. 

\subsection{Limitations of Fixed Noise in Testing Samples}
\label{sec:test_samples}


\vspace{0.3em}
\noindent \textbf{Setting.}
We hereby randomly select three testing samples from the MNIST and CIFAR-10 datasets, respectively, and evaluate the certification radius of the RAB model \cite{Weber2023RAB} trained with $\sigma=1.0$. The certification radius is computed following Eq. (\ref{eq:robust_radius}), using different noise levels with $\sigma$ values ranging from 0 to 1.0 in increments of 0.2. All other experimental settings remain as described in Section \ref{sec:sample_analy}.

\vspace{0.3em}

\noindent \textbf{Result.}  As shown in Figures \ref{fig:test_sample_mnist} and \ref{fig:test_sample_cifar}, the three samples from both MNIST and CIFAR-10 datasets exhibit a trend where the certification radius first increases and then decreases as the noise magnitude increases. Notably, although the model was trained with $\sigma=1.0$, the optimal certification radius does not occur at this noise level. Taking the MNIST dataset as an example, sample 1 reaches its maximum certification radius of approximately 1.5 at $\sigma=0.8$, sample 2 peaks at about 1.3 when $\sigma$ approaches 0.9, while sample 3 maintains a relatively stable low value. This result suggests that the optimal certification performance is not necessarily achieved by using the same noise magnitude during testing as in training. Therefore, \emph{the $\sigma$ value should be optimized for each sample} to achieve the maximum certification radius.

\subsection{Limitations of Fixed Noise in Training Samples}


\begin{figure*}[ht!]
    \vspace{-1.7em}
    \centering
    \includegraphics[width=0.98\textwidth]{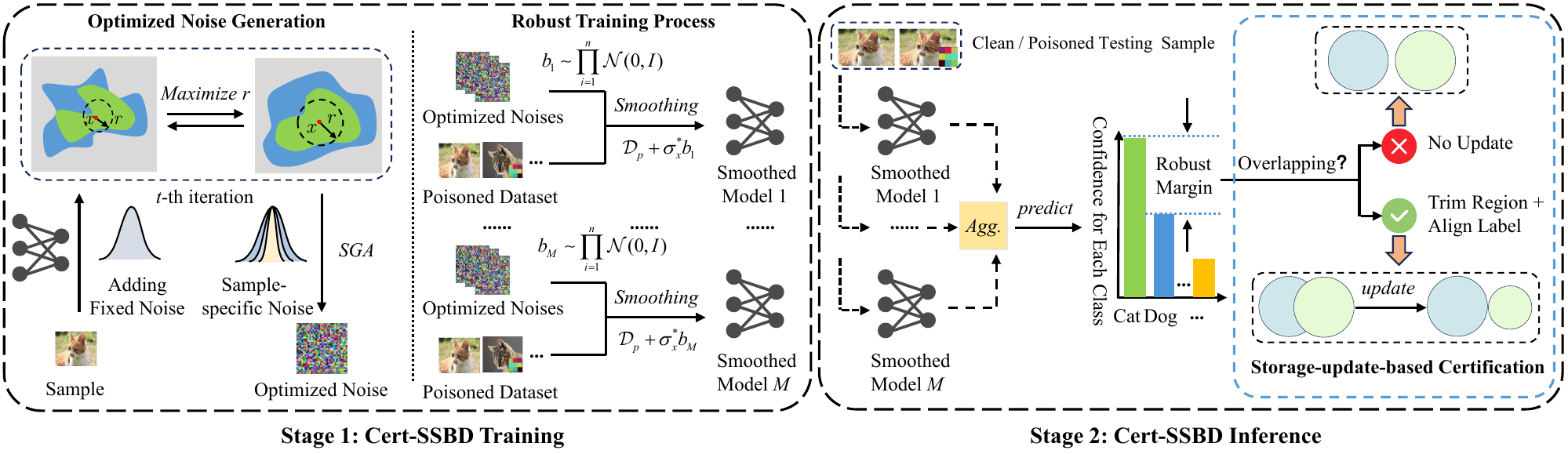}
    \vspace{-0.3em}
    \caption{The main pipeline of our Cert-SSBD consists of two stages. In the first stage, we adopt a stochastic gradient ascent (SGA) to iteratively optimize the noise to maximize the certification radius $r$, thereby obtaining optimal sample-specific noise to train $M$ smoothed models. In the second stage, the $M$ smoothed models trained in the first stage are aggregated to generate the final prediction. In particular, we propose a storage-update-based certification method to ensure non-overlapping certification regions and consistent predictions under the sample-specific noise setting (see Figure \ref{fig:certification} for more details).}
    \label{fig:MAIN}
    \vspace{-1.3em}
\end{figure*}

\vspace{0.3em}
\noindent \textbf{Setting.}
We randomly select three training samples from the MNIST and CIFAR-10 datasets, respectively, and train multiple models with different noise levels. Specifically, we apply noise with standard deviations of $\sigma \in \{0, 0.2, 0.4, 0.6, 0.8, 1.0\}$ during training. During the testing phase, we evaluate each model using the same noise level as in its training phase. That is, for a model trained with $\sigma=0.5$, its certification radius is also computed using $\sigma=0.5$. All other experimental settings remain as described in Section \ref{sec:sample_analy}.

\vspace{0.3em}

\noindent \textbf{Result.} 
As shown in Figures~\ref{fig:train_sample_mnist} and~\ref{fig:train_sample_cifar}, the certification radii of different samples exhibit distinct trends as the noise level varies. Overall, while some samples achieve larger certification radii at appropriate noise levels, others are more sensitive to noise, showing instability or even misclassification at higher noise values. For example, Sample 1 shows a continuously increasing certification radius as the noise $\sigma$ increases, indicating that its robustness remains stable even at higher noise levels. Sample 2 exhibits a stable certification radius in the range of $\sigma = 0.6$ to $0.8$, without significant changes as the noise level further increases. This suggests that this noise level may be optimal for this sample. In this case, increasing the noise further may not improve the certification radius and could even negatively impact classification accuracy. Therefore, for this sample, a trade-off must be made between accuracy and robustness. In contrast, Sample 3 experiences a gradual decrease in certification radius as the noise level increases and eventually undergoes misclassification at higher noise levels. The ``$\times $" markers in the figure indicate misclassified points. This result suggests that \emph{the noise level used for training should be optimized based on the characteristics of individual samples} rather than using a fixed value to achieve better certification performance.

   \section{Methodology} 
   \label{sec:method}
\subsection{Threat Model and the Goal of Certified Defense}
	\subsubsection{Threat Model} 
This work focuses on defending against poison-only backdoor attacks. Adversaries can manipulate the training data but cannot modify other training components, such as the loss function or model architecture. Defenders have full control over the training process but cannot detect whether the data is poisoned, nor do they know the trigger pattern.
    
\subsubsection{Goal of Certified Defense}\label{sec:Goal_Defense}
The primary goal is to defend against poison-only backdoor attacks by obtaining a robustness threshold $r$ through analysis, ensuring that if the total backdoor modification satisfies $\sqrt{\sum_{i=1}^{n} \left \| \bm{\Delta} _i  \right \|_2^2 }<r $, the classifier's predictions on testing samples containing backdoor triggers remain unaffected by whether the model was trained on poisoned or clean data. In other words, the model's predictions should be consistent, expressed as: $f(\bm{x}+\mathcal{B}_{\bm{x}}, \mathcal{D}_p)=
f(\bm{x}+\mathcal{B}_{\bm{x}}, \mathcal{D})$, where $\mathcal{D}_p = \mathcal{D}_m(\bm\delta,\hat{y}) \cup \mathcal{D}_b$ is the poisoned training set
and $\mathcal{D}$ is the clean training set (\ie, $\bm{\Delta}_i=\mathbf{0}$ for all $i$).

\subsection{Overview of the Proposed Method}

As demonstrated in Section \ref{sec:revisiting}, existing randomized smoothing-based certified backdoor defenses exhibit suboptimal certification performance, regardless of whether fixed noise is applied to training or testing samples. This is because each sample has a different distance to the decision boundary. To address this issue, we propose a sample-specific certified backdoor defense method, dubbed Cert-SSBD, in which the noise level is adaptively adjusted for each individual sample. 

As shown in Figure~\ref{fig:MAIN}, our method consists of two main stages:  \textbf{(1)} Cert-SSBD training stage, and \textbf{(2)} Cert-SSBD inference stage. In the training stage, we apply stochastic gradient ascent to iteratively solve for the optimal noise level $\sigma_{\bm{x}}^*$ that maximizes the certification radius. Once the sample-specific noise $\sigma_{\bm{x}}^*$ is obtained, it is injected into the poisoned training set to train $M$ smoothed models. In the inference stage, we aggregate the predictions of these $M$ smoothed models to generate the final output. Intuitively, as long as there exists a non-trivial gap between the predicted probabilities of the most likely class and the runner-up class, a non-zero certification radius can be obtained, and the model can be considered certifiably robust. However, under this sample-specific noise setting, traditional certification methods become inapplicable, as they typically assume a uniform noise level across all inputs. To overcome this limitation, we introduce a storage-update-based certification method. This method categorizes certification regions (\ie, regions defined by the certified radius of each input) to ensure that these regions remain non-overlapping across different inputs and maintain prediction consistency within each region (see Figure \ref{fig:certification} for details). The technical details are as follows.


\subsection{Cert-SSBD Training: Train Models with Optimized Noises}
\label{sec:cert-ssbd_train}

\looseness=-1
In this stage, we describe the training stage of Cert-SSBD, which consists of two sequential steps: 
\textbf{1)} optimizing a sample-specific noise scale $\sigma_{\bm{x}}^{*}$ via stochastic gradient ascent (SGA) to maximize the certified radius, and 
\textbf{2)} training an ensemble of smoothed models using the resulting optimized noise scales. 
In general, the certified radius is computed from the class probabilities of a smoothed classifier trained with a fixed initialized noise level.

\subsubsection{Optimized Sample-Specific Noise Generation}
Given a base smoothed classifier with a fixed noise level $\sigma_0$ 
(\ie, a predefined initialized noise scale), 
our goal is to construct a new smoothed classifier 
$g(\bm{x}, \mathcal{D}_p, \sigma_{\bm{x}}^{*})$
based on a set of optimized, sample-specific noise scales 
$\{\sigma_{\bm{x}_i}^{*}\}_{i=1}^n$, where each $\sigma_{\bm{x}_i}^{*}$ 
corresponds to a single training sample.
Note that when $\sigma_0 = 0$, the base smoothed classifier degenerates to the base classifier $f(\bm{x}, \mathcal{D}_p)$.
The new classifier should ensure that for all training samples $\bm{x}$, 
the predictions of the two smoothed classifiers (with $\sigma_0$ and $\sigma_{\bm{x}}^{*}$) 
remain identical, while also maximizing the certification radius for each sample. Formally, we define $y_A$ as the most probable (top-1) predicted class under the fixed noise level $\sigma_0$, and $y_B$ as the runner-up class with the second-highest predicted probability, \ie,
\begin{equation}\label{eq:y_ya}
 y_A=\arg\max_y \mathcal{P}_{\bm{\epsilon }(Z,D)}[f^y(\bm{x} + \mathcal{B}_{\bm{x}}+Z,\mathcal{D} + \bm{\delta}+D)],  
\end{equation}
  where $Z\sim \mathcal{N}(0,\sigma_0^2 I)$ is assumed to be independent, and $D\sim \mathcal{N}(0,\sigma_0^2 I) $ consists of $n$ \emph{i.i.d.} random variables $D^{(i)}$, each added to a training instance in $\mathcal{D}$. The optimized noise $\sigma_{\bm{x}}^{*}$ is obtained by solving a sample-specific
optimization problem that maximizes the certification radius $r(\bm{x}, \sigma)$ in Eq. (\ref{eq:robust_radius}):
\begin{equation}\label{eq:maxsigma}
		\begin{array}{ll}
			\sigma _{\bm{x}}^{*} =\arg\max _ \sigma \frac{\sigma }{2} ( \Phi^{-1}(P_A(\bm{x},\mathcal{D}, \sigma)) - \Phi^{-1}(P_B(\bm{x}, \mathcal{D},\sigma)) ),
		\end{array}
	\end{equation}
{\small
where $P_A(\bm{x},\mathcal{D}, \sigma)=\mathcal{P}_{\bm{\epsilon }(Z,D) }[f^{y_A}(\bm{x} + \mathcal{B}_{\bm{x}}+Z,\mathcal{D} + \bm{\delta}+D )]$, $P_B(\bm{x}, \mathcal{D},\sigma)=\max_{y_B\ne y_A}\mathcal{P}_{\bm{\epsilon }(Z,D) }[f^{y}(\bm{x} + \mathcal{B}_{\bm{x}}+Z,\mathcal{D} + \bm{\delta}+D)]$.}

\vspace{0.3em}
In practice, we solve Eq. (\ref{eq:maxsigma}) using stochastic gradient ascent, where the probabilities of predicting class $y_A$ and $y$ are estimated via Monte Carlo approximation. Specifically, we introduce noise multiple times, record the output count for these two classes, and approximate the probability distribution using their relative frequencies. Formally, the gradient of the objective at the $t$-th iteration is approximated as follows:
  
\begin{equation}\label{eq:sigma_monte}
		\begin{array}{ll}
			&\bigtriangledown_{\sigma^t} \{\frac{\sigma^t}{2} \cdot [\Phi^{-1}(\frac{1}{J} {\textstyle \sum_{j=1}^{J}f^{y_A}(\bm{x} + \mathcal{B}_{\bm{x}}+Z_i,\mathcal{D} + \bm{\delta}+D_i  )}   ) \\
			&-\Phi^{-1}(\max_{y_B\ne y_A}\frac{1}{J} {\textstyle \sum_{j=1}^{J}f^{y}(\bm{x} + \mathcal{B}_{\bm{x}}+Z_i,\mathcal{D} + \bm{\delta}+D_i )}   )]\},
		\end{array}
	\end{equation}
where $Z_1, \dots, Z_J \sim \mathcal{N}(0, (\sigma^t)^2 I)$ as well as $D_1, \dots, D_J \sim \mathcal{N}(0, (\sigma^t)^2 I)$ are independently sampled at each iteration.

However, since the probabilities depend on the optimization variable $\sigma$, and $\sigma$ parameterizes the smoothed distribution $\mathcal{N} (0,\sigma^2I)$ \cite{williams1992simple}, any change in $\sigma$ affects the underlying distribution, which can result in high variance in the gradient estimation method. To address this problem, we adopt the reparameterization technique proposed by Kingma \textit{et al.} \cite{Kingma2014Auto-encoding} and Rezende \textit{et al.} \cite{rezende2014stochastic}, which allows for a lower-variance gradient estimation of the objective in Eq. (\ref{eq:sigma_monte}). Specifically, we reparameterize the noise as $Z = \sigma \hat{Z}$ and $D = \sigma \hat{D}$, where $\hat{Z}$ and $\hat{D}$ are sampled from a standard normal distribution, \ie, $\hat{Z}, \hat{D} \sim \mathcal{N}(0, I)$. This transformation allows us to reformulate the objective in Eq.~(\ref{eq:maxsigma}) as follows:
\begin{equation}\label{eq:sigma_variance}
			\sigma _{\bm{x}}^{*} =\arg\max _ \sigma \frac{\sigma }{2} ( \Phi^{-1}(\hat{P_A}(\bm{x},\mathcal{D}, \sigma)) - \Phi^{-1}(\hat{P_B}(\bm{x}, \mathcal{D},\sigma)) ),
	\end{equation}  
where $\hat{P_A}(\bm{x},\mathcal{D}, \sigma)=\mathcal{P}_{\bm{\epsilon }(\hat{Z},\hat{D})}[f^{y_A}(\bm{x} + \mathcal{B}_{\bm{x}}+\sigma \hat{Z},\mathcal{D} + \bm{\delta}+\sigma \hat{D} )]$, $\hat{P_B}(\bm{x}, \mathcal{D},\sigma)=\max_{y_B\ne y_A}\mathcal{P}_{\bm{\epsilon }(\hat{Z},\hat{D}) }[f^{y}(\bm{x} + \mathcal{B}_{\bm{x}}+\sigma \hat{Z},\mathcal{D} + \bm{\delta}+\sigma \hat{D})]$. Note that under this reparameterization, the distributions $\hat{Z}$ and $\hat{D}$ are no longer dependent on the optimization variable $\sigma$. As a result, Eq. (\ref{eq:sigma_variance}) typically yields lower-variance gradient estimates compared to the original formulation in Eq. (\ref{eq:maxsigma}). This optimization yields a set of sample-specific noise scales $\{\sigma_{\bm{x}_i}^*\}_{i=1}^n$, which are then used in the subsequent robust training stage to construct an ensemble of smoothed models.

\subsubsection{Robust Training Process}
Once the optimized sample-specific noise 
$\{\sigma_{\bm{x}_i}^*\}_{i=1}^n$ is obtained, we incorporate it into the training process to enhance robustness. Specifically, we first sample $M$ sets of noise vectors $b_1, \cdots, b_M$ from the distribution $D \sim {\textstyle \prod_{i=1}^{n}} \mathcal{N}(0, I)$, where each set contains $n = |\mathcal{D}|$ \emph{i.i.d.} vectors corresponding to the size of the training dataset. For each sampled noise set $b_m$, we construct a perturbed (poisoned) training dataset $\mathcal{D}_p^{(m)} \triangleq \mathcal{D}_p + \{\sigma_{\bm{x}_i}^* b_{m,i}\}_{i=1}^n$ by perturbing each data point in $\mathcal{D}_p$ with its optimized noise scale $\sigma_{\bm{x}_i}^*$. Here, $\mathcal{D}_p$ denotes the poisoned training dataset consisting of both poisoned
and benign samples, as defined in Section \ref{sec:pre}. Next, we train $M$ smoothed models on these perturbed datasets, denoted as $g_1(\bm{x}, \mathcal{D}_p^{(1)}, \sigma_{\bm{x}}^*), \ldots, g_M(\bm{x}, \mathcal{D}_p^{(M)}, \sigma_{\bm{x}}^*)$. To maintain consistency between the noise distributions used during training and inference,
for each trained model $g_m$, we deterministically sample and store a  \emph{unit-scale base noise vector} 
$\mu_m \sim \mathcal{N}(0, I_d)$ using a random seed derived from a hash value of the trained model parameters, where $I_d$ denotes the $d$-dimensional identity matrix corresponding to the input space 
(\ie, $d$ is the dimensionality of the input feature vector). 
This base noise vector is stored together with the model parameters and reused during inference. The detailed inference-time noise application will be described in Section~\ref{sec:cert_ssbd_inference}. By introducing noise perturbations during both training and inference, we ensure that the ensemble of smoothed models $\{g_1, \ldots, g_M\}$ avoids performance degradation when classifying clean inputs. See Algorithm~\ref{alg:train_robust} in our Appendix~\ref{appA} for training details.

\subsection{ Cert-SSBD Inference: Storage-update-based Certification}\label{sec:cert_ssbd_inference}

Building upon the ensemble of smoothed models trained in Section~\ref{sec:cert-ssbd_train}, we now present the inference and certification procedure of Cert-SSBD. 
At inference stage, we aggregate the ensemble outputs under the optimized, sample-specific noise scale $\sigma_{\bm{x}}^{*}$ to obtain the final prediction. 
Since certification is no longer performed with a single fixed noise parameter, existing certification methods are not directly applicable. 
To address this, we propose a novel storage-update-based certification method. By introducing a `storage' mechanism, this method dynamically adjusts certification regions to ensure they are non-overlapping across inputs while preserving prediction consistency for each individual sample.


Formally, given trained models  $\{(g_m, \mu_m)\}_{m=1}^M$ and a testing input $\bm{x}_i$, the prediction is obtained via majority voting under the optimized, sample-specific noise scale $\{\sigma_{\bm{x}_i}^*\}_{i=1}^n$. Concretely, for each model $g_m$, we evaluate the prediction on the perturbed input $\bm{x} + \sigma_{\bm{x}}^* \mu_m$, where $\mu_m \sim \mathcal{N}(0, I_d)$ is the unit-scale base noise vector deterministically sampled and stored during training, and the model is associated with the corresponding perturbed poisoned training dataset $\mathcal{D}_p^{(m)}$ (defined in Section~\ref{sec:cert-ssbd_train}). The resulting vote frequency over the ensemble serves as an unbiased empirical estimate of the class probabilities under the smoothed classifier: $\mathcal{P}_{\bm{\epsilon}}\!\left(g(\bm{x}, \{\mathcal{D}_p^{(m)}\}_{m=1}^{M}, \sigma_{\bm{x}}^*) = y\right)
= \frac{1}{M} \sum_{m=1}^{M}
\mathbb{I}\!\left\{
g_m\!\left(
\bm{x} + \sigma_{\bm{x}}^* \mu_m,\;
\mathcal{D}_p^{(m)}
\right) = y
\right\}$, where  $\mathbb{I}\{\cdot\}$ denotes the indicator function. To account for the statistical uncertainty induced by the finite ensemble size $M$,
we estimate one-sided $(1-\alpha)$-binomial confidence bounds on the class probability estimates. Specifically, let $cnts[y]$ denote the number of votes received by class $y$, and let $(y_A, y_B)$ be the two classes
with the largest vote counts. Based on $cnts[y_A]$ and $cnts[y_B]$, we compute a lower confidence bound
$P_A$ for the target class $y_A$ and an upper confidence bound $P_B$ for the runner-up class
at confidence level $\alpha$. If $P_A > P_B$, the prediction $y_A$ is certified with confidence $1-\alpha$,
which further enables the computation of a certified robust radius. 
The following theorem establishes the robustness guarantee.

\begin{figure}[ht!]
    \vspace{-1.3em}
    \centering
    \includegraphics[width=0.5\textwidth]{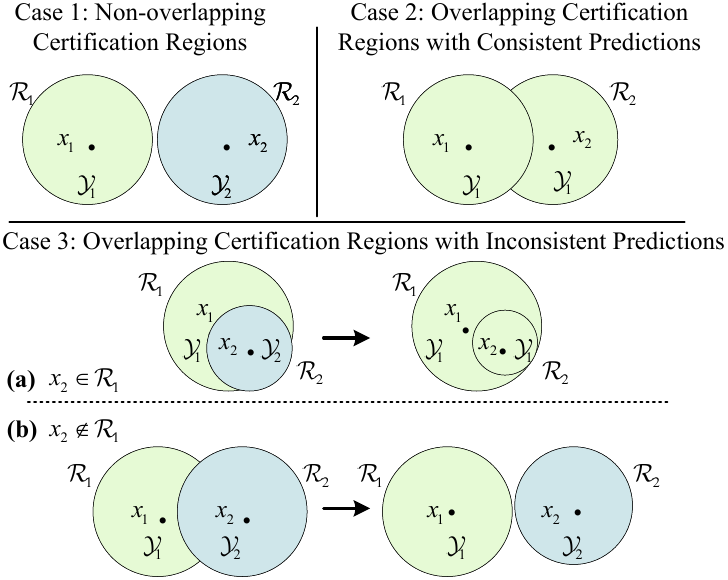}
    \vspace{-0.8em}
    \caption{Storage-update-based certification. 
Given two inputs $\bm{x}_1$ and $\bm{x}_2$ with certified regions $\mathcal{R}_1$ and $\mathcal{R}_2$, and predicted labels $\mathcal{Y}_1$ and $\mathcal{Y}_2$, respectively, the certification process may fall into the three configurations illustrated in the figure, as described in Remark~\ref{re:region}.
\textbf{Case~1}: the storage set remains unchanged.
\textbf{Case~2}: the new input together with its certified region can be directly added to the storage set.
\textbf{Case~3}: the process is further divided into two sub-cases, \textbf{(a)} $\bm{x}_2 \in \mathcal{R}_1$ (inside) and \textbf{(b)} $\bm{x}_2 \notin \mathcal{R}_1$ (outside). In this case, the certified region of the new input is shrunk to a subset (either restricted to the overlapping part or to the largest non-overlapping subset), so that the certification remains valid while conflicts in the storage set are removed.}
     \vspace{-0.5em}
    \label{fig:certification}
    \vspace{-1.0em}
\end{figure}

\begin{theorem}[Certified Robustness of Cert-SSBD]
\label{theorem:robust_cert}
 Let $\mathcal{B}_{\bm{x}} \in \mathbb{R}^d$ denote a backdoor trigger applied to the test input, and $\bm{\delta} \triangleq (\bm{\Delta}_1,\bm{\Delta}_2,\ldots,\bm{\Delta}_n)$ denote the collection of training-set perturbations, where
$\bm{\Delta}_i \in \mathbb{R}^d$ and $\bm{\Delta}_i=\bm{0}$ for benign training samples (as defined in Section~\ref{sec:pre}).  Let $\mathcal{D}$ be a training set and let smoothing noise $\hat{Z} \sim \mathcal{N}(0,I)$, $\hat{D} \sim \mathcal{N}(0,I)$. 
 Let $ y_A \in \mathcal{Y} $, such as $y_A = g(\bm{x} + \mathcal{B}_{\bm{x}},\mathcal{D} + \bm{\delta}) $ with class probabilities satisfying $\mathcal{P}_{\bm{\epsilon}(\hat{Z},\hat{D})}(f(\bm{x} + \mathcal{B}_{\bm{x}}+\sigma_{\bm{x}}^*\hat{Z},\mathcal{D} + \bm{\delta}+\sigma_{\bm{x}}^*\hat{D}) = y_A) \ge P_A \ge P_B \ge \max_{y_B \neq y_A}  \mathcal{P}_{\bm{\epsilon}(\hat{Z},\hat{D})}(f(\bm{x} + \mathcal{B}_{\bm{x}}+\sigma_{\bm{x}}^*\hat{Z},\mathcal{D} + \bm{\delta}+\sigma_{\bm{x}}^*\hat{D}) = y)$. Then, we have $g(\bm{x} + \mathcal{B}_{\bm{x}},\mathcal{D} )=g(\bm{x} + \mathcal{B}_{\bm{x}},\mathcal{D} +\bm{\delta})= y_A$ for all training-set perturbations $\bm{\delta}$ satisfying $ \sqrt{\sum_{i=1}^{n} \left \| \bm{\Delta} _i  \right \|_2^2 }\le r (g;\sigma_{\bm{x}}^*)$, where the certified robust radius $r$ is given by
		\begin{equation}\label{eq:radius}
			\begin{array}{ll}
                r(g;\sigma_{\bm{x}}^*) = \frac{\sigma_{\bm{x}}^* }{2} \left ( \Phi ^{-1}(P_A(\sigma_{\bm{x}}^*)  )-\Phi^{-1} (P_B (\sigma_{\bm{x}}^*) ) \right ).
			\end{array}
		\end{equation}
		
	\end{theorem}

Compared to RAB~\cite{Weber2023RAB}, our method achieves a better trade-off between robustness and accuracy by replacing the fixed smoothing noise $\sigma$ with optimized sample-specific noise $\sigma_{\bm{x}_i}^*$. This advantage is further supported by experimental results presented later. The formal proof is provided in Appendix~\ref{appB}.

\looseness=-1
Notably, unlike prior randomized smoothing–based certification methods that rely on a fixed noise level (\ie, a single  initialized $\sigma_0$ shared by all inputs), Cert-SSBD performs inference under sample-specific optimized noise scales $\sigma_{\bm{x}}^*$.
Under this setting, the certified region associated with each input is determined by its own noise scale rather than a unified initialized parameter, and therefore the certified regions of different inputs are no longer guaranteed to be globally non-overlapping. When certified regions corresponding to different predicted labels overlap, this may introduce ambiguities in the certification process. Motivated by this understanding, we introduce a storage-update-based certification method to ensure the reliability and soundness of certification under sample-specific noise. To formalize the above issue, we first rigorously define the notions of ``overlapping'' and ``non-overlapping'' certification regions.

\begin{definition}[Overlapping and Non-overlapping of Certification Regions]
 \label{def:inter_disjoint}
 Let $g$ be a sample-specific smoothed classifier, and let $r(\sigma_{\bm{x}_1}^*)$ denote the certification radius of $g$ at input $\bm{x}_1$. For any other input $\bm{x}_2$, if $\|\bm{x}_1 - \bm{x}_2\|_2 \leq r(\sigma_{\bm{x}_1}^*)$, the certification regions of $ \bm{x}_1 $ and $\bm{x}_2$ are said to be \emph{overlapping}; otherwise, they are said to be \emph{non-overlapping}.
 \end{definition}
\begin{remark}\label{re:region}
In general, under sample-specific noise, certifying a new input may lead to three possible configurations of certification regions (see Figure~\ref{fig:certification}): \emph{Case~1} (non-overlapping certification regions), \emph{Case~2} (overlapping certification regions with consistent predictions), and \emph{Case~3} (overlapping certification regions with inconsistent predictions). Among them, \emph{Case~3} may lead to ambiguity in certification and therefore requires explicit conflict resolution to ensure soundness.
\end{remark}

\begin{table*}[!t]
  \vspace{-1.8em}
  \captionsetup{font=small}
  \caption{Certified performance (\ie, ERA and AER) of Cert-SSBD and RAB on MNIST, CIFAR-10, and ImageNette under the all-to-one setting with representative attacks (\ie, one-pixel, four-pixel, and blending). At each radius, we report the best result over different noise levels; the best results are marked in boldface.
  } 
  
  \vspace{-0.5em}
  \centering
  \setlength{\tabcolsep}{6pt} 
  \renewcommand{\arraystretch}{1.2} 
  \scalebox{0.95}{
    \begin{tabular}{c|c|c|c|cccccccc} 
      \toprule
      \multirow{2}{*}{Dataset$\downarrow$} & \multirow{2}{*}{Attack Setting$\downarrow$, Metric$\xrightarrow{}$ } & \multirow{2}{*}{Method$\downarrow$}& \multirow{2}{*}{AER} & \multicolumn{8}{c}{Radius (ERA$\uparrow$)} \\ 
      \cline{5-12}
      & & & & 0 & 0.25 & 0.5 & 0.75 & 1.0 & 1.25 & 1.5 & 1.75 \\
      \hline
      \multirow{6}{*}{MNIST} 
      & \multirow{2}{*}{One-pixel}
      & RAB & 1.48&  \textbf{100} & 99.91 & 99.76 & 99.43 & 99.05 & 97.73 & 55.79 & 0 \\
      & &Cert-SSBD &\textbf{1.65} & 99.95 & \textbf{99.91} & \textbf{99.81} & \textbf{99.62} & \textbf{99.34} & \textbf{98.82} & \textbf{86.53} & \textbf{42.98} \\
      \cline{2-12}
      & \multirow{2}{*}{Four-pixel}
      & RAB &1.49& 99.95 & 99.86 & 99.72 & 99.39 & 99.01 & 97.78 & 56.12 & 0 \\
      & &Cert-SSBD&\textbf{1.69} & \textbf{99.95} & \textbf{99.86} & \textbf{99.72} & \textbf{99.57} & \textbf{99.20} & \textbf{98.63} & \textbf{81.94} & \textbf{42.98} \\
      \cline{2-12}
      & \multirow{2}{*}{Blending}
      & RAB &1.46& \textbf{100} & 99.86 & 99.67 & 99.39 & 99.05 & 97.35 & 42.03 & 0 \\
      & &Cert-SSBD&\textbf{1.70} & 99.95 & \textbf{99.86} & \textbf{99.76} & \textbf{99.72} & \textbf{99.20} & \textbf{98.72} & \textbf{72.15} & \textbf{42.84} \\
 \hline

\multirow{6}{*}{CIFAR-10} 
      & \multirow{2}{*}{One-pixel}
      & RAB & 0.55&  \textbf{87.80} & 69.70 & 56.70 & 38.30 & 16.55 & 2.60 & 0 & 0 \\
      & &Cert-SSBD &\textbf{0.62} & 86.55 & \textbf{71.90} & \textbf{60.75} & \textbf{46.30} & \textbf{26.10} & \textbf{11.50} & \textbf{1.45} & 0 \\
      \cline{2-12}
      & \multirow{2}{*}{Four-pixel}
      & RAB &0.56& \textbf{88.70} & 69.50 & 55.70 & 36.60 & 14.15 & \textbf{2.25} & \textbf{0.05} & 0 \\
      & &Cert-SSBD& \textbf{0.65} & 86.40 & \textbf{70.30} & \textbf{59.50} & \textbf{43.55} & \textbf{20.90} & \textbf{1.60} & 0 & 0 \\
      \cline{2-12}
      & \multirow{2}{*}{Blending}
      & RAB &0.56& \textbf{88.00} & 69.80 & 56.25 & 36.95 & 15.00 & \textbf{2.35} & 0 & 0 \\
      & &Cert-SSBD&\textbf{0.64} & 86.15 & \textbf{73.40} & \textbf{61.55} & \textbf{46.55} & \textbf{27.25} & 0.05 & 0 & 0 \\

      \hline

\multirow{6}{*}{ImageNette} 
      & \multirow{2}{*}{One-pixel}
      & RAB & 0.49&94.62 & 74.18 & 52.60 & 35.42 & 14.60 & 0 & 0  & 0 \\
      & &Cert-SSBD &\textbf{0.64} &\textbf{95.20} & \textbf{86.36} & \textbf{72.50} & \textbf{45.08} & \textbf{32.10} & \textbf{17.36} & \textbf{5.08} & 0 \\
      \cline{2-12}
      & \multirow{2}{*}{Four-pixel}
      & RAB &0.48& 94.80 & 73.94 & 52.26 & 33.36 & 13.26 & 0 & 0 & 0 \\
      & &Cert-SSBD&\textbf{0.67} &\textbf{94.90} & \textbf{86.82} & \textbf{77.00} & \textbf{55.22} & \textbf{34.52} & \textbf{20.22} & \textbf{5.76} & 0 \\
      \cline{2-12}
      & \multirow{2}{*}{Blending}
      & RAB &0.47& 94.78 & 74.32 & 51.44 & 33.02 & 12.62 & 0 & 0 & 0 \\
      & & Cert-SSBD &\textbf{0.64} & \textbf{94.94} & \textbf{83.46} & \textbf{58.66} & \textbf{46.30} & \textbf{34.52} & \textbf{20.22} & \textbf{5.76} & 0 \\
      \bottomrule
    \end{tabular}
 }
  \label{table:All-to-one_ERA}
\end{table*}

\begin{table*}[!t]
  \captionsetup{font=small}
  \caption{Certified performance (\ie, CRA and ACR) of Cert-SSBD and RAB on MNIST, CIFAR-10, and ImageNette under the all-to-one setting with representative attacks (\ie, one-pixel, four-pixel, and blending). At each radius, we report the best result over different noise levels; the best results are marked in boldface.}
  \vspace{-0.8em}
  \centering
  \setlength{\tabcolsep}{6pt} 
  \renewcommand{\arraystretch}{1.2} 
  \scalebox{0.95}{
    \begin{tabular}{c|c|c|c|cccccccc} 
      \toprule
      \multirow{2}{*}{Dataset$\downarrow$} & \multirow{2}{*}{Attack Setting$\downarrow$, Metric$\xrightarrow{}$ } & \multirow{2}{*}{Method$\downarrow$}& \multirow{2}{*}{ACR} & \multicolumn{8}{c}{Radius (CRA$\uparrow$)} \\ 
      \cline{5-12}
      & & & & 0 & 0.25 & 0.5 & 0.75 & 1.0 & 1.25 & 1.5 & 1.75 \\
      \hline
      \multirow{6}{*}{MNIST} 
      & \multirow{2}{*}{One-pixel}
      & RAB & 0.69&  \textbf{46.37} &46.24 & 46.10 & 45.01 & 45.91 & 45.49 & 42.51 & 0 \\
      & &Cert-SSBD &\textbf{0.84} & 46.29 & \textbf{46.24} & \textbf{46.24} & \textbf{46.10} & \textbf{45.96} & \textbf{45.91} & \textbf{45.20} & \textbf{42.88} \\
      \cline{2-12}
      & \multirow{2}{*}{Four-pixel}
      & RAB &0.68& \textbf{46.34} &\textbf{46.24} & 46.10 & \textbf{46.72} & 45.91 & 45.63 & 41.23 & 0 \\
      & &Cert-SSBD&\textbf{0.87} & 46.29 & \textbf{46.24} & \textbf{46.24} & 46.10 & \textbf{46.01} & \textbf{45.91} & \textbf{45.67} & \textbf{43.88} \\
      \cline{2-12}
      & \multirow{2}{*}{Blending}
      & RAB &0.69& \textbf{46.34} & 46.24 & 46.10 & 45.01 & 45.91 & 45.49 & 42.46 & 0 \\
      & &Cert-SSBD&\textbf{0.87} & \textbf{46.34} & \textbf{46.29} & \textbf{46.24} & \textbf{46.19} & \textbf{45.96} & \textbf{45.91} & \textbf{45.63} & \textbf{44.30} \\
 \hline

\multirow{6}{*}{CIFAR-10} 
      & \multirow{2}{*}{One-pixel}
      & RAB & 0.32&  48.30 & 39.40 & 30.40 & 20.05 & \textbf{8.35} & 0.55 & 0 & 0 \\
      & &Cert-SSBD &\textbf{0.33} & \textbf{52.65} & \textbf{41.60} & \textbf{34.65} & \textbf{21.30} & 3.50 & 0 & 0 & 0 \\
      \cline{2-12}
      & \multirow{2}{*}{Four-pixel}
      & RAB &0.33& 48.90 & 41.00 & 32.05 & 21.35 & 9.65 & \textbf{0.65} & 0 & 0 \\
      & &Cert-SSBD& \textbf{0.35} & \textbf{56.55} & \textbf{44.00} & \textbf{35.90} & \textbf{26.30} & \textbf{10.30} & 0 & 0 & 0 \\
      \cline{2-12}
      & \multirow{2}{*}{Blending}
      & RAB &\textbf{0.32}&48.40 & 40.70 & 31.55 & 20.75 & \textbf{8.90} & \textbf{0.65} & 0 & 0 \\
      & &Cert-SSBD&\textbf{0.32} & \textbf{58.55} & \textbf{42.05} & \textbf{35.30} & \textbf{24.70} & 0.95 & 0& 0 & 0 \\

      \hline

\multirow{6}{*}{ImageNette} 
      & \multirow{2}{*}{One-pixel}
      & RAB & 0.27&48.40 &  39.78 & 30.30 & 20.30 & 8.22 & 0 & 0  & 0 \\
      & &Cert-SSBD &\textbf{0.36} &\textbf{48.70} & \textbf{43.96} & \textbf{38.06} & \textbf{26.66} & \textbf{16.58} & \textbf{9.48} & \textbf{3.42} & 0 \\
      \cline{2-12}
      & \multirow{2}{*}{Four-pixel}
      & RAB &0.26& 48.68 & 40.10 & 29.00 & 18.32 & 7.10 & 0 & 0 & 0 \\
      & &Cert-SSBD&\textbf{0.49} &\textbf{49.00} & \textbf{42.48} & \textbf{36.76} & \textbf{26.26} & \textbf{19.08} & \textbf{11.44} & \textbf{4.00} & 0 \\
      \cline{2-12}
      & \multirow{2}{*}{Blending}
      & RAB &0.27& 48.72 & 40.14 & 29.62 & 19.02 & 7.16 & 0 & 0 & 0 \\
      & &Cert-SSBD&\textbf{0.48} & \textbf{49.10} & \textbf{42.86} & \textbf{38.96} & \textbf{34.48} & \textbf{27.62} & \textbf{19.06} & \textbf{6.74} & 0 \\
      \bottomrule
    \end{tabular}
  }
  \label{table:All-to-one_CRA}
   \vspace{-1.5em}
\end{table*}

To address the potential overlap of certification regions defined above, we introduce a storage-update-based certification strategy, which enforces non-overlapping certification regions across inputs with different predicted labels while maintaining prediction consistency. Specifically, we maintain a triplet storage set
$\mathcal{S} = \{(\bm{x}_i, \mathcal{Y}_i, \mathcal{R}_i)\}_{i=1}^n$, where each triplet records a previously certified input $\bm{x}_i$, its predicted label $\mathcal{Y}_i$, and its associated certification region $\mathcal{R}_i$. The strategy requires the storage set to satisfy the following key property: for any $i \neq j$,
whenever $\mathcal{Y}_i \neq \mathcal{Y}_j$, it must hold that
$\mathcal{R}_i \cap \mathcal{R}_j = \emptyset$. This property is necessary to ensure the soundness of the certification process. In particular, given a newly certified triplet $(\bm{x}_{n+1}, \mathcal{Y}_{n+1}, \mathcal{R}_{n+1})$, if there exists a stored certification region $\mathcal{R}_i$ such that $\mathcal{R}_{n+1} \cap \mathcal{R}_i \neq \emptyset$ and $\mathcal{Y}_{n+1} \neq \mathcal{Y}_i$
(corresponding to Case~3 in Figure~\ref{fig:certification}),
we resolve the conflict according to the following two cases:
if $\bm{x}_{n+1} \in \mathcal{R}_i$, we update the prediction at $\bm{x}_{n+1}$ to $\mathcal{Y}_i$ and refine $\mathcal{R}_{n+1}$ to the largest subset consistent with $\mathcal{R}_i$; otherwise, if $\bm{x}_{n+1} \notin \mathcal{R}_i$, we refine $\mathcal{R}_{n+1}$ to the largest subset that does not intersect with $\mathcal{R}_i$.
This procedure is applied sequentially over all elements in the storage set, after which the updated triplet is added to $\mathcal{S}$.
As a result, certification regions associated with different predicted labels remain non-overlapping within the storage, ensuring the well-definedness and soundness of certification under sample-specific noise. See Algorithm~\ref{alg:cert_infer_ssb} for the overall Cert-SSBD inference procedure, which invokes the storage-update mechanism detailed in Algorithm~\ref{alg:storage_update} (Appendix~\ref{appA}).

Although a storage-update-based method is theoretically necessary to guarantee the soundness of certification under sample-specific noise, we did not observe any cases in our experiments where certified regions associated with different predictions overlap. This phenomenon can be attributed to the high dimensionality of image inputs and the moderate magnitude of the optimized noise scales in our evaluated datasets. Nevertheless, in some rare yet realistic scenarios, such overlaps may still occur, particularly when the underlying data distribution includes atypical or ambiguous samples (\eg, label noise or annotation errors, boundary-adjacent inputs with small classification margins, or near-duplicate and highly similar instances). In these cases, our method acts as a conservative and general safeguard: upon the emergence of potential conflicts, it systematically resolves ambiguities by appropriately adjusting the certified regions, thereby preserving the well-definedness and soundness of the resulting certification. Its potential benefits, complete formalization, and additional details are provided in Appendix~\ref{app:stor_cert}. 


\section{Experiments}
\label{sec:experiments}

 \begin{table*}[!t]
  \captionsetup{font=small}
   \vspace{-1.8em}
  \caption{Certified performance (\ie, ERA and AER) of Cert-SSBD and RAB on MNIST, CIFAR-10, and ImageNette under the all-to-all setting with representative attacks (\ie, one-pixel, four-pixel, and blending). At each radius, we report the best result over different noise levels; the best results are marked in boldface.}
  \vspace{-0.5em}
  \centering
  \setlength{\tabcolsep}{6pt} 
  \renewcommand{\arraystretch}{1.2} 
  \scalebox{0.96}{
    \begin{tabular}{c|c|c|c|cccccccc}
      \toprule
      \multirow{2}{*}{Dataset$\downarrow$} & \multirow{2}{*}{Attack Setting$\downarrow$, Metric$\xrightarrow{}$ } & \multirow{2}{*}{Method$\downarrow$}& \multirow{2}{*}{AER} & \multicolumn{8}{c}{Radius (ERA$\uparrow$)} \\ 
      \cline{5-12}
      & & & & 0 & 0.25 & 0.5 & 0.75 & 1.0 & 1.25 & 1.5 & 1.75 \\
      \hline
      \multirow{6}{*}{MNIST} 
      & \multirow{2}{*}{One-pixel}
      & RAB & 1.46&  99.95 & 99.81 & 99.62 & 99.48 & 98.77 & 95.93 & 61.94 & 0 \\
      & &Cert-SSBD &\textbf{1.67} &\textbf{99.95} & \textbf{99.86} & \textbf{99.72} & \textbf{99.53} & \textbf{99.11} & \textbf{97.87} & \textbf{92.11} & \textbf{11.11}  \\
      \cline{2-12}
      & \multirow{2}{*}{Four-pixel}
      & RAB &1.44& \textbf{99.95} & \textbf{99.86} & 99.62 & 87.61 & 98.72 & 95.41 & 46.24 & 0 \\
      & &Cert-SSBD&\textbf{1.66} & 99.91 & 99.81 & \textbf{99.76} & \textbf{99.57} & \textbf{99.11} & \textbf{98.35} & \textbf{92.77} & \textbf{5.21} \\
      \cline{2-12}
      & \multirow{2}{*}{Blending}
      & RAB &1.46& 99.91 & 99.86 & 99.67 & 99.34 & 98.72 & 95.56 & 60.57 & 0 \\
      & &Cert-SSBD&\textbf{1.66} & \textbf{99.95 }& \textbf{99.91} & \textbf{99.77} & \textbf{99.72} & \textbf{99.05} & \textbf{97.97} & \textbf{92.25} & \textbf{16.17} \\
 \hline

\multirow{6}{*}{CIFAR-10} 
      & \multirow{2}{*}{One-pixel}
      & RAB & 0.54&  86.50 & 69.70 & 55.90 & 36.05 & 14.11 & 2.85 & 0.05 & 0 \\
      & &Cert-SSBD &\textbf{0.62} & \textbf{86.55} & \textbf{74.25} & \textbf{61.50} & \textbf{42.35} & \textbf{21.25} & \textbf{5.55} & \textbf{1.80} & \textbf{0.5} \\
      \cline{2-12}
      & \multirow{2}{*}{Four-pixel}
      & RAB &0.55& \textbf{87.70} & 69.70 & 56.80 & 38.15 & 17.05 & 3.10 & 0.05 & 0 \\
      & &Cert-SSBD& \textbf{0.73} & 85.55 & \textbf{74.75} & \textbf{68.35} & \textbf{58.15} & \textbf{39.95} & \textbf{10.65} & \textbf{1.20} & \textbf{ 0.05} \\
      \cline{2-12}
      & \multirow{2}{*}{Blending}
      & RAB &0.50& 87.40 & 49.50 & 24.10 & 2.65 & 0 & 0 & 0 & 0 \\
      & &Cert-SSBD&\textbf{0.67}&\textbf{86.90} & \textbf{68.10} & \textbf{50.10} & \textbf{22.35} & \textbf{22.35} & \textbf{0.45} & 0 & 0 \\

      \hline

\multirow{6}{*}{ImageNette} 
      & \multirow{2}{*}{One-pixel}
      & RAB & 0.49&94.56 &  73.36 & 52.86 & 35.04 & 14.24 & 0 & 0  & 0 \\
      & &Cert-SSBD &\textbf{0.74} &\textbf{94.62} & \textbf{81.06} & \textbf{61.46} & \textbf{50.84} & \textbf{38.84} & \textbf{25.08} & \textbf{10.28} & 0 \\
      \cline{2-12}
      & \multirow{2}{*}{Four-pixel}
      & RAB &0.48& \textbf{94.44} & 73.66 & 51.48 & 33.24 & 13.46 & 0 & 0 & 0 \\
      & &Cert-SSBD&\textbf{0.65} &\textbf{94.00} & \textbf{77.78} & \textbf{59.64} & \textbf{44.80} & \textbf{28.36} & \textbf{14.40} & \textbf{5.68} & 0 \\
      \cline{2-12}
      & \multirow{2}{*}{Blending}
      & RAB &0.48& \textbf{94.66} & 74.28 & \textbf{51.56 }& 33.60 & 13.28 & 0 & 0 & 0 \\
      & &Cert-SSBD&\textbf{0.70} & 93.32 & \textbf{78.16} & 42.48 & \textbf{52.34} & \textbf{38.26} & \textbf{17.28} & \textbf{1.40} & 0 \\
      \bottomrule
    \end{tabular}
  }
  \label{table:All-to-all_ERA}
\end{table*}

 \begin{table*}[!t]

  \captionsetup{font=small}
  \caption{Certified performance (\ie, CRA and ACR) of Cert-SSBD and RAB on MNIST, CIFAR-10, and ImageNette under the all-to-all setting with representative attacks (\ie, one-pixel, four-pixel, and blending). At each radius, we report the best result over different noise levels; the best results are marked in boldface.}
  \vspace{-0.8em}
  \centering
  \setlength{\tabcolsep}{6pt} 
  \renewcommand{\arraystretch}{1.2} 
  \scalebox{0.96}{
    \begin{tabular}{c|c|c|c|cccccccc}
      \toprule
      \multirow{2}{*}{Dataset$\downarrow$} & \multirow{2}{*}{Attack Setting$\downarrow$, Metric$\xrightarrow{}$ } & \multirow{2}{*}{Method$\downarrow$}& \multirow{2}{*}{ACR} & \multicolumn{8}{c}{Radius (CRA$\uparrow$)} \\ 
      \cline{5-12}
      & & & & 0 & 0.25 & 0.5 & 0.75 & 1.0 & 1.25 & 1.5 & 1.75 \\
      \hline
      \multirow{6}{*}{MNIST} 
      & \multirow{2}{*}{One-pixel}
      & RAB & 0.01&  0.19 & 0.14 & 0.10 & 0.10 & 0.10 & 0 & 0 & 0 \\
      & &Cert-SSBD &\textbf{0.77} &\textbf{46.34} & \textbf{46.24} & \textbf{46.15} & \textbf{45.96} & \textbf{45.91} & \textbf{45.34} & \textbf{43.36} & \textbf{6.71}  \\
      \cline{2-12}
      & \multirow{2}{*}{Four-pixel}
      & RAB &0& 0.10 & 0.10 & 0 & 0 & 0 & 0 & 0 & 0 \\
      & &Cert-SSBD&\textbf{0.76} & \textbf{46.29} & \textbf{46.24} & \textbf{46.05} & \textbf{45.91} & \textbf{45.86} & \textbf{45.01} & \textbf{42.36} & \textbf{4.49} \\
      \cline{2-12}
      & \multirow{2}{*}{Blending}
      & RAB &0.01& 0.52 & 0.52 & 0.52 & 0.426 & 0.28 & 0.14 & 0.14 & 0 \\
      & &Cert-SSBD&\textbf{0.77} & \textbf{46.29 }& \textbf{46.24} & \textbf{46.10} & \textbf{45.96} & \textbf{45.82} & \textbf{45.34} & \textbf{42.93} & \textbf{5.39} \\
 \hline

\multirow{6}{*}{CIFAR-10} 
      & \multirow{2}{*}{One-pixel}
      & RAB &0.04&  12.6 & 0 & 0 & 0 & 0 & 0 & 0 & 0 \\
      & &Cert-SSBD &\textbf{0.24} & \textbf{51.55} & \textbf{38.65} & \textbf{26.30} & \textbf{10.20} & \textbf{1.75} & \textbf{0.10} & \textbf{0} & \textbf{0} \\
      \cline{2-12}
      & \multirow{2}{*}{Four-pixel}
      & RAB &0.04& 10.90 & 6.80 & 3.60 & 1.20 & 0.10 & 0 & 0 & 0 \\
      & &Cert-SSBD& \textbf{0.30} & \textbf{48.65} & \textbf{42.60} & \textbf{35.05} & \textbf{19.50} & \textbf{0.10} & 0 & 0 & 0 \\
      \cline{2-12}
      & \multirow{2}{*}{Blending}
      & RAB &0.04& 11.80 & 6.90 & 3.60 &1.20 & 0 & 0 & 0 & 0 \\
      & &Cert-SSBD&\textbf{0.29}&\textbf{47.70} & \textbf{36.90} & \textbf{29.50} & \textbf{21.30} & \textbf{9.65} & 0 & 0 & 0 \\

      \hline

\multirow{6}{*}{ImageNette} 
      & \multirow{2}{*}{One-pixel}
      & RAB & 0.01&7.76 &  3.88 & 1.68 & 0.04 & 0 & 0 & 0  & 0 \\
      & &Cert-SSBD &\textbf{0.44} &\textbf{52.72} & \textbf{46.86} & \textbf{38.24} & \textbf{30.64} & \textbf{22.54} & \textbf{13.64} & \textbf{4.26} & 0 \\
      \cline{2-12}
      & \multirow{2}{*}{Four-pixel}
      & RAB &0.01& 6.52 & 2.84 & 1.04 & 0.48 & 0 & 0 & 0 & 0 \\
      & &Cert-SSBD&\textbf{0.41} &\textbf{56.88} & \textbf{49.42} & \textbf{39.58} & \textbf{28.40} & \textbf{20.78} & \textbf{11.86} & \textbf{3.28} & 0 \\
      \cline{2-12}
      & \multirow{2}{*}{Blending}
      & RAB &0.01& 6.64 & 3.20 & 1.36& 0.32 & 0 & 0 & 0 & 0 \\
      & &Cert-SSBD&\textbf{0.36} & \textbf{50.58} & \textbf{42.02} & \textbf{32.24} & \textbf{24.02} & \textbf{16.58} & \textbf{8.92} & \textbf{2.70} & 0 \\
      \bottomrule
    \end{tabular}
  }
  \label{table:All-to-All_CRA}
   \vspace{-1em}
\end{table*}

\subsection{Main Settings}
\label{sec:main_setting}

\subsubsection{Datasets and Models}
We conduct experiments on MNIST \cite{deng2012mnist}, CIFAR-10 \cite{krizhevsky2009learning}, and ImageNette \cite{Howard2020imagenette}, using a simple CNN model \cite{Gu2017BadNets}, a lightweight ResNet-like model \cite{Cohen2019Certified}, and standard ResNet-18 model \cite{he2016deep}, respectively.

\subsubsection{Training Settings}
We adopt a sample-specific smoothing approach during training. In this stage, we set the number of sampled Gaussian noise vectors (\ie, augmented datasets) to $M = 1,000$ for MNIST and CIFAR-10, and $M = 200$ for ImageNette, resulting in ensembles of 1,000 and 200 models, respectively. Following previous works \cite{Cohen2019Certified,zhai2020macer}, the added noise follows a Gaussian distribution with mean $\mu=0$ and a fixed noise level $\sigma_{0}$, set as follows: for MNIST and CIFAR-10,
$\sigma_{0} \in \{0.12, 0.25, 0.5, 1.0\}$; for ImageNette,
$\sigma_{0} \in \{0.25, 0.5, 1.0\}$.  Additionally, we set the number of stochastic gradient ascent iterations to $T = 1$, the number of Monte Carlo samples to $J = 1$, and the learning rate to $\alpha = 10^{-4}$ (we use $T=100$ during inference unless otherwise specified). During optimization, the sample-specific noise $\sigma_{\bm{x}}^*$ is initialized at the fixed noise level $\sigma_0$ and is iteratively updated via stochastic gradient ascent.

\subsubsection{Attack Settings}
We evaluate the certified performance of Cert-SSBD against three representative backdoor attacks: one-pixel pattern, four-pixel pattern, and random but fixed noise patterns blended across the entire image \cite{chen2017Targeted}. The perturbation magnitude of the attack is controlled by the $\ell_2$-norm of the backdoor patterns, with $\left \| \Delta  \right \| _2 =0.1$.  Following prior work \cite{Weber2023RAB}, we inject 10\% poisoned samples into the MNIST dataset and 5\% into the CIFAR-10 and ImageNette datasets. The goal of these attacks is to induce the model to misclassify inputs as `0' in MNIST, `airplane' in CIFAR-10, and `tench' in ImageNette. In addition to the all-to-one attack, we also consider an all-to-all attack objective \cite{Gu2017BadNets}, where the compromised model alters its predictions based on the original labels. We hereby primarily focus on the perturbation magnitude and the number of injected backdoor samples without considering specific backdoor patterns.

\begin{figure*}[!t]
 \vspace{-1.3em}
    \begin{minipage}[t]{0.5\linewidth}
        \centering
    {
		\includegraphics[width=0.95\linewidth]{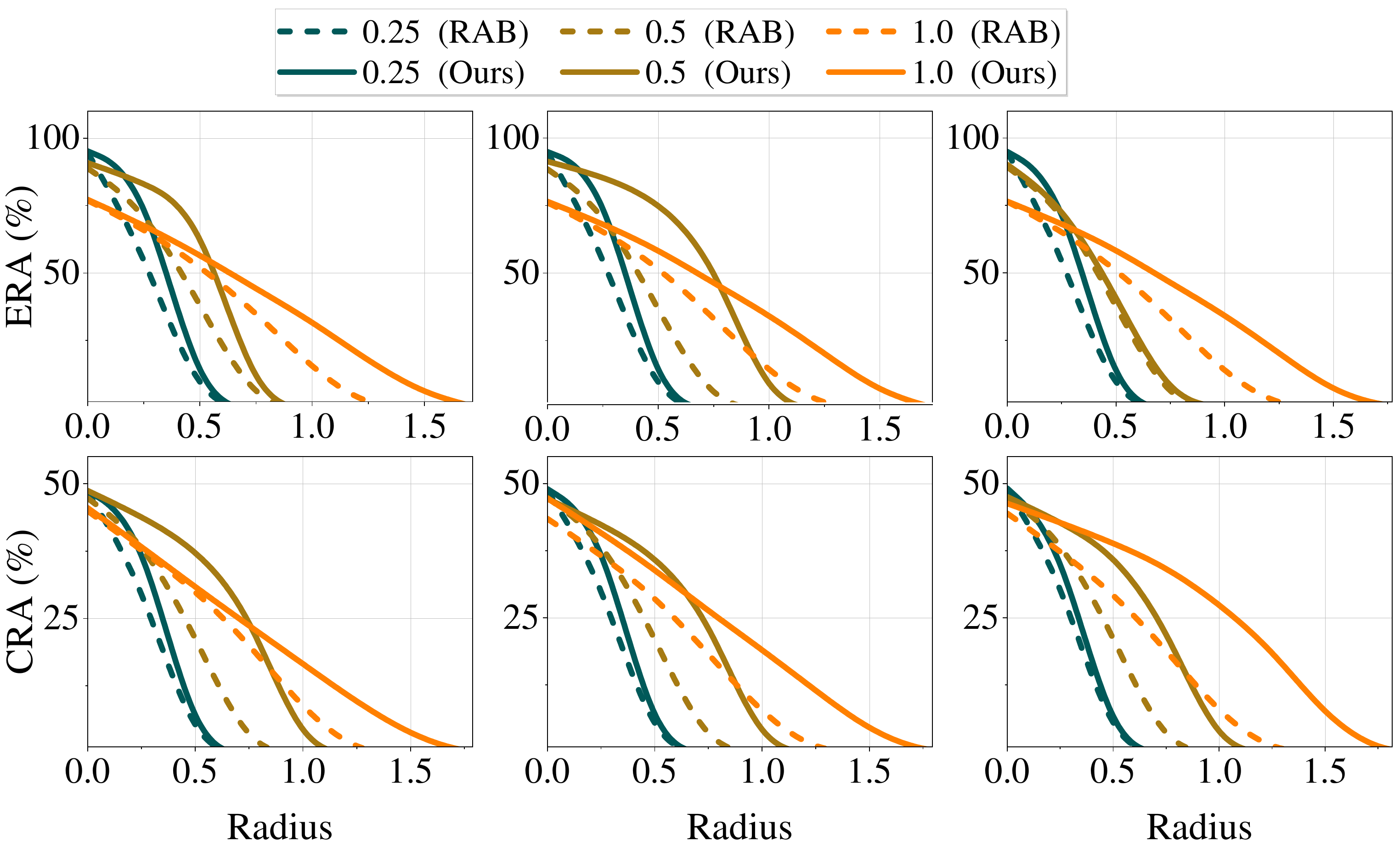}}\hspace{0.6em}
        \vspace{-0.3em}
	\caption{Certified performance (\ie, ERA and CRA) under different certification radii on the ImageNette dataset in the all-to-one setting with various noise levels (0.25, 0.5, and 1.0). The first column corresponds to the one-pixel attack, the second to the four-pixel attack, and the third to the blending attack.}
     \vspace{-0.5em}
     \label{fig:cert_curve_ato}
    \end{minipage}\hspace{0.5em}
    \begin{minipage}[t]{0.5\linewidth}
        \centering
    {
		\includegraphics[width=0.95\linewidth]{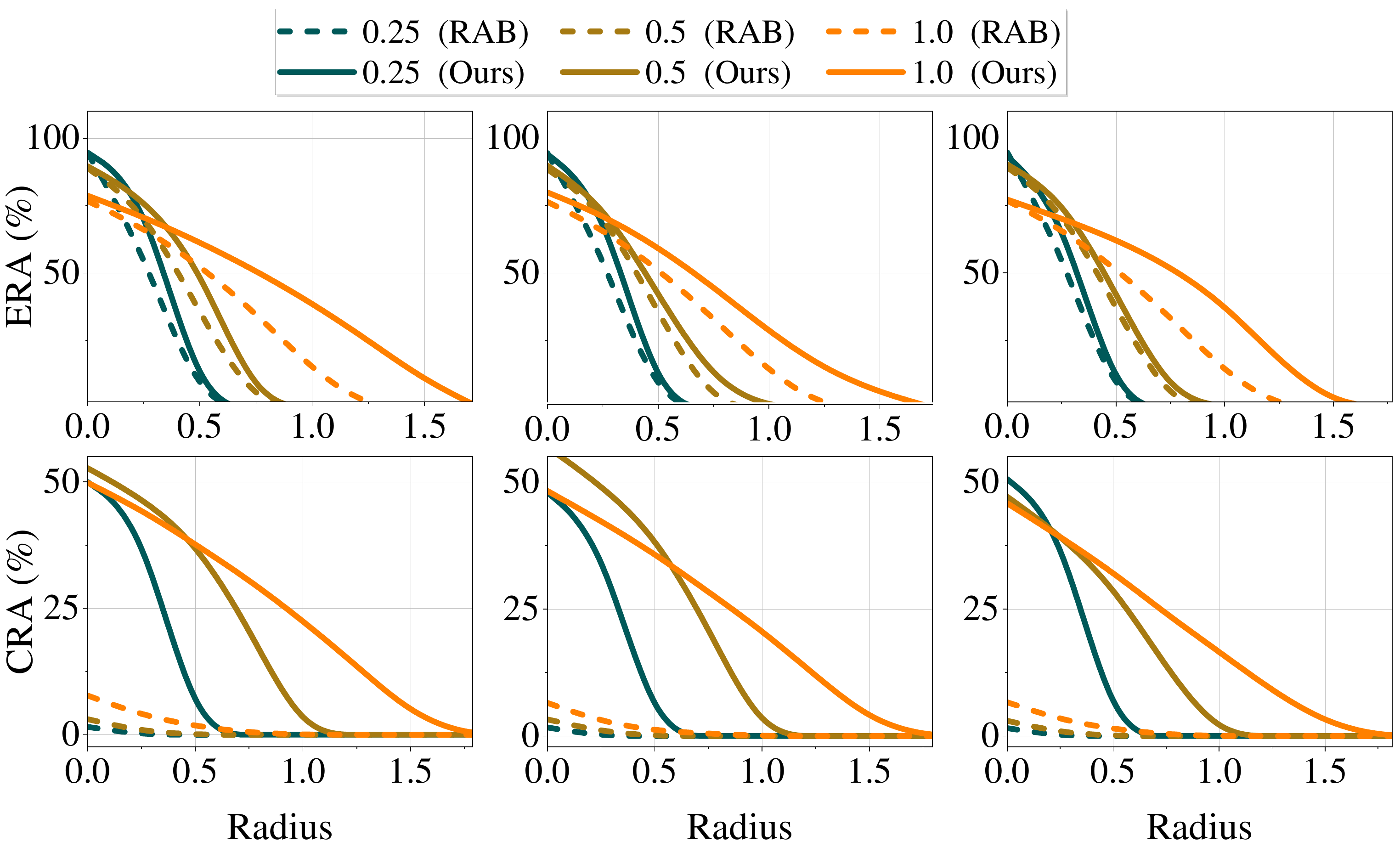}}\hspace{0.6em}
        \vspace{-0.3em}
	\caption{Certified performance (\ie, ERA and CRA) under different certification radii on the ImageNette dataset in the all-to-all setting with various noise levels (0.25, 0.5, and 1.0). The first column corresponds to the one-pixel attack, the second to the four-pixel attack, and the third to the blending attack.}
 \label{fig:cert_curve_ata}
    \end{minipage}%
\end{figure*}

 \begin{figure}[ht!]
    \vspace{-1em}
    \centering
    \includegraphics[width=0.5\textwidth]{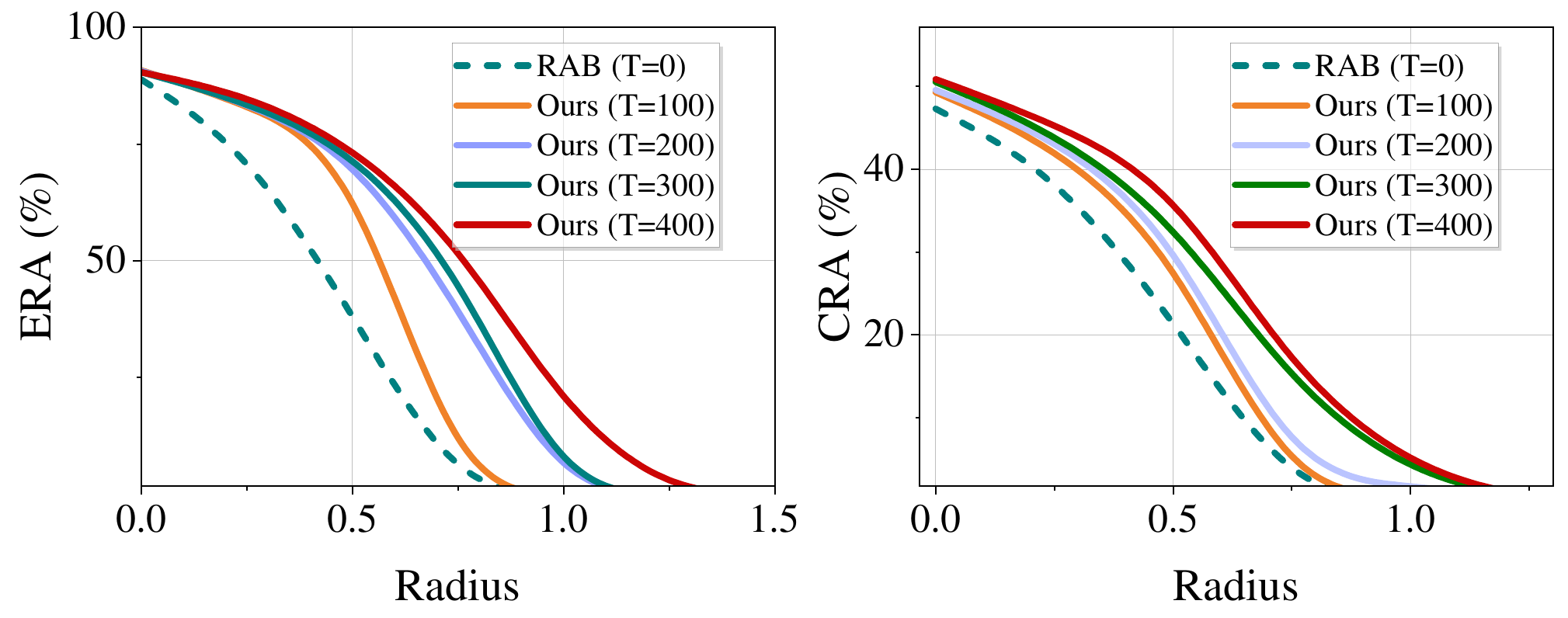}
    \vspace{-1.0em}
    \caption{Effect of stochastic gradient ascent iterations $T$.}
    \vspace{-1.5em}
    \label{fig:vary T}
    
\end{figure}

\subsubsection{Evaluation Metrics}

Following previous works \cite{Cohen2019Certified,Weber2023RAB}, we evaluate the effectiveness of our method using empirical robust accuracy (ERA), certified robust accuracy (CRA), average empirical radius (AER), and average certified radius (ACR). Specifically, ERA refers to the classification accuracy on clean samples (serving as the upper bound for CRA), while CRA denotes the robust accuracy for backdoor samples within the certified radius $r$ (\ie, the predictions are provably invariant within $r$ and correct). AER is the average empirical radius over clean samples, while ACR is the average certified radius over backdoor samples. In general, higher values of ERA, CRA, AER, and ACR indicate better certification performance. In particular, we present certification curves (see Figure~\ref{fig:cert_curve_ato}–\ref{fig:cert_curve_ata}) to intuitively compare certified performance (\ie, ERA and CRA) under different noise levels.

\subsection{Main Results under the All-to-One Setting} \label{main_result_ato}

As shown in Tables \ref{table:All-to-one_ERA}-\ref{table:All-to-one_CRA}, our Cert-SSBD achieves the best performance under the all-to-one setting across three datasets and three attack types (one-pixel, four-pixel, and blending). For instance, on the MNIST dataset, at a radius of 1.5, ERA exceeds 72\% (an improvement of approximately 30\%), while CRA surpasses 45\% (an increase of around 3\%). Even on the more challenging ImageNette dataset, at a radius of 0.75, ERA exceeds 45\% (an improvement of nearly 15\%), and CRA is above 26\% (an increase of 10\%). In both cases, AER and ACR also improve by approximately 0.2. These experimental results validate the effectiveness of our certification method.

 
As shown in Figure \ref{fig:cert_curve_ato}, our method achieves significantly higher ERA and CRA across various noise levels (\eg, 0.25, 0.5, and 1.0) on ImageNette compared to traditional methods, validating its superior performance. Notably, the trade-off between accuracy and robustness is more pronounced in this context: stronger noise tends to degrade performance at smaller radii while improving it at larger ones. The certification curves for CIFAR-10 and MNIST are provided in Appendix~\ref{appC}.

 \subsection{Main Results under the All-to-All Setting} \label{main_result_ata}
 
 As shown in Tables \ref{table:All-to-all_ERA}-\ref{table:All-to-All_CRA}, our Cert-SSBD method also achieves the best performance under the all-to-all setting across three datasets and three attack types (one-pixel, four-pixel, and blending). For example, on the MNIST dataset, at a radius of 1.5, ERA exceeds 92\% (an improvement of approximately 30\%), while CRA surpasses 42\% (an increase of about 40\%). Even on the more challenging ImageNette dataset, at a radius of 0.75, ERA exceeds 40\% (an improvement of nearly 15\%), and CRA is above 20\% (an increase of 20\%). In both cases, AER improves by approximately 0.2, while ACR increases by 0.7 on MNIST and 0.4 on ImageNette. These results validate the effectiveness of our method.

\looseness=-1

As shown in Figure \ref{fig:cert_curve_ata}, our method achieves significantly higher ERA and CRA on ImageNette under various noise levels (\eg, 0.25, 0.5, and 1.0) compared to traditional methods. Furthermore, the trade-off between model accuracy and robustness remains consistent with the all-to-one setting. The curves for CIFAR-10 and MNIST are provided in Appendix~\ref{appC}.

\begin{table*}[!t]
  \vspace{-0.5em}
  \captionsetup{font=small}
  \caption{Certified performance (\ie, ERA and AER) of Cert-SSBD and RAB on MNIST under the all-to-one setting with diverse trigger designs (\ie, BadNets, WaNet, SIG, and an adaptive trigger).
At each radius, we report the best result over different noise levels; the best results are marked in boldface.
}
  \vspace{-0.5em}
  \centering
  \setlength{\tabcolsep}{6pt}
  \renewcommand{\arraystretch}{1.2}
  \scalebox{0.96}{
    \begin{tabular}{c|c|c|cccccccc}
      \toprule
      \multirow{2}{*}{Attack Setting$\downarrow$, Metric$\xrightarrow{}$} 
      & \multirow{2}{*}{Method$\downarrow$} 
      & \multirow{2}{*}{AER} 
      & \multicolumn{8}{c}{Radius (ERA$\uparrow$)} \\
      \cline{4-11}
      & & & 0 & 0.25 & 0.5 & 0.75 & 1.0 & 1.25 & 1.5 & 1.75 \\
      \hline

      \multirow{2}{*}{BadNets}
      & RAB & 1.56 & 99.95 & 99.86 & 99.76 & 99.67 & 99.43 & 98.82 & 74.80 & 0 \\
      & Cert-SSBD & \textbf{1.81} & \textbf{100} & \textbf{100} & \textbf{100}  & \textbf{100} & \textbf{100} & \textbf{99.57} & \textbf{99.15}& \textbf{98.49} \\
      \hline

      \multirow{2}{*}{WaNet}
      & RAB & 1.49 & \textbf{100} & \textbf{99.91} & 99.67 & \textbf{99.57} & \textbf{98.96} & 50.83 & 0 & 0 \\
      & Cert-SSBD & \textbf{1.56} & \textbf{100} & 99.86 & \textbf{99.81} & 99.34 & 98.91 & \textbf{97.07} & \textbf{78.16} & \textbf{39.67} \\
      \hline

      \multirow{2}{*}{SIG}
      & RAB & 1.54 & \textbf{99.95} & \textbf{99.86} & 99.76 & \textbf{99.67} & \textbf{99.48} & 93.48 & 0 & 0 \\
      & Cert-SSBD & \textbf{1.74} & \textbf{99.95} & \textbf{99.86} & \textbf{99.86} & \textbf{99.67} & 99.34 & \textbf{98.82} & \textbf{96.97} & \textbf{88.09} \\
      \hline

      \multirow{2}{*}{Adaptive Trigger}
      & RAB & 1.51 & 99.95 & \textbf{99.86} & 99.76 & 99.57& \textbf{99.20} & 98.44 & 79.43  & 0 \\
      & Cert-SSBD & \textbf{1.65} & \textbf{100} & \textbf{99.86} & \textbf{99.81} & \textbf{99.62} & \textbf{99.20} & \textbf{98.53} & \textbf{93.52} & \textbf{41.70} \\
      \bottomrule
    \end{tabular}
  }
  \label{table:MNIST_All_to_One_Trigger_clean}
  \vspace{0.6em}
\end{table*}

\begin{table*}[!t]
  \vspace{-1em}
  \captionsetup{font=small}
  \caption{Certified performance (\ie, CRA and ACR) of Cert-SSBD and RAB on MNIST under the all-to-one setting with diverse trigger designs (\ie, BadNets, WaNet, SIG, and an adaptive trigger).
At each radius, we report the best result over different noise levels; the best results are marked in boldface.
}
  \vspace{-0.5em}
  \centering
  \setlength{\tabcolsep}{6pt}
  \renewcommand{\arraystretch}{1.2}
  \scalebox{0.96}{
    \begin{tabular}{c|c|c|cccccccc}
      \toprule
      \multirow{2}{*}{Attack Setting$\downarrow$, Metric$\xrightarrow{}$} 
      & \multirow{2}{*}{Method$\downarrow$} 
      & \multirow{2}{*}{ACR} 
      & \multicolumn{8}{c}{Radius (CRA$\uparrow$)} \\
      \cline{4-11}
      & & & 0 & 0.25 & 0.5 & 0.75 & 1.0 & 1.25 & 1.5 & 1.75 \\
      \hline

      \multirow{2}{*}{BadNets}
      & RAB & 1.49 & \textbf{99.95} & \textbf{99.86} & 99.72 & 99.48 & 99.01 & 94.52 & 0 & 0 \\
      & Cert-SSBD & \textbf{1.74} & \textbf{99.95} & \textbf{99.86} & \textbf{99.86}  & \textbf{99.62} & \textbf{99.34} & \textbf{98.77} & \textbf{96.88}& \textbf{87.09} \\
      \hline

      \multirow{2}{*}{WaNet}
      & RAB & 0.68 & \textbf{91.96} & \textbf{46.24} & 46.01 & \textbf{46.01} & \textbf{45.86} & \textbf{45.25} & 39.72 & 0 \\
      & Cert-SSBD & \textbf{0.79} & 46.29 & \textbf{46.24} & \textbf{46.15} & 45.91 & 45.82 & 45.01 & \textbf{43.22} & \textbf{37.92} \\
      \hline

      \multirow{2}{*}{SIG}
      & RAB & 1.53 & \textbf{99.95} & \textbf{99.91} & \textbf{99.86} & \textbf{99.72} & 99.34 & 98.72 & 90.26 & 0 \\
      & Cert-SSBD & \textbf{1.73} & \textbf{99.95} & 99.86 & \textbf{99.86} & \textbf{99.72} & \textbf{99.43} & \textbf{98.82} & \textbf{96.64} & \textbf{86.95} \\
      \hline

      \multirow{2}{*}{Adaptive Trigger}
      & RAB & 0.68 & 46.76 & 46.29 & 46.01 & \textbf{46.01} & \textbf{45.91} & 45.58 & 41.89 & 0 \\
      & Cert-SSBD & \textbf{0.81} & \textbf{99.81} & \textbf{97.26} & \textbf{46.34} & \textbf{46.01} & \textbf{45.91} & \textbf{45.67} & \textbf{44.44} & \textbf{41.23} \\
      \bottomrule
    \end{tabular}
  }
  \label{table:MNIST_All_to_One_Trigger_poisoned}
  \vspace{-0.5em}
\end{table*}

\subsection{Ablation Study}
\label{sec:Ablation}

\looseness=-1
In this section, we conduct ablation studies to analyze the effects of key design choices in Cert-SSBD. Unless otherwise specified, experiments are conducted using the one-pixel attack on the ImageNette dataset under the all-to-one setting. More ablation results are provided in Appendix~\ref{app:model_count}.

\vspace{0.3em}

\noindent \textbf{Effect of Stochastic Gradient Ascent Iterations $\bm{T}$}. 
As shown in Figure \ref{fig:vary T}, both the empirical robust accuracy and certified robust accuracy consistently increase as $T$ increases, particularly at larger certification radii. The underlying reason is that a larger $T$ allows for a more optimized smoothing parameter $\sigma_{\bm{x}}^*$ for each input $\bm{x}$, thereby expanding the certified radius and leaving room for further improvements in strong defense methods. However, excessively increasing $T$ also leads to higher computational costs. Therefore, defenders must choose an appropriate $T$ based on specific requirements.

   \vspace{0.3em}
\noindent \textbf{Effect of Trigger Diversity on Certified Robustness}. We hereby evaluate Cert-SSBD under a more diverse set of backdoor trigger settings, including BadNets~\cite{Gu2017BadNets}, WaNet~\cite{nguyen2021wanet}, SIG~\cite{SIG}, and an adaptive trigger~\cite{qi2023revisiting}. We hereby use the all-to-one setting on MNIST as an example for discussion. All other training and certification settings follow Section~\ref{sec:main_setting} to ensure fair comparisons. Specifically, we adopt a fixed patch trigger placed at the bottom-right corner for BadNets; use the smooth geometric warping-based trigger of WaNet; inject a globally diffused low-amplitude sinusoidal signal as in SIG; and employ an input-aware adaptive trigger whose pattern and placement are jointly optimized with respect to the target model. As shown in Tables~\ref{table:MNIST_All_to_One_Trigger_clean}-\ref{table:MNIST_All_to_One_Trigger_poisoned}, Cert-SSBD consistently outperforms RAB across all four trigger designs, with more pronounced advantages at larger certification radii. Taking BadNets as an example, at radius $r=1.5$, Cert-SSBD achieves an ERA of 99.15\% with a corresponding CRA of 96.88\%, whereas RAB attains an ERA of 74.80\% and its CRA drops to 0. Meanwhile, the AER increases from 1.56 to 1.81, and the ACR improves from 1.49 to 1.74. For the adaptive trigger, at a larger radius $r=1.75$, both the ERA and CRA of RAB drop to 0, while Cert-SSBD still maintains an ERA of 41.70\% and a CRA of 41.23\%. In addition, the AER improves from 1.51 to 1.65, and the ACR increases from 0.68 to 0.81. These results demonstrate that Cert-SSBD remains robust and consistent under more diverse and challenging trigger settings, especially at larger certification radii.

\begin{figure}[ht!]
 \vspace{-1em}
		\centering
		\subfigure[\,clean image]{
			\includegraphics[scale =0.225]
            {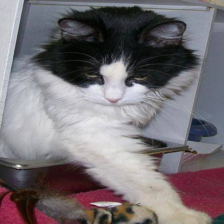}\label{fig4a} } 
		\subfigure[\,clean image]{
			\includegraphics[scale =0.225]{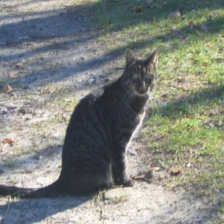}\label{fig4b} }
        \subfigure[\,clean image]{
			\includegraphics[scale =0.225]{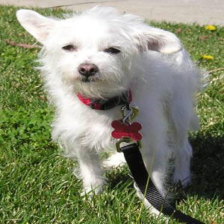}\label{fig4c} } 
		\subfigure[\,clean image]{
			\includegraphics[scale =0.225]{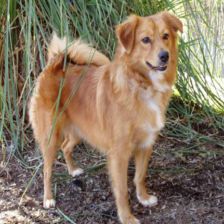}\label{fig5a} }
        
        \subfigure[\,$\sigma_{\bm{x}}^*=0.311$]{
			\includegraphics[scale =0.225]{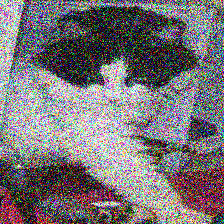}\label{fig5b} }
        \subfigure[\,$\sigma_{\bm{x}}^*=0.248$]{
			\includegraphics[scale =0.2255]{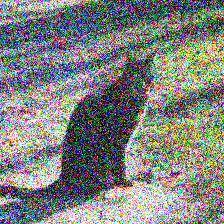}\label{fig5c} }
		\subfigure[\,$\sigma_{\bm{x}}^*=0.236$]{
			\includegraphics[scale =0.225]{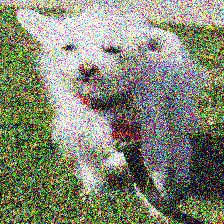}\label{fig6a} }
		\subfigure[\,$\sigma_{\bm{x}}^*=0.299$]{
			\includegraphics[scale =0.225]{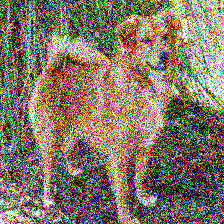}\label{fig6b} }
		\caption{Examples of clean and perturbed images using optimized noise $\sigma_{\bm{x}}^*$ (initialized from $\sigma_0 = 0.25$).}
		\label{fig:visua}
           \vspace{-0.5em}
	\end{figure}

\begin{figure}[ht!]
    \vspace{-0.5em}
    \centering
    \includegraphics[width=0.5\textwidth]{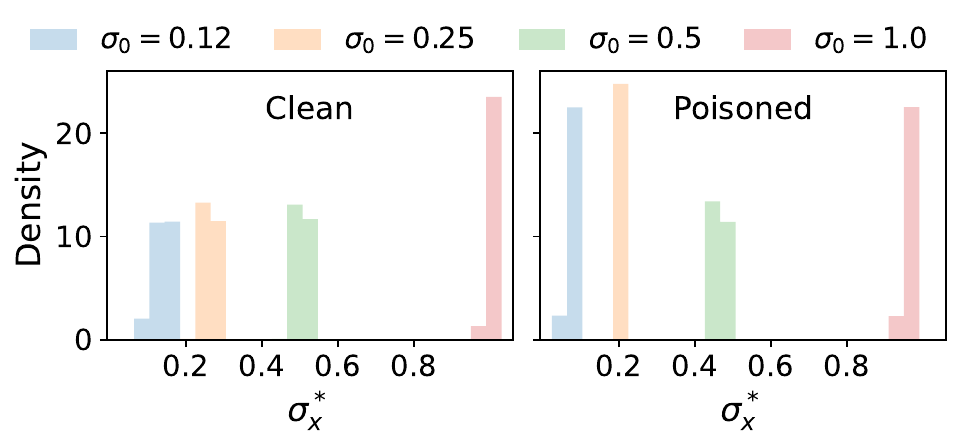}
    \vspace{-1.2em}
    \caption{Distribution of optimized noise $\sigma_x^*$ on the MNIST testing set under different fixed noise levels $\sigma_0$ for clean and poisoned testing samples. The left panel shows clean testing samples, while the right panel shows poisoned testing samples under the one-pixel attack.}
    \vspace{-1.5em}
   \label{fig:sigma_clean}
    
\end{figure}

\subsection{Discussions}
\label{sec:discuss}
\looseness=-1
In this section, we provide further analysis and discussion to better understand the behavior, interpretability, and robustness of the optimized noise $\sigma_{\bm{x}}^*$ under diverse conditions.

\subsubsection{Visualization of Optimized Noise}
We hereby randomly select two input images from two categories, respectively, and perform noise optimization for each input image $\bm{x}$, starting from a fixed initialized noise level of $\sigma_0 = 0.25$, to obtain the optimal noise $\sigma_{\bm{x}}^*$ that maximizes the certified radius. As shown in Figure \ref{fig:visua}, the optimized noise values vary significantly across different inputs, with some being larger and others smaller. Notably, even within the same category, there exist considerable differences among the optimized results. These findings further demonstrate the necessity of adaptively optimizing the noise for each individual input.

\subsubsection{Distribution of Optimized Noise}
We hereby analyze the distribution of the optimized per-sample noise $\sigma_{\bm x}^*$, considering both clean test samples and poisoned test samples generated by a one-pixel attack. For discussion, we use the MNIST dataset under the all-to-one setting as an illustrative example. We report the results under different fixed noise levels, with $\sigma_0 \in \{0.12, 0.25, 0.5, 1.0\}$. As shown in Figure~\ref{fig:sigma_clean}, the optimized noise exhibits a non-uniform distribution across the dataset under all noise levels, indicating that different samples are assigned distinct noise magnitudes. Moreover, as the fixed noise level $\sigma_0$ increases, the overall range of $\sigma_{\bm x}^*$ correspondingly expands. Nevertheless, even under the same fixed noise level, substantial variability persists among individual samples. Therefore, the proposed optimization procedure adaptively adjusts $\sigma_{\bm x_i}^*$ for each sample, rather than learning a single fixed noise level.

\begin{table*}[!t]
  \vspace{-1.5em}
  \captionsetup{font=small}
  \caption{Certified performance (\ie, ERA and AER) of Cert-SSBD on MNIST under the all-to-one setting against margin-aware adaptive poisoning (MAP) combined with representative attacks (\ie, one-pixel, four-pixel, and blending). At each radius, we report the best result over different noise levels; the best certification results are marked in boldface.}
  \vspace{-0.5em}
  \centering
  \setlength{\tabcolsep}{5pt}
  \renewcommand{\arraystretch}{1.2}
  \scalebox{0.96}{
    \begin{tabular}{c|c|ccccccccc}
      \toprule
      \multirow{2}{*}{Attack Setting$\downarrow$, Metric$\xrightarrow{}$} 
      & \multirow{2}{*}{AER} 
      & \multicolumn{9}{c}{Radius (ERA$\uparrow$)} \\
      \cline{3-11}
      & & 0 & 0.25 & 0.5 & 0.75 & 1.0 & 1.25 & 1.5 & 1.75 & 2.0 \\
      \hline
      One-pixel & 1.65 & \textbf{99.95} & \textbf{99.91} & 99.81 & \textbf{99.62} & \textbf{99.34} & \textbf{98.82} & \textbf{86.53} & 42.98 & 0 \\
      One-pixel + MAP & \textbf{1.78} & 99.91 & 99.86 & 99.86 & 99.53 & 98.63 & 95.70 & 85.39 & \textbf{61.23} & \textbf{24.21} \\
      \hline
      Four-pixel & 1.69 & \textbf{99.95} & \textbf{99.86} & 99.72 & \textbf{99.57} & \textbf{99.20} & \textbf{98.63} & 81.94 & 42.98 & 0 \\
      Four-pixel + MAP & \textbf{1.75} & 99.91 & \textbf{99.86} & \textbf{99.81} & 99.39 & 97.97 & 94.14 & \textbf{82.08} & \textbf{56.69} & \textbf{23.64} \\
      \hline
      Blending & 1.70 & \textbf{99.95} & 99.86 & 99.76 & \textbf{99.72} & \textbf{99.20} & \textbf{98.72} & 72.15 & 42.84 & 0 \\
      Blending + MAP & \textbf{1.83} & \textbf{99.91} & 99.91 & \textbf{99.86} & 99.62 & 98.06 & 94.47 & \textbf{84.54} & \textbf{65.96} & \textbf{36.93} \\
      \bottomrule
    \end{tabular}
  }
  \label{table:MNIST_MAP_clean}
\end{table*}

\begin{table*}[!t]
  \vspace{-0.3em}
  \captionsetup{font=small}
  \caption{Certified performance (\ie, CRA and ACR) of Cert-SSBD on MNIST under the all-to-one setting against margin-aware adaptive poisoning (MAP) combined with representative attacks (\ie, one-pixel, four-pixel, and blending). At each radius, we report the best result over different noise levels; the best certification results are marked in boldface.}
  \vspace{-0.5em}
  \centering
  \setlength{\tabcolsep}{5pt}
  \renewcommand{\arraystretch}{1.2}
  \scalebox{0.96}{
    \begin{tabular}{c|c|ccccccccc}
      \toprule
      \multirow{2}{*}{Attack Setting$\downarrow$, Metric$\xrightarrow{}$} 
      & \multirow{2}{*}{ACR} 
      & \multicolumn{9}{c}{Radius (CRA$\uparrow$)} \\
      \cline{3-11}
      & & 0 & 0.25 & 0.5 & 0.75 & 1.0 & 1.25 & 1.5 & 1.75 & 2.0 \\
      \hline
      One-pixel & 0.84 & 46.29 & 46.24 & 46.24 & 46.10 & \textbf{45.96} & \textbf{45.91} & \textbf{45.20} & \textbf{42.88} & 0 \\
      One-pixel + MAP & \textbf{0.90} & \textbf{90.00} & \textbf{90.40} & \textbf{47.16} & \textbf{46.15} & 45.91 & 45.77 & 44.78 & 40.80 & 24.16 \\
      \hline
      Four-pixel & 0.87 & 46.29 & 46.24 & 46.24 & \textbf{46.10} & \textbf{46.01} & \textbf{45.91} & \textbf{45.67} & \textbf{43.88} & 0 \\
      Four-pixel + MAP & \textbf{0.89} & \textbf{93.99} & \textbf{92.01} & \textbf{85.20} & \textbf{46.24} & 45.86 & 45.62 & 44.21 & 38.91 & \textbf{23.12} \\
      \hline
      Blending & 0.87 & 46.34 & 46.29 & 46.24 & \textbf{46.19} & \textbf{45.96} & \textbf{45.91} & \textbf{45.63} & \textbf{44.30} & 0 \\
      Blending + MAP & \textbf{0.95} & \textbf{92.67} & \textbf{87.85} & \textbf{47.47} & \textbf{46.19} & 45.91 & 45.82 & 45.20   & 43.36 & \textbf{36.40} \\
      \bottomrule
    \end{tabular}
  }
  \label{table:MNIST_MAP_poisoned}
  \vspace{-1.0em}
\end{table*}

\subsubsection{Result to Potential Adaptive Attack on Optimized Noise}
Assuming an adaptive adversary who is fully aware of our defense mechanism, the attacker may attempt to weaken the certified robustness of Cert-SSBD in an \emph{indirect} manner. Specifically, in Cert-SSBD, the noise scale $\sigma_{\bm{x}}$ for each sample $\bm{x}_i$ is optimized via stochastic gradient ascent (SGA), and its optimal value $\sigma_{\bm{x}_i}^*$ is closely related to the sample’s proximity to the model’s decision boundary. Therefore, an adaptive attacker can perform targeted poisoning of the training data to push the decision boundary closer to selected target samples, thereby indirectly reducing the statistical advantage (\eg, $P_A - P_B$) and ultimately shrinking the certified radius. 
To evaluate the robustness of Cert-SSBD under such adaptive adversary scenarios, we design a \emph{Margin-Aware Adaptive Poisoning} (MAP) attack to simulate this capability. Unlike standard poisoning attacks that randomly select training samples, MAP strategically selects poisoning samples based on their proximity to a set of \emph{vulnerable} testing samples, where vulnerability is identified via the logit margin. It is worth emphasizing that MAP uses the same trigger patterns as standard attacks (\eg, one-pixel, four-pixel, and blending); its key distinction lies solely in the poisoning sample selection strategy rather than the trigger pattern itself. Specifically, MAP adopts a two-stage heuristic strategy: \textbf{(1)} identifying vulnerable testing samples with small logit margins (\ie, close to the decision boundary), and \textbf{(2)} selecting training samples with the smallest feature distances to these vulnerable samples for poisoning. The detailed formalization 
and implementation of the MAP attack are provided in Appendix~\ref{app:MAP}.

As shown in Tables~\ref{table:MNIST_MAP_clean}-\ref{table:MNIST_MAP_poisoned}, Cert-SSBD demonstrates good robustness against the MAP adaptive attack. The adaptive poisoning strategy does not significantly degrade the certification performance, and some metrics even show improvements. On the clean testing set (Table~\ref{table:MNIST_MAP_clean}), ERA exhibits slight fluctuations at intermediate radii but remains overall stable, with a notable improvement at $r=2.0$, increasing from 0\% to 24.21\%--36.93\%. AER improves to varying degrees under all attacks, with blending + MAP reaching 1.83 (compared to 1.70 for standard poisoning). On the poisoned 
testing set (Table~\ref{table:MNIST_MAP_poisoned}), changes are more pronounced at small and large radii. At $r=0$, CRA shows substantial growth (\eg, one-pixel + MAP increases from 46.29\% to 90.00\%); at $r=2.0$, CRA improves from 0\% to 23.12\%--36.40\%. ACR improves overall, with blending + MAP reaching 0.95 (compared to 0.87 for standard poisoning). These results indicate that the sample-specific noise optimization mechanism of Cert-SSBD exhibits a degree of inherent robustness: even when adaptive poisoning indirectly perturbs the model parameters, the SGA-based optimization of $\sigma_{\boldsymbol{x}}^*$ can still adaptively adjust the noise scale, thereby empirically maintaining relatively stable certified defense performance.

\subsection{The Analysis of Computational Complexity}
\label{sec:Com_Complexity}
In this section, we analyze the computational complexity of Cert-SSBD under an experimental setup running Ubuntu 22.04, equipped with an Intel Xeon Silver 4214 CPU, a Tesla V100-PCIE-32GB GPU, and CUDA 12.0. We particularly focus on the computational costs of the noise optimization and storage-update-based certification processes. A more detailed runtime analysis is provided in Appendix~\ref{appendix:runtime}.

\vspace{0.3em}
\noindent \textbf{The Complexity of Noise Optimization.} Let $N$ and $T$ denote the number of training samples and the number of
stochastic gradient ascent (SGA) iterations used for noise optimization,
respectively. Since Cert-SSBD performs $T$ rounds of SGA-based noise optimization for \emph{each} training sample to obtain sample-specific noise scales,
the overall complexity of the noise optimization phase is $\mathcal{O}(N \cdot T)$. Furthermore, Cert-SSBD supports parallel processing, as the optimization process for each sample is independent. For instance, with $T{=}100$, the per-sample optimization time is approximately 0.01 seconds for MNIST and 0.05 seconds for CIFAR-10.  
Therefore, the additional computational overhead of our method in the noise optimization phase is acceptable.

\vspace{0.3em}
\looseness=-1
\noindent \textbf{The Complexity of Storage-update-based Certification.} 
In this stage, the defender adopts a storage-update-based method to dynamically update the certification process, resolving potential overlaps among certification regions across different samples. For time complexity, let $p$ be the probability that the certified region of a new sample $x_{n+1}$ overlaps with an existing certification region, and let $c$ denote the cost of computing a single certification region. The expected running time is $\mathcal{O}(np + (1-p)(2n+c))$. Specifically, $n$ comparisons are required to check whether the certified region of the new sample overlaps with an existing region; if no overlap occurs, a new region is computed (cost $c$) and the storage is updated (cost $n$). In practice, when overlaps are rare (\ie, $p \approx 0$), the complexity simplifies to $\mathcal{O}(2n+c)$; our experiments confirm that this condition holds across all datasets. For instance, on MNIST and CIFAR-10, executing storage-update-based certification on the entire testing set takes only approximately 5 seconds, which is negligible compared to the certified performance gains. Besides, storing $n$ certified triplets incurs $\mathcal{O}(n)$ memory overhead in practice.

\section{Potential Limitations and Future Directions}
\label{sec:limitations}

As one of the early studies on certified backdoor defenses, we admit that our method may exhibit certain potential limitations, which warrant further investigation in future work. 

Firstly, Cert-SSBD introduces additional computational overhead compared to standard randomized-smoothing-based methods. This overhead primarily arises from two components: sample-specific noise optimization and storage-update-based certification. Specifically, the noise optimization procedure is a one-time offline preprocessing step that can be efficiently parallelized and therefore does not affect deployment-time inference efficiency. The storage-update-based certification mechanism only triggers updates when potential conflicts are detected, resulting in limited overhead in practice. As detailed in Section~\ref{sec:Com_Complexity}, the overall computational cost remains controllable relative to the achieved certified performance gains. Nevertheless, future work may further explore strategies to accelerate and streamline our approach.

Secondly, Cert-SSBD maintains additional storage to support storage-update-based certification, which incurs extra memory and storage overhead. In practice, this overhead is typically manageable and primarily functions as a conservative safeguard to ensure certification consistency. In future work, we will investigate more compact storage representations as well as more efficient update and retrieval mechanisms to further reduce storage costs and improve scalability. 

Thirdly, Cert-SSBD is currently evaluated primarily on image classification tasks and has not yet been systematically extended to other settings, such as text, speech, multimodal learning, or generative models. Arguably, the underlying principles of our approach are broadly applicable. In future work, we plan to extend sample-specific noise learning and the consistency-based certification mechanism to a wider range of modalities and task settings, and to evaluate their effectiveness under more complex attack scenarios and data distributions.

Finally, Cert-SSBD currently adopts isotropic scalar noise and adapts only the noise magnitude on a per-sample basis, without explicitly modeling direction-dependent decision-boundary geometry. This design choice is consistent with common practice in randomized-smoothing-based certification frameworks (see~\cite{maho2022randomized}) and does not affect our main conclusions regarding sample-specific noise learning and consistency-based certification. In future work, we plan to explore sample-specific anisotropic noise (\eg, ellipsoidal certification schemes) to more accurately characterize local decision-boundary geometry.

\section{Conclusion}
\label{sec:conclusion}
In this paper, we revisited existing randomized smoothing-based certified backdoor defense methods and revealed that using fixed noise for all samples led to suboptimal certification performance. To address this issue, we proposed a sample-specific certified backdoor defense method (\ie, Cert-SSBD), which employed stochastic gradient ascent to iteratively optimize sample-specific noise in order to maximize the certification radius. The optimized noise was then injected into the poisoned training set to retrain multiple smoothed models, whose predictions are aggregated to obtain the final robust prediction. Since existing certification methods typically assumed a fixed noise level and thus did not apply to our setting, we further introduced a storage-update-based certification approach to improve certification accuracy and reliability. Extensive experiments on multiple benchmark datasets demonstrated that Cert-SSBD significantly outperformed existing methods in terms of certification performance. We hope this work inspires future exploration of how sample-specific noise relates to model decision boundaries for better personalized certification.

\bibliographystyle{IEEEtran}

\bibliography{main}

\newpage
\appendix

\setcounter{page}{1}

\setcounter{equation}{0}
\setcounter{theorem}{0}
\setcounter{figure}{0}
\setcounter{remark}{0}

\begin{algorithm}[!ht]
\caption{Cert-SSBD Training: Train the Model with Optimized Noise}
\label{alg:train_robust}
\begin{algorithmic}[1]
\STATE \textbf{Input}: Stochastic gradient ascent iterations $T$, poisoned training dataset $\mathcal{D}_p$ (consisting of both poisoned and benign samples), initialized noise scale $\sigma_0$, number of models $M$, learning rate $\alpha$
\STATE \textbf{Output}: Model collection $\{(g_1, \mu_1), \dots, (g_M, \mu_M)\}$
\FOR{$m = 1, \dots, M$}
    \STATE \textbf{Step 1: Optimize Sample-Specific Noise $\sigma_{\bm{x}}^*$}
    \STATE Initialize $\sigma_{\bm{x}}^0 = \sigma_0$
    \FOR{$t = 0, \dots, T-1$}
        \STATE Sample $\hat{Z _1},..., \hat{Z _J}(\hat{D _1},..., \hat{D _J})\sim \mathcal{N} (0,I)$
        \STATE Compute class probabilities:
        
        $\varphi (\sigma _{\bm{x}}^{j} )=\frac{1}{J} {\textstyle \sum_{j=1}^{J}} f(  (\bm{x} + \mathcal{B}_{\bm{x}}+\sigma_x^t \hat{Z}_j,\mathcal{D} + \bm{\delta}+\sigma_x^t \hat{D}_j ) )$

        \STATE Define $F_A(\sigma _x^t)=\max_y\varphi _y , y_A= \arg \max _y\varphi _y$, and $F_B(\sigma _{\bm{x}}^t)=\max_{y_B\ne y_A}\varphi _y $
        \STATE Compute certified radius:
        
        $r(\sigma_{\bm{x}}^t  )=\frac{\sigma_{\bm{x}}^t }{2}(\Phi ^{-1}(F_A)-\Phi ^{-1}(F_B))$
        \STATE Update $\sigma_{\bm{x}}^{t+1}  =\sigma_x^t +\alpha \bigtriangledown _{\sigma_{\bm{x}}^t}r(\sigma_{\bm{x}}^t)$
    \ENDFOR
    \STATE Set $\sigma_{\bm{x}_i}^* = \sigma_{\bm{x}}^T$ for all $t$

    \STATE \textbf{Step 2: Robust Training Process}
    \STATE Sample noise vectors $b_ {m_1},..., b_ {m_n}\sim  {\textstyle \prod_{i=1}^{n}} \mathcal{N} (0,I)$
    \STATE Construct augmented dataset:
    
   $\mathcal{D}_p^{(m)} \triangleq \mathcal{D}_p + \{\sigma_{\bm{x}_i}^* b_{m,i}\}_{i=1}^n$
    \STATE Train model $g_m(\bm{x}, \mathcal{D}_p^{(m)}, \sigma_{\bm{x}}^*) = \text{train\_model}(\mathcal{D}_p^{(m)})$
    \STATE Deterministically sample and store a \emph{unit-scale base noise vector} 
    $\mu_m\sim\mathcal{N}(0,I_d)$ using a random seed derived from $\textit{hash}(g_m(\bm{x}, \mathcal{D}_p^{(m)}, \sigma_{\bm{x}}^*))$ 
    
\ENDFOR

\end{algorithmic}
\end{algorithm}

\subsection{The Detailed Algorithm for Cert-SSBD}
\label{appA}

We hereby provide the complete algorithmic description of Cert-SSBD, including its training and inference procedures, as well as the storage-update mechanism used during certification.
Specifically, Algorithm~\ref{alg:train_robust} details the training procedure of Cert-SSBD with optimized sample-specific noise.
Algorithm~\ref{alg:cert_infer_ssb} presents the overall inference and certification pipeline, which invokes a storage-update mechanism to resolve potential conflicts between certified regions.
The storage-update strategy itself is formalized in Algorithm~\ref{alg:storage_update}, which is called as a subroutine during inference.

\begin{algorithm}[!t]
\caption{Cert-SSBD Inference: Storage-update-based Certification.}
\label{alg:cert_infer_ssb}
\begin{algorithmic}[1]
\STATE \textbf{Input:} 
Test sample $\bm{x}$, sample-specific noise scale $\sigma_{\bm{x}}^*$,
poisoned training datasets $\{\mathcal{D}_p^{(m)}\}_{m=1}^{M}$,
models $\{(g_m,\mu_m)\}_{m=1}^M$,
confidence level $\alpha$,
storage set $\mathcal{S}$
\STATE \textbf{Output:} Prediction $\mathcal{Y}$ and certified region $\mathcal{R}$ (after storage-update); or \textbf{ABSTAIN}
\STATE Compute vote counts: 

$cnts[y] = \left|\left\{m: g_m(\bm{x}+\sigma_{\bm{x}}^*\mu_m,\ \mathcal{D}_p^{(m)})=y\right\}\right|$,
$\forall y \in \{1,\ldots,K\}$
\STATE Identify the two classes $y_A$ and $y_B$ with the largest vote counts according to $cnts$
\STATE Compute the lower confidence bound $P_A$ for $y_A$ and the upper confidence bound $P_B$ for $y_B$ using one-sided binomial confidence bounds at level $1-\alpha$. 

\IF{$P_A \le P_B$}
    \STATE \textbf{return} \textbf{ABSTAIN}.
\ENDIF
\STATE Compute the certified robust radius $r (g;\sigma_{\bm{x}}^*)$ according to Eq.~(\ref{eq:radius}) in Theorem~\ref{theorem:robust_cert} 

\STATE Construct the certified region:

$\mathcal{R} \leftarrow \textsc{Ball}(\bm{x},\, r (g;\sigma_{\bm{x}}^*))$
\triangledComment{$\textsc{Ball}(\bm{x}, r)$ constructs an $\ell_2$-ball centered at $\bm{x}$ with radius $r$}

\STATE Initialize the new certification triplet: 

$(\bm{x}_{\text{new}},\mathcal{Y}_{\text{new}},\mathcal{R}_{\text{new}}) \leftarrow (\bm{x},y_A,\mathcal{R})$
\STATE Apply the storage-update-based certification strategy: 

$((\bm{x}_{\text{new}}, y_{\text{new}}, \mathcal{R}_{\text{new}}), \mathcal{S}) \leftarrow \textsc{StorageUpdate}((\bm{x}_{\text{new}}, \bm{x}_{\text{new}}, \mathcal{R}_{\text{new}}), \mathcal{S}$) 

\STATE \textbf{return} $\mathcal{Y}_{\text{new}}$, $\mathcal{R}_{\text{new}}$ 

\end{algorithmic}
\end{algorithm}

\begin{algorithm}[!t]
\caption{\textsc{StorageUpdate} (Storage-update-based Certification).}
\label{alg:storage_update}
\begin{algorithmic}[1]
\STATE \textbf{Input:} New triplet $(x_{n+1}, y_{n+1}, R_{n+1})$ and storage set 
$\mathcal{S} = \{(x_i, y_i, R_i)\}_{i=1}^n$,
\emph{where $S$ is maintained to satisfy the non-overlapping property
for inputs with different predicted labels.}
\STATE \textbf{Output:} Updated $(\bm{x}_{n+1},\mathcal{Y}_{n+1},\mathcal{R}_{n+1})$ and updated $\mathcal{S}$

\vspace{0.2em}
\STATE \textbf{Case 1: Non-overlapping Certification Regions.}
\IF{$\forall i \neq j$, whenever $\mathcal{Y}_i \neq \mathcal{Y}_j$, it holds that $\mathcal{R}_i \cap \mathcal{R}_j = \emptyset$}

 \STATE Keep existing triplets in $\mathcal{S}$ unchanged
\ENDIF

\vspace{0.2em}
\STATE \textbf{Case 2: Overlapping Certification Regions with Consistent Predictions.}
\IF{$\exists i \in \{1,\ldots,n\}$ such that $\mathcal{R}_i \cap \mathcal{R}_{n+1} \neq \emptyset$ \textbf{ and } $\mathcal{Y}_{n+1} = \mathcal{Y}_i$}
\STATE Add $(\bm{x}_{n+1},\mathcal{Y}_{n+1},\mathcal{R}_{n+1})$ directly to $\mathcal{S}$
     \STATE \textbf{return} $(\bm{x}_{n+1},\mathcal{Y}_{n+1},\mathcal{R}_{n+1})$, $\mathcal{S}$
\ENDIF

\vspace{0.2em}
\STATE \textbf{Case 3: Overlapping Certification Regions with Inconsistent Predictions.}
\IF{$\exists i \in \{1,\ldots,n\}$ such that $\mathcal{R}_i \cap \mathcal{R}_{n+1} \neq \emptyset$ \textbf{ and } $\mathcal{Y}_{n+1} \neq \mathcal{Y}_i$}
    \STATE Choose one such conflicting index $i$
    \IF{$\bm{x}_{n+1} \in \mathcal{R}_i$}
        \STATE Let $\tilde{\mathcal{R}}_{n+1}$ be the largest subset such that 
        $\tilde{\mathcal{R}}_{n+1}\subseteq \mathcal{R}_{n+1}$ and $\tilde{\mathcal{R}}_{n+1}\subseteq \mathcal{R}_i$
        \STATE $\mathcal{R}_{n+1}\leftarrow \tilde{\mathcal{R}}_{n+1}$; \quad $\mathcal{Y}_{n+1}\leftarrow \mathcal{Y}_i$
    \ELSE
        \STATE Let $\mathcal{R}'_{n+1}$ be the largest subset such that
        $\mathcal{R}'_{n+1}\subseteq \mathcal{R}_{n+1}$ and $\mathcal{R}'_{n+1}\cap \mathcal{R}_i=\emptyset$
        \STATE $\mathcal{R}_{n+1}\leftarrow \mathcal{R}'_{n+1}$
    \ENDIF
\ENDIF

\vspace{0.2em}
\STATE Add the final (possibly updated) triplet 

$(\bm{x}_{n+1},\mathcal{Y}_{n+1},\mathcal{R}_{n+1})$ to $\mathcal{S}$
\STATE \textbf{return} $(\bm{x}_{n+1},\mathcal{Y}_{n+1},\mathcal{R}_{n+1})$, $\mathcal{S}$
\end{algorithmic}
\end{algorithm}

\subsection{Proof of Theorem \ref{theorem:robust_cert}}
\label{appB}

Here we provide the proof for Theorem \ref{theorem:robust_cert_0}. As the proof is based on statistical hypothesis testing, we begin by defining the type-I and type-II error probabilities. Formally, we denote the type-I error probability under the null hypothesis $H_0$ as $\alpha (\phi )=\alpha (\phi;H_0 )$ and the type-II error probability under the alternative hypothesis $H_1$ as $\beta(\phi )=\beta(\phi;H_1)$. To facilitate the proof of Theorem \ref{theorem:robust_cert_0}, we first state and apply Lemma \ref{lemma:RAB}, which establishes a key robustness condition based on hypothesis testing. This result ensures that the classifier's decision remains stable under specified probability constraints, even in the presence of perturbations.

	\begin{lemma}[\cite{Weber2023RAB}]\label{lemma:RAB}
		Let $g$ be the sample-specific smoothed classifier defined as $g(\bm{x},\mathcal{D}, \sigma)= \arg\max_{y} \mathcal{P}_{\bm{\epsilon}(Z,D)} \big( f(\bm{x} + Z , \mathcal{D} + D) = y \big)$, where the smoothing distribution is given by $X \triangleq (Z,D)$, with $Z$ taking values in $\mathbb{R}^d$ and $D$ being a collection of $n$ independent $\mathbb{R}^d$-valued random variables: $D = (D^{(1)},\cdots,D^{(n)}) = (\sigma_{\bm{x}_1}^* \bm{\epsilon}^1, \cdots, \sigma_{\bm{x}_n}^* \bm{\epsilon}^n)$, where $\bm{\epsilon} \sim \mathcal{N}(0, I)$. Let $\mathcal{B}_{\bm{x}} \in R^d $ and let $\bm{\delta}=(\bm{\Delta}_1, \bm{\Delta}_2,...,\bm{\Delta}_n)$ for backdoor patterns  $\bm{\Delta}_i \in R^d $. Let $ y_A \in \mathcal{Y} $  and let $P_A$, $P_B \in [0,1]$ such that $ y_A=g(\bm{x},\mathcal{D},\sigma) $ and
        \begin{equation}
				\mathcal{P}_{\bm{\epsilon}}(g(\bm{x}, \mathcal{D},\sigma ) = y_A) \ge P_A \ge P_B \ge \max_{y \neq y_A}  \mathcal{P}_{\bm{\epsilon}}(g(\bm{x} ,\mathcal{D},\sigma ) = y),
		\end{equation}
		If the optimal type II errors, for testing the null $X\sim H_0$ against the alternative  $X+(\mathcal{B}_{\bm{x}},\bm{\delta})\sim H_1$, satisfy
		\begin{equation}
				\beta ^*(1-P_A;H_1)+\beta ^*(P_B;H_1)>1,
		\end{equation}
		then it is guaranteed that $y_A = g(\bm{x} + \mathcal{B}_{\bm{x}}, \mathcal{D} + \bm{\delta}, \sigma)$.
	\end{lemma}
    
	Building upon Lemma~\ref{lemma:RAB}, we derive Theorem~\ref{theorem:robust_cert_0}, which formally guarantees robustness by providing an explicit certified radius within which the classifier’s prediction remains unchanged. The key idea is to ensure that the likelihood ratio test satisfies the probability bounds established earlier.

\begin{theorem}[Certified Robustness of Cert-SSBD]
\label{theorem:robust_cert_0}
 Let $\mathcal{B}_{\bm{x}} \in \mathbb{R}^d$ denote a backdoor trigger applied to the test input, and let $\bm{\delta} \triangleq (\bm{\Delta}_1,\bm{\Delta}_2,\ldots,\bm{\Delta}_n)$ denote the collection of training-set perturbations, where
$\bm{\Delta}_i \in \mathbb{R}^d$ and $\bm{\Delta}_i=\bm{0}$ for benign training samples (as defined in Section~\ref{sec:pre}),  and let $\mathcal{D}$ be a training set, and let smoothing noise $\hat{Z} \sim \mathcal{N}(0,I)$, $\hat{D} \sim \mathcal{N}(0,I)$. 
 Let $ y_A \in \mathcal{Y} $, such as $y_A = g(\bm{x} + \mathcal{B}_{\bm{x}},\mathcal{D} + \bm{\delta}) $ with class probabilities satisfying $\mathcal{P}_{\bm{\epsilon}(\hat{Z},\hat{D})}(f(\bm{x} + \mathcal{B}_{\bm{x}}+\sigma_{\bm{x}}^*\hat{Z},\mathcal{D} + \bm{\delta}+\sigma_{\bm{x}}^*\hat{D}) = y_A) \ge P_A \ge P_B \ge \max_{y_B \neq y_A}  \mathcal{P}_{\bm{\epsilon}(\hat{Z},\hat{D})}(f(\bm{x} + \mathcal{B}_{\bm{x}}+\sigma_{\bm{x}}^*\hat{Z},\mathcal{D} + \bm{\delta}+\sigma_{\bm{x}}^*\hat{D}) = y)$. Then, we have $g(\bm{x} + \mathcal{B}_{\bm{x}},\mathcal{D} )=g(\bm{x} + \mathcal{B}_{\bm{x}},\mathcal{D} +\bm{\delta})= y_A$ for all training-set perturbations $\bm{\delta}$ satisfying $ \sqrt{\sum_{i=1}^{n} \left \| \bm{\Delta} _i  \right \|_2^2 }\le r (g;\sigma_{\bm{x}}^*)$, where the certified robust radius $r$ is given by
		\begin{equation}\label{eq:radius_0}
			\begin{array}{ll}
                r(g;\sigma_{\bm{x}}^*) = \frac{\sigma_{\bm{x}}^* }{2} \left ( \Phi ^{-1}(P_A(\sigma_{\bm{x}}^*)  )-\Phi^{-1} (P_B (\sigma_{\bm{x}}^*) ) \right ).
			\end{array}
		\end{equation}
		
	\end{theorem}

	\begin{proof}
    We prove this theorem by directly applying Lemma \ref{lemma:RAB}. Consider the smoothing noise jointly distributed as $X=(Z,D)$ and define the perturbed and unperturbed input distributions as follows: $\tilde{X}=(\mathcal{B}_{\bm{x}},\bm{\delta} ) +X, \quad
     \tilde{X'} \triangleq (\mathcal{B}_{\bm{x}},0) + X$.
Correspondingly, the probability of the smoothed classifier can be expressed as: $\mathcal{P}_\epsilon(\tilde{g}(\bm{x},\mathcal{D} ) = y) = \mathcal{P}_\epsilon(g(\bm{x} + \mathcal{B}_{\bm{x}},\mathcal{D} + \bm{\delta}) = y)$. By assumption, the classifier satisfies: 
    \begin{equation}
\mathcal{P}_\epsilon(\tilde{g}(\bm{x} ,\mathcal{D} ) = y_A)\ge  P_A, \quad
    \max_{y_B \neq y_A}  \mathcal{P}_\epsilon(\tilde{g}(\bm{x} ,\mathcal{D} ) = y ) \le  P_B. 
    \end{equation}

Applying Lemma \ref{lemma:RAB}, it follows that if $\beta (\phi_a ) + \beta (\phi_b ) > 1$, then the classifier output remains unchanged under perturbations, ensuring: $\tilde{g}(\bm{x}, \mathcal{D} ) = \tilde{g}(\bm{x},\mathcal{D}-\bm{\delta}) =y_A$. To verify this condition, we analyze the likelihood ratio between $\tilde{X}$ and $\tilde{X'}$ at $z=(\bm{x},b)$, given by $\Lambda (z)=\exp\{  {\textstyle \sum_{i=1}^{n}}(-\frac{\left \| \bm{\Delta}_i \right \|^2}{2{(\sigma _{\bm{x}}^*)}^2} +\frac{b_i^T\bm{\Delta}_i}{{(\sigma _{\bm{x}}^*)}^2} ) \}$.
Since Gaussian distributions assign probability density rather than discrete probabilities, any likelihood ratio test takes the form:
		\begin{equation}
			\begin{array}{ll}
				\phi _t(z)=\left\{\begin{matrix}
					1& \Lambda (z)\ge t,\\
					0&\Lambda (z)< t.
				\end{matrix}\right.
			\end{array}
		\end{equation}
        
      To compute the error probabilities, the threshold for $P\in [0,1]$ is given by: $t_P \triangleq \exp(\Phi ^{-1}(P) \frac{\sqrt{\sum_{i=1}^{n}\left \|\bm{\Delta}_i \right \| ^2_2}}{\sigma _{\bm{x}}^*}
      - \frac{\sum_{i=1}^{n}\left \|\bm{\Delta}_i \right \| ^2_2}{2(\sigma _{\bm{x}}^*)^2})$ and note that $ \alpha (\phi(t_P) )=1-P$ since $\alpha (\phi(t_P) )=1-\Phi (\frac{\log(t_P)+\frac{1}{2}  \frac{\sum_{i=1}^{n}\left \|\bm{\Delta}_i \right \| ^2_2}{(\sigma _{\bm{x}}^*)^2}}
      { {\textstyle \frac{\sqrt{\sum_{i=1}^{n}\left \|\bm{\Delta}_i \right \| ^2_2}}{\sigma _{\bm{x}}^*}}})$, where $\Phi$ is the CDF of the standard normal distribution. For the test $\phi_a=\phi_{t_a}$ with $t_a\equiv t_{P_A}$, the type I error probability satisfies: $\alpha (\phi_a)=1-P_A$. Similarly, for $\phi_b=\phi_{t_b}$ with $t_b\equiv t_{1-P_B}$, we have:  $\alpha (\phi_a)=P_B$. Evaluating the type II error probabilities, we obtain: $\beta (\phi_a)=\Phi (\Phi ^{-1}(P_A)-\frac{\sqrt{\sum_{i=1}^{n}\left \|\bm{\Delta}_i \right \| ^2_2}}{\sigma _{\bm{x}}^*})$, $\beta (\phi_b)=\Phi (\Phi ^{-1}(1-P_B)-\frac{\sqrt{\sum_{i=1}^{n}\left \|\bm{\Delta}_i \right \| ^2_2}}{\sigma _{\bm{x}}^*})$. 
      Substituting these into condition $\beta (\phi_a ) + \beta (\phi_b ) > 1$, we conclude that the inequality holds if and only if: $\sqrt{\sum_{i=1}^{n} \left \| \bm{\Delta} _i  \right \|_2^2 }  < \frac{\sigma_{\bm{x}}^*}{2} \left ( \Phi ^{-1}(P_A )-\Phi ^{-1}(P_B  ) \right )$.  
    Rearranging, the certified robust radius is obtained as: $r(g;\sigma_{\bm{x}}^*) = \frac{\sigma_{\bm{x}}^* }{2} \left ( \Phi ^{-1}(P_A  )-\Phi^{-1} (P_B ) \right )$.
  Thus, the classifier $g$ remains robust against backdoor patterns,  ensuring: $\tilde{g}(\bm{x}, \mathcal{D} ) = \tilde{g}(\bm{x},\mathcal{D}-\bm{\delta}) =y_A$.
  \end{proof}

  \begin{figure*}[!t]
    \vspace{-1em}
    \begin{minipage}[t]{0.5\linewidth}
        \centering
    {
		\includegraphics[width=0.95\linewidth]{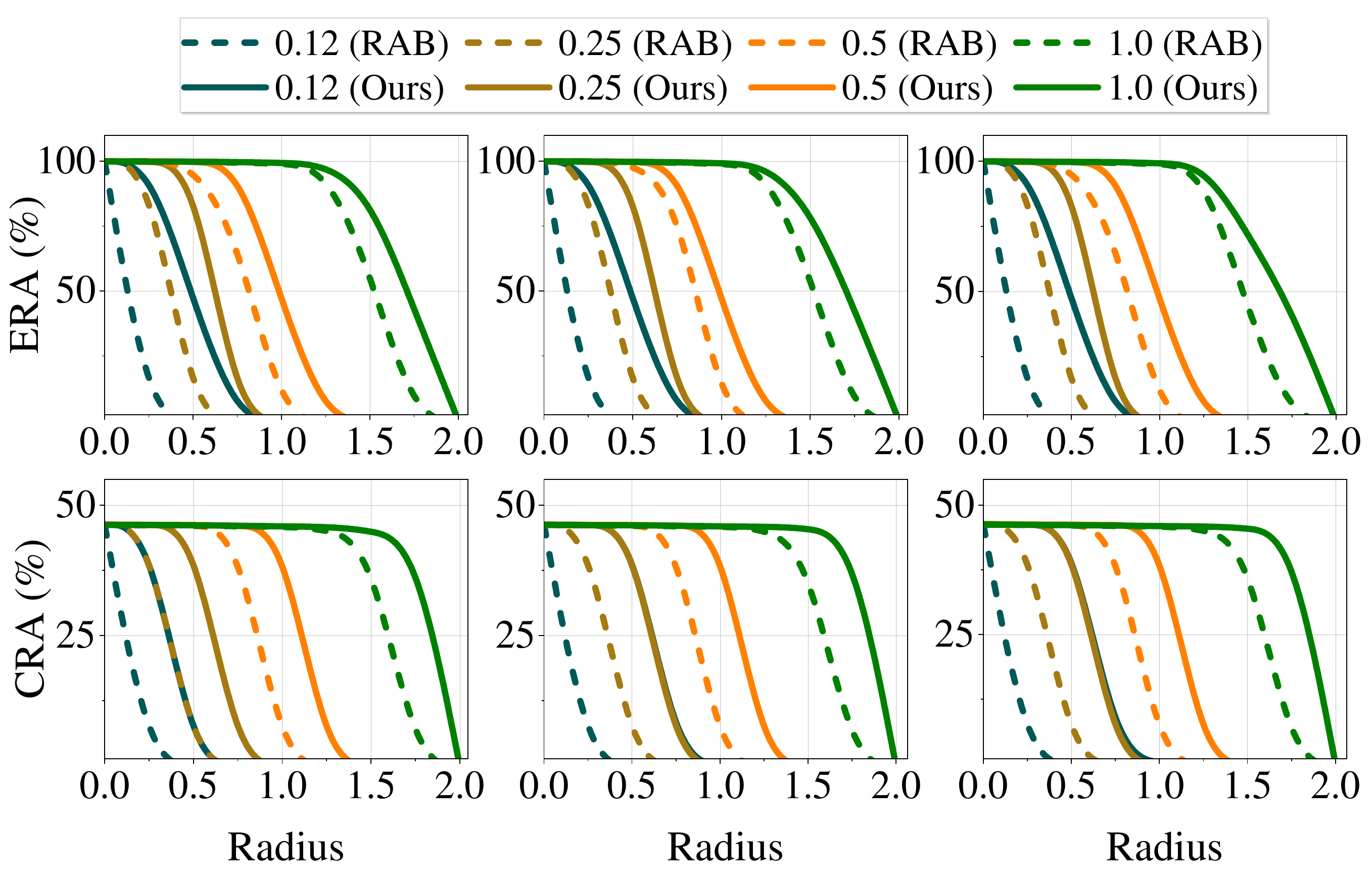}}\hspace{0.5em}
        \vspace{-0.5em}
	\caption{Certified performance (\ie, ERA, CRA) under different certification radii on the MNIST dataset in the all-to-one setting with various noise levels (0.12, 0.25, 0.5, and 1.0). The first column corresponds to the one-pixel attack, the second to the four-pixel attack, and the third to the blending attack.}
     \label{fig:cert_curve_ato_MNIST}
    \end{minipage}\hspace{0.5em}
    \begin{minipage}[t]{0.5\linewidth}
        \centering
    {
		\includegraphics[width=0.95\linewidth]{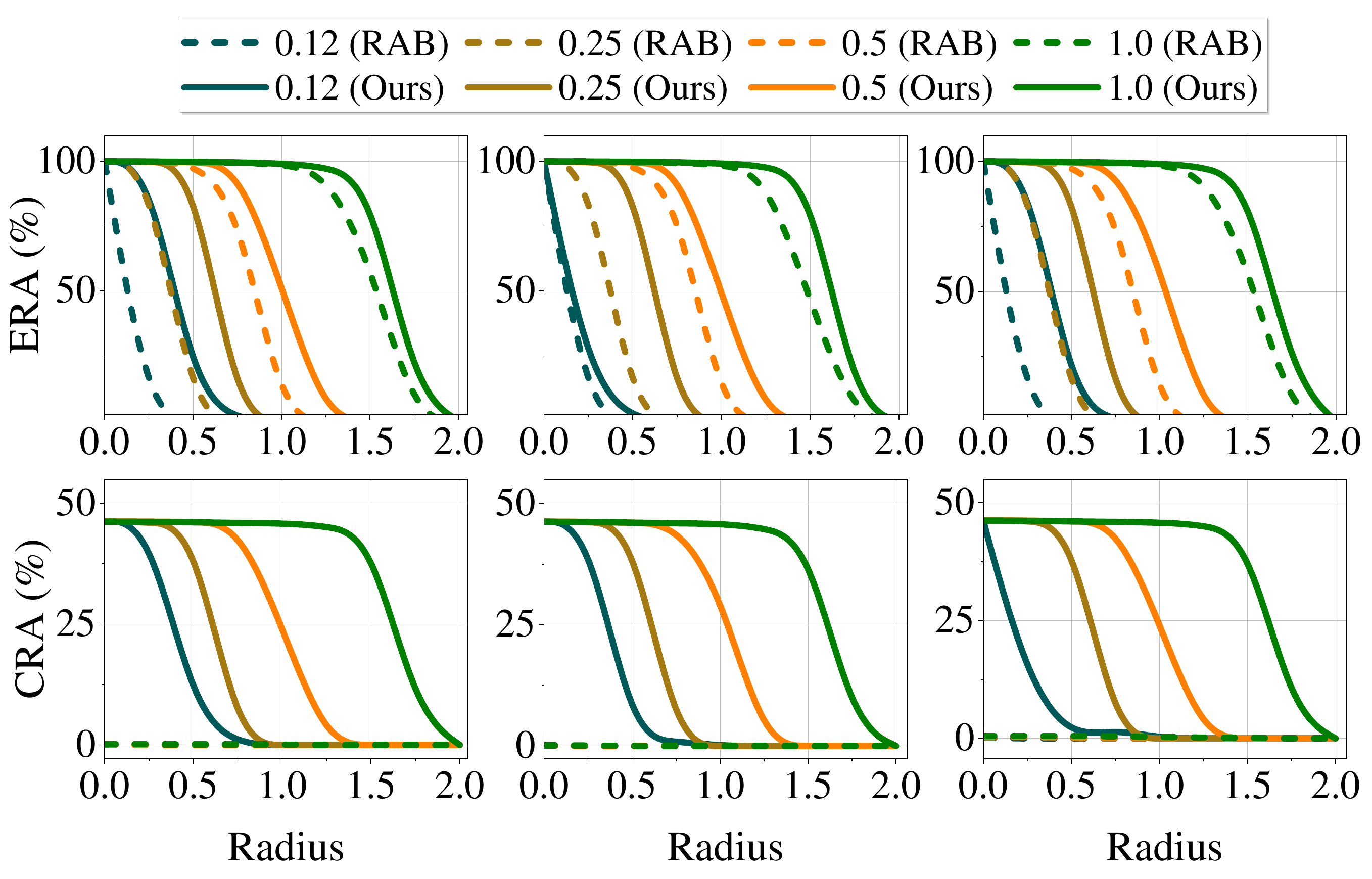}}\hspace{0.5em}
        \vspace{-0.5em}
	\caption{Certified performance (\ie, ERA, CRA) under different certification radii on the MNIST dataset in the all-to-all setting with various noise levels (0.12, 0.25, 0.5, and 1.0). The first column corresponds to the one-pixel attack, the second to the four-pixel attack, and the third to the blending attack.}
 \label{fig:cert_curve_ata_MNIST}
    \end{minipage}%
    
    \vspace{0.3em}
    \begin{minipage}[t]{0.5\linewidth}
        \centering
    {
		\includegraphics[width=0.95\linewidth]{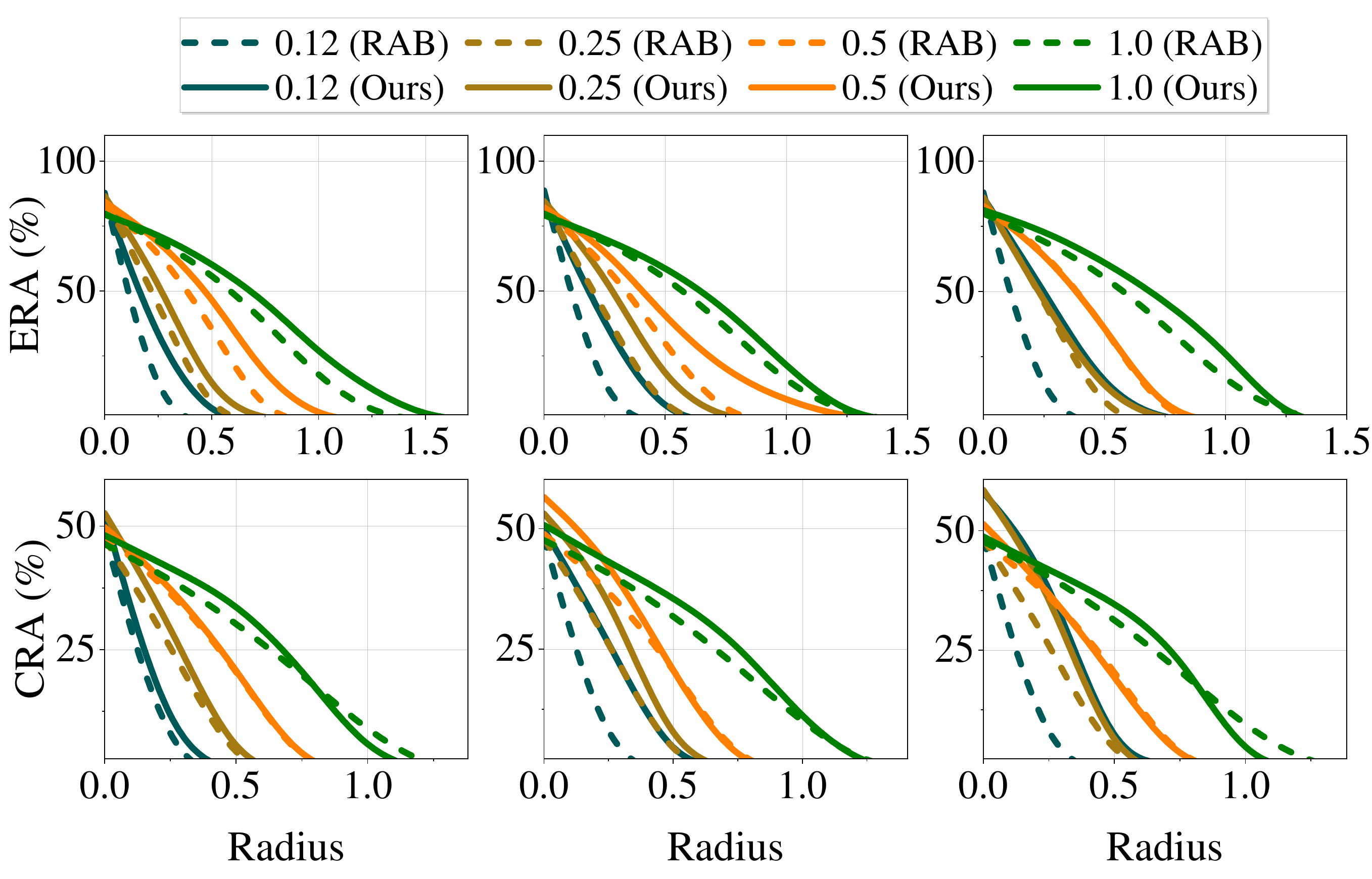}}\hspace{0.5em}
        \vspace{-0.5em}
	\caption{Certified performance (\ie, ERA, CRA) under different certification radii on the CIFAR-10 dataset in the all-to-one setting with various noise levels (0.12, 0.25, 0.5, and 1.0). The first column corresponds to the one-pixel attack, the second to the four-pixel attack, and the third to the blending attack.}
     \label{fig:cert_curve_ato_CIFAR}
    \end{minipage}\hspace{0.5em}
    \vspace{0.3em}
    \begin{minipage}[t]{0.5\linewidth}
        \centering
    {
		\includegraphics[width=0.95\linewidth]{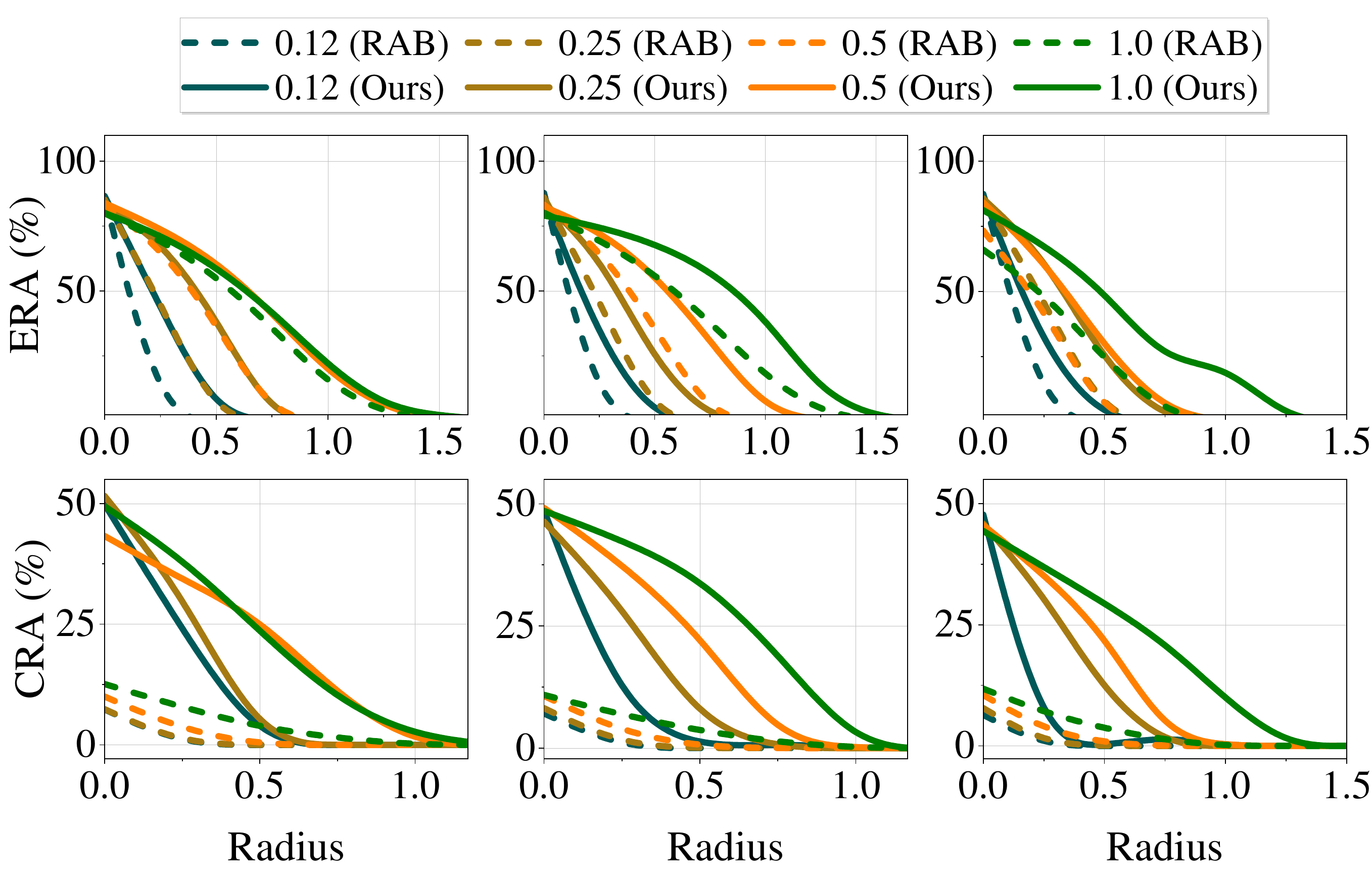}}\hspace{0.5em}
        \vspace{-0.5em}
	\caption{Certified performance (\ie, ERA, CRA) under different certification radii on the CIFAR-10 dataset in the all-to-all setting with various noise levels (0.12, 0.25, 0.5, and 1.0). The first column corresponds to the one-pixel attack, the second to the four-pixel attack, and the third to the blending attack.}
 \label{fig:cert_curve_ata_CIFAR}
    \end{minipage}%
\end{figure*}

 \subsection{Details of Storage-Update-Based Certification}
\label{app:stor_cert}

We hereby provide further details of the proposed storage-update-based certification method.
Under the sample-specific noise setting considered in this paper, traditional certification methods are no longer directly applicable. This limitation mainly stems from two core assumptions: \textbf{(1)} they typically assume that the certification regions associated with all inputs are globally non-overlapping; and \textbf{(2)} they rely on an initialized shared smoothing parameter (\ie, a fixed noise level $\sigma$ applied to all inputs). To address these issues, we propose a storage-update-based certification strategy that relaxes the above assumptions while still guaranteeing the reliability and soundness of the certification process. Before formally introducing this method, we first build upon the formal definitions of “overlapping” and “non-overlapping” certification regions given in Definition~\ref{def:inter_disjoint}, and systematically classify the possible relationships between certification regions in Definition~\ref{def:certification_region_classification}. Based on this classification, we then provide a complete formalization and analysis of the proposed storage-update-based certification mechanism (see Proposition~\ref{pro:Storage}).

\begin{definition}[Classification Criteria of Certification Regions]
\label{def:certification_region_classification}
Let the triplet storage set $\mathcal{S} = \{(\bm{x}_i, \mathcal{Y}_i, \mathcal{R}_i)\}_{i=1}^n$ store all previously predicted inputs $\bm{x}_i$, their corresponding predictions $\mathcal{Y}_i$, and their associated certification regions $\mathcal{R}_i$. Here, $\mathcal{R}_i$ denotes the certification region centered at $\bm{x}_i$, characterized by the certification radius $r_i$. The certification regions $\mathcal{R}_i$ for different inputs are classified as follows (see Figure~\ref{fig:certification}):
\begin{itemize}
\item \textbf{Case 1: Non-overlapping Certification Regions.}  
All certification regions are non-overlapping, \ie, $\forall i \neq j, \mathcal{R}_i \cap \mathcal{R}_j = \emptyset$, and the corresponding predictions are different, \ie, $\mathcal{Y}_i \neq \mathcal{Y}_j$.

\item \textbf{Case 2: Overlapping Certification Regions with Consistent Predictions.} The certification region $\mathcal{R}_{n+1}$ of a new input $\bm{x}_{n+1}$ overlaps with an existing certification region $\mathcal{R}_i$, and their predictions are consistent, \ie, $\exists  i$ such that $\mathcal{R}_i \cap \mathcal{R}_{n+1} \neq \emptyset$ and $\mathcal{Y}_{n+1} = \mathcal{Y}_i$.

\item \textbf{Case 3: Overlapping Certification Regions with Inconsistent Predictions.}  
The certification region $\mathcal{R}_{n+1}$ of a new input $\bm{x}_{n+1}$ overlaps with an existing region $\mathcal{R}_i$, but their predictions differ, \ie, $\mathcal{Y}_{n+1} \neq \mathcal{Y}_i$. This case can be further divided into two subcases:
\begin{itemize}
    \item The new input lies \emph{inside} the existing certification region, \ie, $\bm{x}_{n+1} \in \mathcal{R}_i$ and $\mathcal{R}_{n+1} \cap \mathcal{R}_i \neq \emptyset$.
    \item The new input lies \emph{outside} the existing certification region, \ie, $\bm{x}_{n+1} \notin \mathcal{R}_i$ but $\mathcal{R}_{n+1} \cap \mathcal{R}_i \neq \emptyset$.
\end{itemize}
\end{itemize}

\end{definition}

Based on the classification in Definition \ref{def:certification_region_classification}, we propose a storage-update-based certification method, that enforces non-overlapping certification regions while maintaining prediction consistency (\ie,  $\forall i \neq j, \mathcal{R}_i \cap \mathcal{R}_j = \emptyset, \mathcal{Y}_i \neq \mathcal{Y}_j$).  
In this way, the certification regions of inputs with different predicted labels do not overlapping within the storage set $\mathcal{S}$. This is a key property of a sound certification process.

Now, we introduce Proposition \ref{pro:Storage}, which formalizes the proposed storage-update-based certification method.

 \begin{table}[!t]
  \captionsetup{font=small}
  \caption{
  Abstain rate (\%) under different noise levels in the all-to-one setting.
Results are reported separately on \emph{clean} and \emph{poisoned} test samples.
We evaluate three types of attacks (\ie, one-pixel, four-pixel, and blending) on the
MNIST, CIFAR-10, and ImageNette datasets. The best (lowest) abstain rates are
highlighted in bold.
  }
  \vspace{-0.5em}
  \centering
  \setlength{\tabcolsep}{5.5pt}
  \renewcommand{\arraystretch}{1.15}
  \scalebox{0.85}{
    \begin{tabular}{c|c|c|ccc|ccc}
      \toprule
      \multirow{3}{*}{Dataset}
      & \multirow{3}{*}{Attack}
      & \multirow{3}{*}{Method}
      & \multicolumn{6}{c}{$\sigma$ (Abstain Rate$\downarrow$)} \\
      \cline{4-9}
      & & & \multicolumn{3}{c|}{Clean Test}
            & \multicolumn{3}{c}{Poisoned Test} \\
      \cline{4-9}
      & & & 0.25 & 0.5 & 1.0 & 0.25 & 0.5 & 1.0 \\
      \hline

      \multirow{6}{*}{MNIST}
      & \multirow{2}{*}{One-pixel}
      & RAB & 0.28  & \textbf{0.09}  & \textbf{0.14} & 0.33 & \textbf{0.09}  & \textbf{0.09} \\
      & & Ours &\textbf{0.14}  &\textbf{0.09}  &\textbf{0.14}  &\textbf{0.14}  & \textbf{0.09} & \textbf{0.09}  \\
      \cline{2-9}

      & \multirow{2}{*}{Four-pixel}
      & RAB & 0.33 &\textbf{0.14}  & 0.14 & 0.47 &\textbf{0.14}  &0.14  \\
      & & Ours & \textbf{0.14} & 0.19 &\textbf{0.05}  &\textbf{0.19}  &0.19  & \textbf{0.05} \\
      \cline{2-9}

      & \multirow{2}{*}{Blending}
      & RAB &0.43  &\textbf{0.09}  &\textbf{0.14}  &0.43  &\textbf{0.09}  & \textbf{0.09} \\
      & & Ours &\textbf{0.24}  &0.14  & \textbf{0.14} & \textbf{0.24} &0.14  &\textbf{0.09}  \\

      \hline

      \multirow{6}{*}{CIFAR-10}
      & \multirow{2}{*}{One-pixel}
      & RAB &14.95  &\textbf{9.05}  &\textbf{5.05}  &14.95  &8.50  &\textbf{5.50}  \\
      & & Ours &\textbf{14.80}  & 9.80 &6.80  &\textbf{10.90 } &\textbf{7.80}  &6.55  \\
      \cline{2-9}

      & \multirow{2}{*}{Four-pixel}
      & RAB &14.45  &9.10  &\textbf{4.95}  &14.65  &9.30  &\textbf{5.15}  \\
      & & Ours &\textbf{13.35}  &\textbf{7.95}  &6.60  &\textbf{13.20 } &\textbf{8.50}  &6.75  \\
      \cline{2-9}

      & \multirow{2}{*}{Blending}
      & RAB &14.84  &\textbf{8.90}  &\textbf{5.05}  &14.60  &9.15 &\textbf{5.10 }  \\
      & & Ours &\textbf{13.80}  &9.95  &5.60  &\textbf{13.90}  & \textbf{8.75} &5.20  \\

      \hline

      \multirow{6}{*}{ImageNette}
      & \multirow{2}{*}{One-pixel}
      & RAB &\textbf{4.26} &4.80  &5.36  &\textbf{4.80}  &5.36  &\textbf{4.24}  \\
      & & Ours &5.20  &\textbf{4.78}  &\textbf{5.22} &4.98  &\textbf{4.84}  & 5.04 \\
      \cline{2-9}

      & \multirow{2}{*}{Four-pixel}
      & RAB & \textbf{4.24} &5.08  &5.48  &\textbf{4.24}  &5.04  &5.48  \\
      & & Ours &4.76  &\textbf{4.98}  &\textbf{5.28}  &4.76  &\textbf{4.26}  &\textbf{4.84}  \\
      \cline{2-9}

      & \multirow{2}{*}{Blending}
      & RAB &\textbf{4.40}  &4.72  &5.54  &\textbf{4.40}  &4.72  &5.54  \\
      & & Ours &4.88  &\textbf{4.56}  &\textbf{5.50}  &4.88  &\textbf{4.48}  &\textbf{4.94}  \\

      \bottomrule
    \end{tabular}
  }
  \label{table:All-to-one_abstain_combined}
\end{table}
 
\begin{proposition}[Storage-update-based Certification]
\label{pro:Storage}
Based on Definition~\ref{def:certification_region_classification}, the storage-update-based certification method handles new inputs according to the following cases:
\begin{itemize}
    \item \textbf{Case 1:} If $\forall i \neq j$, $\mathcal{R}_i \cap \mathcal{R}_j = \emptyset$ and $\mathcal{Y}_i \neq \mathcal{Y}_j$, then all existing triplets $(\bm{x}_i, \mathcal{Y}_i, \mathcal{R}_i)$ and $(\bm{x}_j, \mathcal{Y}_j, \mathcal{R}_j)$ in storage remain unchanged.
    
    \item \textbf{Case 2:} If there exists some $i$ such that $\mathcal{R}_{n+1} \cap \mathcal{R}_i \neq \emptyset$ and $\mathcal{Y}_{n+1} = \mathcal{Y}_i$, then the new certification triplet $(\bm{x}_{n+1}, \mathcal{Y}_{n+1}, \mathcal{R}_{n+1})$ can be directly added to the storage.
    
    \item \textbf{Case 3:} If $\mathcal{R}_{n+1} \cap \mathcal{R}_i \neq \emptyset$ and $\mathcal{Y}_{n+1} \neq \mathcal{Y}_i$, the method proceeds as follows\footnote{This process is straightforward when the certification regions are $\ell_2$-balls.} (see Figure~\ref{fig:certification}):
    \begin{itemize}
        \item If $\bm{x}_{n+1} \in \mathcal{R}_i$: The new certification region is updated to the largest subset $\tilde{\mathcal{R}}_{n+1}$ such that $\tilde{\mathcal{R}}_{n+1} \subseteq \mathcal{R}_{n+1}$ and $\tilde{\mathcal{R}}_{n+1} \subseteq \mathcal{R}_i$. Then, $\mathcal{R}_{n+1}$ is replaced by $\tilde{\mathcal{R}}_{n+1}$, and the label $\mathcal{Y}_{n+1}$ is updated to $\mathcal{Y}_i$ to ensure prediction consistency.
        
        \item If $\bm{x}_{n+1} \notin \mathcal{R}_i$: The new certification region is updated to the largest subset $\mathcal{R}_{n+1}'$ such that $\mathcal{R}_{n+1}' \subseteq \mathcal{R}_{n+1}$ and $\mathcal{R}_{n+1}' \cap \mathcal{R}_i = \emptyset$. Then, original $\mathcal{R}_{n+1}$ is replaced by $\mathcal{R}_{n+1}'$.
    \end{itemize}
\end{itemize}

After applying the appropriate case, the final triplet $(\bm{x}_{n+1}, \mathcal{Y}_{n+1}, \mathcal{R}_{n+1})$ (or its updated form) is added to the storage set $\mathcal{S}$ for use in future certification.
\end{proposition}

\begin{remark}[Correctness and Scalability of Storage-update-based Certification]
The update rules preserve certification correctness because they perform necessary and local region shrinking on the newly constructed certified region only when conflicts across different predictions are detected. Concretely, in both subcases of Case~3, the updated region is always a subset of the original certified region, \ie,
$\tilde{\mathcal{R}}_{n+1} \subseteq \mathcal{R}_{n+1}$ or $\mathcal{R}'_{n+1} \subseteq \mathcal{R}_{n+1}$.
Since $\mathcal{R}_{n+1}$ is a certified region obtained from Theorem~\ref{theorem:robust_cert}, any subset of $\mathcal{R}_{n+1}$ remains a valid certified region (although potentially more conservative). Therefore, the update process cannot introduce false certification, while preserving certification tightness as much as possible.
Regarding scalability, the update mechanism only checks overlaps between the new certified region and the stored regions, and applies local region updates for conflicting indices. Since the process does not involve backtracking or global recomputation, it remains scalable in practice. The implementation is summarized in Algorithm~\ref{alg:storage_update}.
\end{remark}

In practice, in our experiments, we did not observe any cases where inputs with different predictions have overlapping certified regions. 
That is, for each input, the certified region stored in $\mathcal{S}$ is essentially determined by the certification radius computed using Eq. (\ref{eq:radius}) for the sample-specific smoothed classifier $g(\bm{x}, \mathcal{D}, \sigma_{\bm{x}}^*)$. 
This can be attributed to two main reasons: \textbf{1)} Due to the high dimensionality of image datasets, the $\ell_2$-norm distance between samples is significantly larger than the certification radius provided by randomized smoothing; 
\textbf{2)} The optimized noise $\sigma_{\bm{x}}^*$ tends to have a moderate value ($\sigma_{\bm{x}}^* \le 1.0$), resulting in relatively small certification regions. For example, the certification region corresponds to an $\ell_2$-ball with a radius of approximately $4\sigma_{\bm{x}}^*$, which is much smaller than the distances between samples in high-dimensional datasets (\eg, ImageNet). 
Nonetheless, overlaps between certified regions with different predicted labels can still arise in rare but realistic situations, especially when the data distribution contains atypical or ambiguous samples, making such overlaps more plausible in principle. Specifically, \textbf{(1)} label noise or annotation errors can cause a mismatch between semantics and labels, making nearby inputs receive different predictions; \textbf{(2)} boundary-adjacent ambiguous samples tend to have small margins and fragile certified regions; and \textbf{(3)} near-duplicate inputs can be extremely close in the input space, making overlaps more likely. In these cases, an explicit conflict-resolution rule is needed to keep the certification outcome unambiguous. Our storage-update-based certification provides this safeguard by appropriately adjusting the certified regions (and the associated predictions when necessary) to resolve potential conflicts, as formalized in Proposition~\ref{pro:Storage}.

 \vspace{0.3em}
\noindent \textbf{Potential Merits of the Storage-Update-Based Certification}. The storage-update-based certification mechanism is designed to resolve potential conflicts and ensure that the certification output remains well-defined under the sample-specific certification setting. Its potential merits include:
\textbf{(1)} As a conservative safeguard for sample-specific certification, it can still guarantee a well-defined and consistent certification output in rare or adversarially constructed cases (\eg, when conflicting certified regions arise); \textbf{(2)} It is a method-agnostic post-processing strategy that does not depend on a specific model architecture or smoothing implementation, but instead provides consistency guarantees under input-adaptive robustness settings. Therefore, it can be naturally integrated as a general module into other certification methods that employ adaptive robustness parameters; \textbf{(3)} It offers forward-looking support for more challenging application scenarios (\eg, label noise/annotation errors, ambiguous boundary-adjacent samples, near-duplicate or highly similar inputs, as well as low-dimensional inputs or larger noise-scale settings). In such scenarios, the mechanism can serve as a reliable safety component to handle potential conflicts in advance and avoid ambiguity in certification outcomes. Overall, storage-update-based certification is a general safety mechanism for edge-case consistency, not a prerequisite for our experimental results.

 \subsection{Additional Experimental Results}
\label{appC}     
 We hereby present additional experimental results, with all experimental settings consistent with those described in Section \ref{main_result_ato}. Figures \ref{fig:cert_curve_ato_MNIST}-\ref{fig:cert_curve_ata_MNIST} illustrate the certification curves for the MNIST dataset under the all-to-one and all-to-all settings, respectively. Figures \ref{fig:cert_curve_ato_CIFAR}-\ref{fig:cert_curve_ata_CIFAR} show the certification curves for the CIFAR-10 dataset under the all-to-one and all-to-all settings. The results are consistent with the conclusions in Sections~\ref{main_result_ato} and~\ref{main_result_ata}, demonstrating that our method maintains strong certification performance (\ie, empirical robust accuracy (ERA) and certified robust accuracy (CRA)) across different datasets. In particular, in the all-to-all setting, our method achieves a significant improvement in certified robust accuracy, further validating its effectiveness and generalization capability.

\subsection{Abstain Rate}
\label{appE}   

Following prior works~\cite{Cohen2019Certified,Weber2023RAB}, we report the abstain rates of Cert-SSBD under the all-to-one setting, evaluated at three noise levels ($\sigma_0 = 0.25, 0.5, 1.0$) and three attack types (one-pixel, four-pixel, and blending), on both clean and poisoned test sets. As shown in Table~\ref{table:All-to-one_abstain_combined}, Cert-SSBD exhibited abstain rates comparable to RAB across the evaluated datasets, attack types, and noise levels, with only slight variations. These results suggest that the performance gains of Cert-SSBD did not come at the cost of a substantially increased abstain rate.

\begin{table}[t]
  \captionsetup{font=small}
  \caption{Total runtime (minutes) of SGA-based noise optimization under different iterations $T$.}
  \vspace{-0.5em}
  \centering
  \setlength{\tabcolsep}{6pt}
  \renewcommand{\arraystretch}{1.1}
  \scalebox{0.96}{
    \begin{tabular}{c|c|c|cccc}
      \toprule
      \multirow{2}{*}{Dataset} 
      & \multirow{2}{*}{Model} 
      & \multirow{2}{*}{Split}
      & \multicolumn{4}{c}{Iterations $T$} \\
      \cline{4-7}
      & & & 10 & 50 & 100 & 150 \\
      \hline
      \multirow{2}{*}{MNIST} 
      & \multirow{2}{*}{CNN} 
      & Train & 0.25 & 1.26 & 2.54 & 3.79 \\
      & & Test  & 0.06 & 0.30 & 0.61 & 0.94 \\
      \hline
      \multirow{2}{*}{CIFAR-10} 
      & \multirow{2}{*}{ResNet-like} 
      & Train & 0.94 & 4.60 & 8.95 & 13.96 \\
      & & Test  & 0.24 & 1.20 & 2.59 & 3.89 \\
      \hline
      \multirow{2}{*}{ImageNette} 
      & \multirow{2}{*}{ResNet-18} 
      & Train & 4.86 & 24.02 & 48.98 & 73.38 \\
      & & Test  & 1.37 & 6.85 & 14.06 & 21.20 \\
      \bottomrule
    \end{tabular}
  }
  \label{tab:sga_T}
\end{table}

\begin{table}[t]
  \captionsetup{font=small}
  \caption{Runtime analysis of Cert-SSBD with different model architectures \cite{he2016deep} on CIFAR-10 dataset. We report noise optimization time on training/testing sets (with $T{=}1$), single model training time, and certification time.}
  \vspace{-0.5em}
  \centering
  \setlength{\tabcolsep}{6pt}
  \renewcommand{\arraystretch}{1.1}
  \scalebox{0.9}{
    \begin{tabular}{c|cc|c|c}
      \toprule
      \multirow{2}{*}{Model} 
      & \multicolumn{2}{c|}{Noise Opt. (seconds)}
      & Train 1 Model
      & Certify Testing Set \\
      \cline{2-3}
      & Train & Test & (seconds) & (minutes) \\
      \hline
      ResNet-18  & 15.46 & 3.00 & 13.51 & 20.27 \\
      ResNet-34  & 20.67 & 4.40 & 21.60 & 32.05 \\
      ResNet-50  & 24.58 & 5.11 & 21.90 & 32.49 \\
      ResNet-101 & 28.06 & 5.94 & 32.24 & 47.84 \\
      \bottomrule
    \end{tabular}
  }
  \label{tab:model_runtime}
\end{table}

\subsection{Detailed Runtime Analysis}
\label{appendix:runtime}

We hereby provide a more detailed runtime analysis of Cert-SSBD, including noise optimization time and training and certification time costs.

\vspace{0.3em}
\looseness=-1
\noindent \textbf{Noise Optimization Time.}
As shown in Table~\ref{tab:sga_T}, the SGA-based noise optimization scales approximately linearly with the iteration number $T$. With $T{=}100$, optimization takes about 9/2.6 minutes on the CIFAR-10 train/test sets and about 49 minutes on the ImageNette training set (224$\times$224). The per-sample optimization time is about 0.01s (MNIST), 0.05s (CIFAR-10), and 0.31s (ImageNette). Noise optimization is a one-time offline preprocessing step, and the optimized $\sigma_{\bm{x}}^*$ can be stored and reused. Table~\ref{tab:model_runtime} further shows that the optimization time increases from 15.46s (ResNet-18, 11M) to 28.06s (ResNet-101, 44M), \ie, 1.8$\times$ for a 4$\times$ parameter increase, suggesting sub-linear growth with respect to parameter count in our tested range.

\vspace{0.3em}
\noindent \textbf{Training and Certification Time.}
We hereby analyze the runtime costs of storage-update, certification inference, and model training.
As shown in Table~\ref{tab:runtime_breakdown}, the storage-update overhead is small (about 0.1 minutes on MNIST/CIFAR-10 and about 22 minutes on ImageNette). Under the same ensemble size $M$, certification inference time is comparable to RAB (e.g., 10.86 vs.\ 9.34 minutes on CIFAR-10, with about 16\% overhead). The single-model training time is nearly identical to RAB (difference $<2\%$), while training $M$ smoothed models dominates the overall cost (e.g., on CIFAR-10, $7.32\text{s}\times 1000 \approx 122$ minutes). Table~\ref{tab:model_runtime} shows that increasing the model from ResNet-18 to ResNet-101 leads to a 2.4$\times$ increase in single-model training time (13.51s$\to$32.24s) and a 2.4$\times$ increase in certification time (20.27$\to$47.84 minutes), which is below the 4$\times$ parameter increase, suggesting good scalability with respect to model size in our tested range.

\begin{table*}[t]
  \captionsetup{font=small}
  \caption{Runtime analysis of Cert-SSBD. We report the training time of a single smoothed model, the certification time on the entire testing set with $M$ ensemble models, and the storage-update-based certification overhead.}
  \vspace{-0.5em}
  \centering
  \setlength{\tabcolsep}{6pt}
  \renewcommand{\arraystretch}{1.1}
  \scalebox{0.9}{
    \begin{tabular}{c|c|c|cc|cc|c}
      \toprule
      \multirow{2}{*}{Dataset} 
      & \multirow{2}{*}{Model} 
      & \multirow{2}{*}{$M$}
      & \multicolumn{2}{c|}{Train 1 Model (seconds)}
      & \multicolumn{2}{c|}{Certify Testing Set (minutes)}
      & Storage-update-based Certification \\
      \cline{4-5} \cline{6-7}
      & & & Fixed $\sigma$ (RAB) & Optimized $\sigma_x^*$ (Ours) 
      & Fixed $\sigma$ (RAB) & Optimized $\sigma_x^*$ (Ours)
      & (minutes) \\
      \hline
      MNIST      & CNN         & 1000 & 1.60 & 1.63 & 2.63 & 3.35 & 0.08 \\
      CIFAR-10   & ResNet-like & 1000 & 7.23 & 7.32 & 9.34 & 10.86 & 0.09 \\
      ImageNette & ResNet-18   & 200  & 32.30 & 32.71 & 12.76 & 35.07 & 22.3 \\
      \bottomrule
    \end{tabular}
  }
  \label{tab:runtime_breakdown}
\end{table*}

\subsection{Details of the MAP Attack}
\label{app:MAP}

We hereby provide the detailed formulation of the Margin-Aware Adaptive Poisoning (MAP) attack introduced in Section~\ref{sec:discuss}. Specifically, given a benign dataset $\mathcal{D} = \{(\boldsymbol{x}_i, y_i)\}_{i=1}^n$, 
a testing set $\mathcal{D}_{test}$, a poisoning rate $\lambda$, and a trigger function $\tau(\cdot)$, the attacker's goal is to select a poisoning sample set $\mathcal{P}$ such that the certification performance on target samples degrades as much as possible:
\begin{equation}
\min_{\mathcal{P} \subset \mathcal{D},\, |\mathcal{P}| \leq \lambda n}
\sum_{\boldsymbol{x} \in \mathcal{T}}
r\!\left(g;\, \sigma_{\boldsymbol{x}}^*(\bm \theta_{\mathcal{P}}),\, 
\bm \theta_{\mathcal{P}}\right),
\end{equation}
\looseness=-1
where $\sigma_{\boldsymbol{x}}^*(\bm \theta_{\mathcal{P}})$ denotes the sample-specific noise scale automatically optimized by the defender via SGA under model parameters $\bm \theta_{\mathcal{P}}$, which is determined by the 
defense mechanism and cannot be directly accessed or manipulated by the attacker; $\mathcal{T} \subset \mathcal{D}_{test}$ is the set of target samples (vulnerable samples); and $\bm \theta_{\mathcal{P}}$ denotes the model 
parameters trained on the poisoned dataset $\mathcal{D}_p$. The attacker can only indirectly influence $\bm \theta_{\mathcal{P}}$ by selecting $\mathcal{P}$, 
which in turn affects the optimization result of $\sigma_{\boldsymbol{x}}^*$.

Since directly optimizing this objective is computationally prohibitive, we adopt a two-stage heuristic strategy.

\textbf{Stage 1: Vulnerable Sample Identification.} We train a base model $f_{\theta_0}$ on the standard poisoned dataset 
$\mathcal{D}_p = \mathcal{D}_m(\boldsymbol{\delta}, \hat{y}) \cup \mathcal{D}_b$, and use the logit margin defined in Definition~\ref{def:boundary_sample} as a proxy measure for the distance to the decision boundary:
\begin{equation}
m(\boldsymbol{x}) = |\phi_y(\boldsymbol{x}; \boldsymbol{w})| =
\left|f_y(\boldsymbol{x}; \boldsymbol{w}) - \max_{y' \neq y} 
f_{y'}(\boldsymbol{x}; \boldsymbol{w})\right|.
\end{equation}

\begin{figure}[ht!]
    \vspace{-0.5em}
    \centering
    \includegraphics[width=0.5\textwidth]{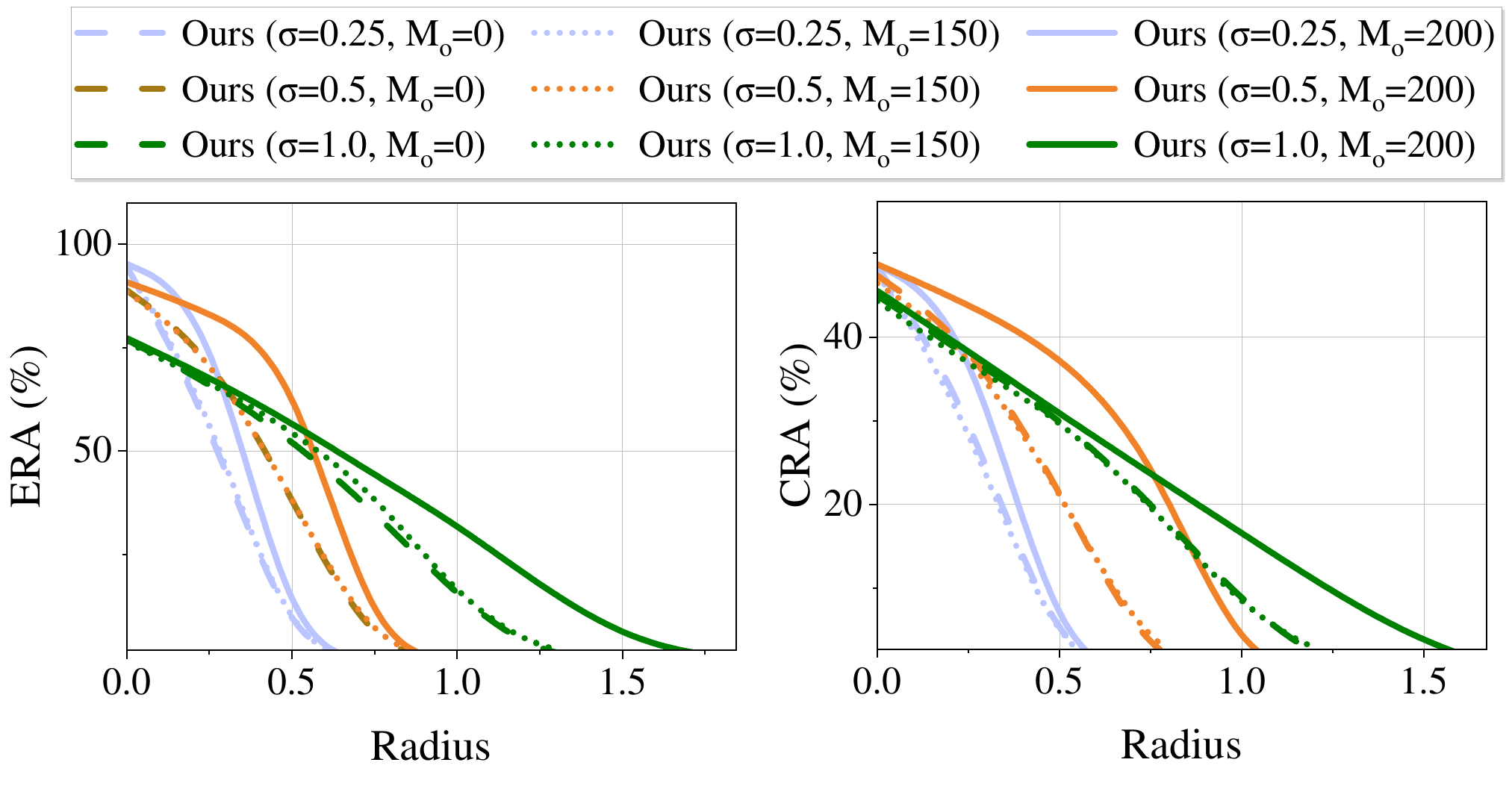}
    \vspace{-1.0em}
    \caption{Effect of Optimized-Noise Model Count $M_o$.}
    \vspace{-1.5em}
    \label{fig:Pre-train}
    
\end{figure}

We select the $k$ testing samples with the smallest margins as the vulnerable target set:
\begin{equation}
\mathcal{T} = \underset{S \subset \mathcal{D}_{test},\, |S|=k}{\arg\min}
\sum_{\boldsymbol{x} \in S} m(\boldsymbol{x}).
\end{equation}

A smaller margin indicates that the sample is closer to the decision boundary, and thus its corresponding sample-specific noise optimization and certification process are more susceptible to boundary perturbations.

\textbf{Stage 2: Poisoning Sample Selection.} Among non-target-class training 
samples $\mathcal{D}^{-} = \{(\boldsymbol{x}_i, y_i) \in \mathcal{D} : y_i \neq 
\hat{y}\}$, we select those with the smallest feature distances to the vulnerable target set for poisoning:
\begin{equation}
\mathcal{P}^* = \underset{\mathcal{P} \subset \mathcal{D}^{-},\, 
|\mathcal{P}|=\lambda n}{\arg\min} \sum_{\boldsymbol{x}_i \in \mathcal{P}}
\min_{\boldsymbol{x}_t \in \mathcal{T}} \|\boldsymbol{x}_i - \boldsymbol{x}_t\|_2.
\end{equation}

Finally, we construct the poisoned dataset:
\begin{equation}
\mathcal{D}_p^{\text{MAP}} = \{(\boldsymbol{x}, y) : \boldsymbol{x} \notin 
\mathcal{P}^*\} \cup \{(\tau(\boldsymbol{x}), \hat{y}) : \boldsymbol{x} \in 
\mathcal{P}^*\},
\end{equation}
where $\hat{y} = G_Y(y)$ is the poisoned label generator specified by the attacker (defined in Section~\ref{sec:pre}), with $G_Y(y) = y_t$ and $y_t \in \mathcal{Y}$ being the target label.

   \subsection{Additional Ablation Study}
\label{app:model_count}
\vspace{0.3em}
\noindent \textbf{Effect of Optimized-Noise Model Count $\bm{M_o}$}. 
Considering that the final prediction is obtained through an ensemble of multiple models, we further investigate the impact of the number of optimized-noise models on certification performance (\ie, ERA and CRA) under different certification radii. Specifically, we trained 50 models with fixed noise $\sigma_0$ and 150 models with optimized noise $\sigma_{\bm{x}}^*$, forming an ensemble of 200 models (\ie, $M_f = 50$, $M_o = 150$). This setup is compared against two baselines: one where all models are trained with fixed noise (\ie, $M_f = 200$, $M_o = 0$), and another where all models are trained with optimized noise (\ie, $M_f = 0$, $M_o = 200$). We evaluate the certification performance of these three settings under various noise levels (\ie, $\sigma = 0.25, 0.5, 1.0$) and across different certification radii. As shown in Figure~\ref{fig:Pre-train}, the ensemble trained entirely with optimized noise achieves significantly higher ERA and CRA at all certification radii, compared to those incorporating a portion of fixed-noise models. These results indicate that increasing the number of optimized-noise models helps improve the robustness of the ensemble, while introducing fixed-noise models may limit the overall performance.

\end{document}